\documentclass{jfm}

\usepackage{amsmath,mathtools,mathabx}
\usepackage{graphicx,color}
\usepackage{esdiff}
\usepackage{epstopdf, epsfig}
\usepackage{breqn}

\shorttitle{Controlling interfacial instabilities in Hele-Shaw flow}
\shortauthor{L. C. Morrow, T. J. Moroney, and S. W. McCue}

\title{Numerical investigation of controlling interfacial instabilities in non-standard Hele-Shaw configurations}

\author{Liam C. Morrow, Timothy J. Moroney,	\and Scott W. McCue
	\corresp{\email{scott.mccue@qut.edu.au}}}

\affiliation{School of Mathematical Sciences, Queensland University of Technology, Brisbane QLD 4001, Australia
}

\begin{document}
	
\maketitle

\begin{abstract}
	Viscous fingering experiments in Hele-Shaw cells lead to striking pattern formations which have been the subject of intense focus among the physics and applied mathematics community for many years.  In recent times, much attention has been devoted to devising strategies for controlling such patterns and reducing the growth of the interfacial fingers.  We continue this research by reporting on numerical simulations, based on the level set method, of a generalised Hele-Shaw model for which the geometry of the Hele-Shaw cell is altered. First, we investigate how imposing constant and time-dependent injection rates in a Hele-Shaw cell that is either standard, tapered or rotating can be used to reduce the development of viscous fingering when an inviscid fluid is injected into a viscous fluid over a finite time period.  We perform a series of numerical experiments comparing the effectiveness of each strategy to determine how these non-standard Hele-Shaw configurations influence the morphological features of the inviscid-viscous fluid interface. Surprisingly, a converging or diverging taper of the plates leads to reduced metrics of viscous fingering at the final time when compared to the standard parallel configuration, especially with carefully chosen injection rates; for the rotating plate case, the effect is even more dramatic, with sufficiently large rotation rates completely stabilising the interface.  Next, we illustrate how the number of non-splitting fingers can be controlled by injecting the inviscid fluid at a time-dependent rate while increasing the gap between the plates. Our simulations compare well with previous experimental results for various injection rates and geometric configurations. We demonstrate how the number of non-splitting fingers agrees with that predicted from linear stability theory up to some finger number; for larger values of our control parameter, the fully nonlinear dynamics of the problem lead to slightly fewer fingers than this linear prediction.
\end{abstract}

\section{Introduction}
	
\begin{figure}
\centering
\includegraphics[width=0.6\linewidth]{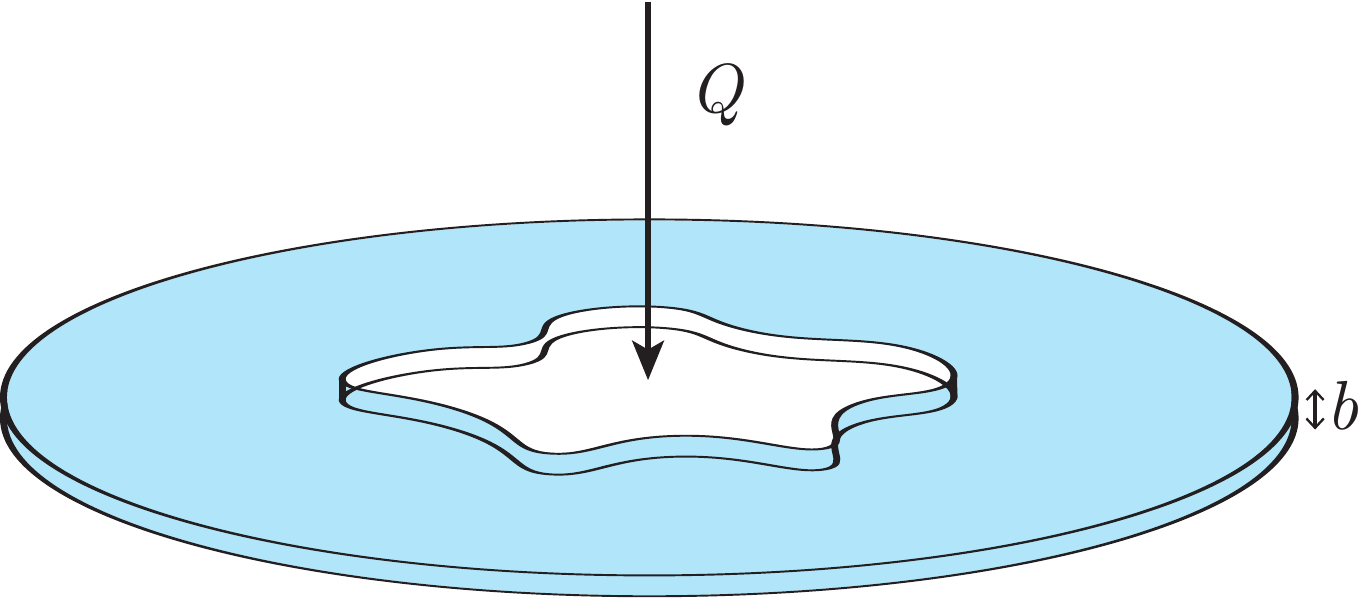}
\caption{Illustration of the standard Hele-Shaw cell experiment with two parallel plates separated by a small gap filled with viscous fluid (shaded). An inviscid fluid (white) is injected into the viscous fluid at a rate $Q$.  The immiscible fluids are separated by a sharp interface, which becomes increasingly unstable as it expands, forming distinct viscous fingering patterns characterised by their branching and tip-splitting morphology.}
\label{fig:Figure1}
\end{figure}

A standard Hele-Shaw cell is an experimental device (figure \ref{fig:Figure1}) consisting of two parallel plates separated by a small gap filled with a viscous fluid.  Fluid flow in this device has received significant attention largely due to the interfacial patterns that form when an inviscid fluid is injected into the viscous fluid.  These viscous fingering patterns form due to the Saffman-Taylor instability \citep{Saffman1958}, and are characterised by their distinctive branching and tip-splitting behaviour.  Closely related interfacial instabilities appear in a wide variety of phenomena, including saturated flow in porous media \citep{Homsy1987}, the growth of bacterial colonies \citep{Ben1992}, crystal solidification \citep{Mullins1988}, and fractal growth due to diffusion limited aggregation \citep{Witten1983} amongst others, and the Hele-Shaw framework is often used as a model to describe these processes \citep{Ben1990,Liang1986,Li2004,Mirzadeh2017}.

In practice, the presence of fingering instabilities may be undesirable, for example in the application of oil recovery.  As such, there is a significant body of research devoted to devising strategies for controlling the pattern formation and/or suppressing the viscous fingering \citep{Rabbani2018}.  The majority of these studies consider injecting the inviscid fluid at a time-dependent flow rate (linearly increasing in time \citep{Dias2012}, piecewise constant \citep{Dias2010} and sinusoidal \citep{Lins2017}), while in recent times researchers have proposed to alter the geometry of the Hele-Shaw cell to affect the fingering pattern.  Examples of such alterations include separating the plates in a time-dependent fashion \citep{Zheng2015,Vaquero2019}, tapering the Hele-Shaw plates so that they are no longer parallel \citep{Al2012,Al2013b,Anjos2018,Bongrand2018,Dias2013,Jackson2017,Lu2018,Stone2017}, and replacing one of the plates with an elastic membrane \citep{Al2013,Lister2013,Pihler2013,Pihler2014,Pihler2018}.  In the present study, we shall report on fully nonlinear simulations of scenarios which combine some of these non-standard configurations, computed using a numerical scheme based on the level set method.

In terms of motivating our numerical scheme, it is worth emphasising that the majority of mathematical studies concerned with the viscous fingering in non-standard Hele-Shaw geometry are performed using linear stability analysis.  While this technique provides a valuable tool for studying nonlinear problems, its relevance is restricted to sufficiently small times such that the interface is near circular and nonlinear effects are negligible.  Thus, we must resort to numerical techniques to study the long term and nonlinear behaviour of the solutions. Typically, numerical solutions to the standard Hele-Shaw problem are found using the boundary integral method, which requires that the pressure of the viscous fluid be harmonic \citep{Dai1993,Degregoria1986,Li2009,Shelley1997}. However, in general this is no longer true when the gap between the plates becomes a function of time and/or space, and thus more flexible schemes are required.  As such, we shall employ a robust numerical scheme based on the level set method, presented in Appendix~\ref{sec:NumericalScheme}.  This scheme is capable of describing the complex interfacial patterns which develop in the Hele-Shaw cell, and produces solutions consistent with experimental results. From these simulations, we provide insight into how altering the time-dependent injection rate and the physical geometry of a Hele-Shaw cell influences the evolution of the inviscid-viscous fluid interface which extend beyond the limitations of linear stability analysis.

In this article, we consider two broad objectives for controlling viscous fingering instabilities in a Hele-Shaw experiment. The first involves injecting a prescribed amount of inviscid fluid over a finite period of time in order to reduce the development of the fingering pattern according to standard metrics of how round (or close to a circle) an interface is.  We allow for both constant and time-dependent injection rates of inviscid fluid and, in addition to the standard Hele-Shaw geometry, we study examples for which the plates are tapered (either converging or diverging in the direction of flow) or rotating.  We present a number of new findings.  For the standard Hele-Shaw cell with parallel plates, we explore the proposed optimal injection rate of \cite{Dias2012} and determine how effective this strategy is over a range of parameter values, including a number of examples in which there is significant fingering (the only numerical example of this strategy provided by \cite{Dias2012} involved a near-circular interface).  For the case of tapered plates, we extend the work of \cite{Al2013b,Bongrand2018}, which involved experiments and linear stability theory, by performing a series of numerical simulations over a wider range of injection rates and taper angles.  We find that our new optimal injection rate appears to noticeably reduce the fingering pattern (via a reduction in the isoperimetric and circularity metrics) for the converging case, producing an atypical fingering pattern with short and stubby fingers (which appear similar to those observed by \cite{Pihler2012,Pihler2013}).  On the other hand, for the diverging geometry, this optimal injection rate also appears to reduce the instability, although with a much less dramatic effect.  Finally, the case of rotating plates with injection of inviscid fluid has not been considered previously in the literature.  Here, we explore cases in which fingers initially develop in the usual way; however, the centrifugal force acts to stabilise the interface so that the bubble ends up tending to a circle in shape in the long time limit. Physically speaking, this effect is due to the centrifugal force propelling the dense fluid outward which stabilises the interface.

The second objective with which we shall be concerned involves adjusting the flow rate and the geometry of the experimental apparatus in an attempt to prevent ongoing tip splitting so that the pattern evolves with a predetermined number of fingers.  For example, numerical, weakly nonlinear and experimental studies indicate that for a standard Hele-Shaw cell, an injection rate with the scaling $Q \sim t^{-1/3}$ can produce $N$-fold symmetric bubbles whose shape is independent of the initial condition, and can be controlled by the strength of the injection rate \citep{Brener1990,Dias2010b,Li2004,Li2009}.  Analogous results are suggested in the study of \citet{Zheng2015}, who apply linear stability analysis and experimental results to provide evidence that a constant number of fingers should develop if the parallel plates are separated via the scaling $b\sim t^{1/7}$. We extend this work by providing numerical evidence confirming that the number of non-splitting fingers can be controlled by implementing a more complicated time-dependent injection rate at the same time as separating the plates, as proposed by \citet{Zheng2015}.  Further, our simulations provide insight into how interactions between neighbouring fingers can influence the evolution of the interface extending beyond linear stability analysis; indeed, the number of non-splitting fingers is observed to be less than that predicted by linear stability analysis for a sufficiently large control parameter.  Our numerical results here are consistent with the  very recent findings of \cite{Vaquero2019}, who use a finite element scheme to also explore the $b\sim t^{1/7}$ scaling.

The outline of this paper is as follows. In \S~\ref{sec:ModelFormulation}, we summarise a generalised model for Hele-Shaw flow in a non-standard geometry, for which the gap between the plates depends on both time and space and the plates are allowed to rotate.  In \S~\ref{sec:LinearStabilityAnalysis}, we compare numerical simulations for the standard Hele-Shaw configuration (parallel stationary plates with constant injection rate) with experimental results and predictions from linear stability analysis.  We show that the numerical simulations agree with predictions from linear stability analysis for small time, and reproduce key morphological features observed in experiments for large time. In \S~\ref{sec:MinisingInstabilities}, we consider the objective of injecting a prescribed amount of fluid over fixed period of time for the different geometric configurations of parallel, tapered or rotating plates.  Subsequently, in \S~\ref{sec:NumberOFingers}, we study the objective of preventing tip-splitting and controlling the number of viscous fingers by carefully altering the time-dependent gap and/or injection rate.  Finally, in \S~\ref{sec:Conclusion} we conclude by discussing the results and suggesting possibilities for future work.

\section{Mathematical model} \label{sec:ModelFormulation}

\subsection{Governing equations}
		
The geometry we consider involves the injection of an incompressible inviscid fluid (with flow rate $Q$) through a small orifice in the centre of a Hele-Shaw cell otherwise filled with a viscous fluid (figure \ref{fig:Figure1}).  We assume the two fluids are immiscible and denote the simply connected domain of inviscid fluid by $\Omega(t)$ and the interface between the two fluids by $\partial\Omega$.  In our model, the viscous fluid is infinite in its extent, and so while the problem is driven by injection of an inviscid fluid at a point, we can also interpret the flow as being driven by a suction of viscous fluid from infinity.  A feature of our model is that we allow the small gap between the plates, $b$, to depend on both space and time.

We use a two-dimensional model of Hele-Shaw flow in a rotating frame that is derived by averaging Stokes flow over the small gap between the plates.  Denoting $\hat{p}$, $\boldsymbol{v}$, $\mu$, and $\rho$ as the pressure, velocity, viscosity, and density of the viscous fluid, the governing field equations modified to incorporate Hele-Shaw plates rotating at angular velocity $\hat{\omega}$ are \citep{Carrillo1999}
\begin{align}
	\boldsymbol{v} &= -\frac{b^2}{12\mu} \left( \nabla \hat{p} - \hat{\omega}^2 \rho r \boldsymbol{e}_r  \right) , &&\boldsymbol{x} \in \mathbb{R}^2 \backslash \Omega(t).                    \label{eq:Model0}
\end{align}
We note that the only effect of the rotation considered is the centrifugal force, and the Coriolis force is neglected. The centrifugal term in \eqref{eq:Model0} is removed by introducing $p = \hat{p} - \hat{\omega}^2 \rho r^2 / 2$, and thus we have
\begin{align}
	\boldsymbol{v} &= -\frac{b^2}{12\mu} \nabla p, &&\boldsymbol{x} \in \mathbb{R}^2 \backslash \Omega(t),                    \label{eq:Model1} \\
	\nabla \cdot \left(  b \boldsymbol{v} \right) &= -\frac{\partial b}{\partial t}.   &&\boldsymbol{x} \in \mathbb{R}^2 \backslash \Omega(t),       \label{eq:Model2}
\end{align}
noting that for all the cases we consider $\partial b / \partial t$ is spatially uniform. Equation \eqref{eq:Model1} is analogous to Darcy's law, which provides an intimate connection between Hele-Shaw flow and porous media flow \citep{Homsy1987}.  Equation \eqref{eq:Model2} ensures that the fluid's volume is conserved, and reduces to the traditional divergence free condition, $\nabla \cdot \boldsymbol{v} = 0$, in the standard configuration for which the plates are parallel and stationary. Note that we shall ignore the pressure gradients in the inviscid domain $\Omega(t)$, which makes this a one-phase Hele-Shaw model. The pressure of the bubble is taken as the reference pressure, so $p = 0$ at all times in the bubble.
	
\begin{figure}
	\centering
	\includegraphics[width=0.5\linewidth]{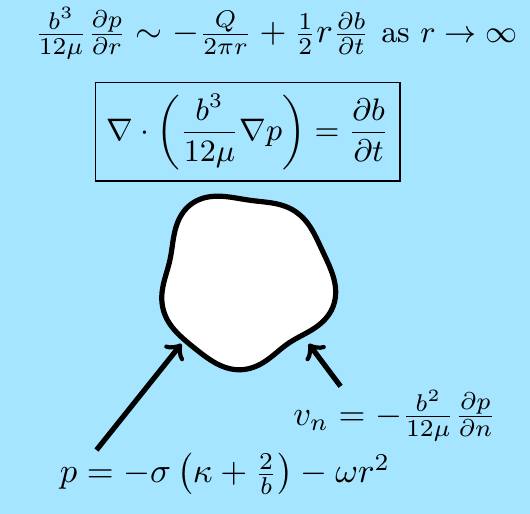}
	\caption{A schematic of our generalised Hele-Shaw model with spatially and/or temporally dependent plate gap thickness and rotating plates \eqref{eq:Model1}-\eqref{eq:Model6}. The viscous fluid is represented by the shaded (blue) region and the inviscid bubble is represented by the white region.}
	\label{fig:Figure2}
\end{figure}

By substituting \eqref{eq:Model1} into \eqref{eq:Model2}, we have the Reynolds lubrication equation 
\begin{align}
	\nabla \cdot \left(  \frac{b^3}{12 \mu} \nabla p \right)  &= \frac{\partial b}{\partial t}, &&\boldsymbol{x} \in \mathbb{R}^2 \backslash \Omega(t)       \label{eq:Model3}.
\end{align}
The boundary conditions on the interface are
\begin{align}
	p &= -\sigma \left(  \kappa + \frac{2}{b} \right) - \omega r^2,  &\boldsymbol{x} &\in \partial \Omega(t)             \label{eq:Model4}, \\
	v_n &= -\frac{b^2}{12 \mu} \frac{\p p}{\p n},       &\boldsymbol{x} &\in \partial \Omega(t)     		 \label{eq:Model5},
\end{align}
where the centrifugal parameter $\omega = \rho \hat{\omega}^2/2$. The dynamic boundary condition \eqref{eq:Model4} incorporates the effects of surface tension via the Young-Laplace equation, where $\sigma$ is the surface tension parameter and $\kappa$ is the signed curvature of the interface in the lateral direction.  The term $2/b$ in \eqref{eq:Model4} represents the curvature in the transverse direction for the case where the fluid is perfectly wetting \citep{Mclean1981}.  The kinematic boundary condition \eqref{eq:Model5} equates the velocity of the interface to the velocity of the viscous fluid on the interface. We note that both viscous stresses and the effect of a thin wetting film left behind by the viscous fluid are ignored.
The far-field boundary condition is
\begin{align}
	\frac{b^3}{12 \mu}\frac{\partial p}{\partial r} &\sim -\frac{Q}{2 \pi r} + \frac{1}{2} r \frac{\partial b}{\partial t} && r \to \infty,					\label{eq:Model6}
\end{align}
where $Q$ is a time-dependent flow-rate at which the inviscid fluid is injected. This form of the far-field boundary condition ensures the rate of change of volume of the inviscid bubble is indeed given by $Q$. Our model is summarised by a schematic in figure~\ref{fig:Figure2}.

We note in passing that the complementary geometry with viscous fluid in $\Omega(t)$ and inviscid fluid in $\mathbb{R}^2 \backslash \Omega(t)$ (that is, the opposite case with the fluids swapped) has attracted interest in the literature.  For that scenario, both the lifting and centrifugal configurations produce fingers which appear to be distinct from traditional Saffman-Taylor fingers.  These problems with the complementary geometry have been studied through a combination of experimental, analytical, and numerical techniques \citep{Alvarez2004,Anjos2017,Carrillo1996,Carrillo1999,Chen2005,Dias2010b,Gadelha2004,Lindner2005,Nase2011,Shelley1997}, but will not be pursued here.

\subsection{Numerical scheme}

Many numerical schemes used to study viscous fingering in a standard Hele-Shaw cell, where the governing equation for pressure \eqref{eq:Model3} reduces to Laplace's equation $\nabla^2 p = 0$, implement a boundary integral method \citep{Dai1993,Degregoria1986,Li2009,Shelley1997}. However, for non-standard Hele-Shaw configurations, i.e.\ when the gap between the plates is spatially and/or temporally dependent, the pressure is no longer harmonic and boundary integral methods become a less desirable option. For our study, we utilise the level set method, proposed by \cite{Osher1988}, which is based around the idea of representing interfaces implicitly as the zero level set of a higher dimensional hypersurface. Other flexible numerical schemes based on front tracking methods have been used to study viscous fingering in non-standard Hele-Shaw cells \citep{Fast2004,Pihler2014}. One advantage of the level set method is that it can be used describe the evolution of complex interfacial patterns using a uniform grid, eliminating the need to generate meshes that adapt as the interface evolves. The level set method has previously been used to study interfacial instabilities in a variety of moving boundary problems, including Hele-Shaw flow \citep{Hou1997,Lins2017} and Stefan problems \citep{Chen1997}.  We summarise the details of our scheme in Appendix~\ref{sec:NumericalScheme}.

\section{Review of standard configuration} \label{sec:LinearStabilityAnalysis}
	
Most mathematical studies investigating the influence of manipulating the geometry of the Hele-Shaw cell on viscous fingering are performed using linear stability analysis.  While this approach provides a useful tool for understanding the qualitative behaviour of solutions, as well as for deriving strategies for controlling viscous finger development, it is only accurate for small time and, as such, does not capture the full nonlinear dynamics of the problem. In this section, we review linear stability analysis for the standard Hele-Shaw problem where the plates are parallel and stationary ($b$ constant) and the inviscid fluid is injected at a constant rate $Q$. Further, we show that our numerical simulations are consistent with predictions made by linear stability analysis when time is sufficiently small, and can accurately reproduce experimental results for longer times.

Considering \eqref{eq:Model1}-\eqref{eq:Model6} in polar coordinates $(r, \theta)$ with $p = p(r, \theta, t)$ and the interface $\partial\Omega$ denoted by $r = s(\theta, t)$, we assume a perturbed circular solution
\begin{align}
	p(r, \theta, t) &= p_0(r,t) + \varepsilon \sum_{n=2}^{\infty} P_n(t) r^{-n} \cos n \theta + \mathcal{O}(\varepsilon^2), \label{eq:LSA1} \\
	s(\theta, t) &= s_0(t) + \varepsilon \sum_{n=2}^{\infty} \gamma_n(t) \cos n \theta + \mathcal{O}(\varepsilon^2),        \label{eq:LSA2}
\end{align}
where $\varepsilon \ll 1 $. The leading order radius of the interface becomes
$$
s_0 = \left(s_0(0)^2 + \frac{Qt}{\pi b}\right)^{1/2}.
$$
The resulting differential equation for the $n$th mode of perturbation is \citep{Paterson1981}
\begin{align} \label{eq:LSAclassic}
	\frac{\dot{\gamma}_n}{\gamma_n} = \frac{n-1}{s_0} \left( \frac{Q}{2 \pi b s_0} - \frac{n (n+1)b^2 \sigma}{12 \mu s_0^2} \right),
\end{align}
where the most unstable mode of perturbation, $n_{\max}$, is predicted to be
\begin{align} \label{eq:UnstableMode}
	n_{\max} = \sqrt{ \frac{1}{3} \left(  1 + \frac{6 \mu Q s_0}{\pi \sigma b^3} \right)  }.
\end{align}
Equation \eqref{eq:UnstableMode} comes from setting $\partial (\dot{\gamma}_n /\gamma_n)/ \partial n = 0$ and solving for $n$.  As such, $n_{\max}$ is not an integer, and so in practice the most unstable mode is the closest integer to $n_{\max}$.  Note that, given $s_0$ is an increasing function of time, then $n_{\max}$ also increases in time, which means the most unstable mode predicted by linear stability is a dynamic property (while not strictly relevant for fully nonlinear pattern formation, this observation is closely related to the ongoing tip-splitting that occurs for longer times).

\begin{figure}
	\centering
	\includegraphics[width=0.28\linewidth]{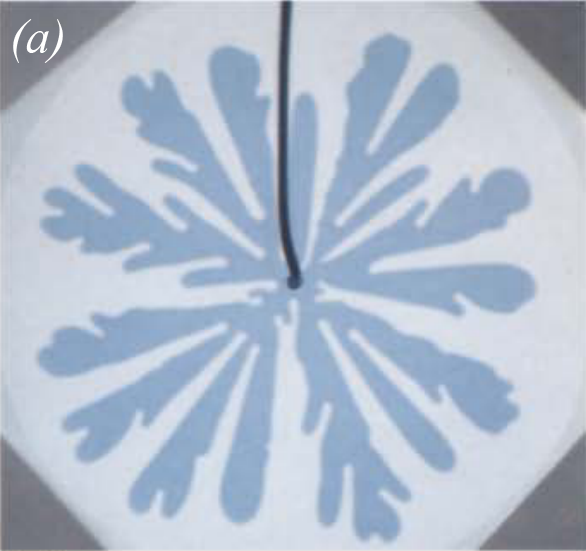}
	\includegraphics[width=0.28\linewidth]{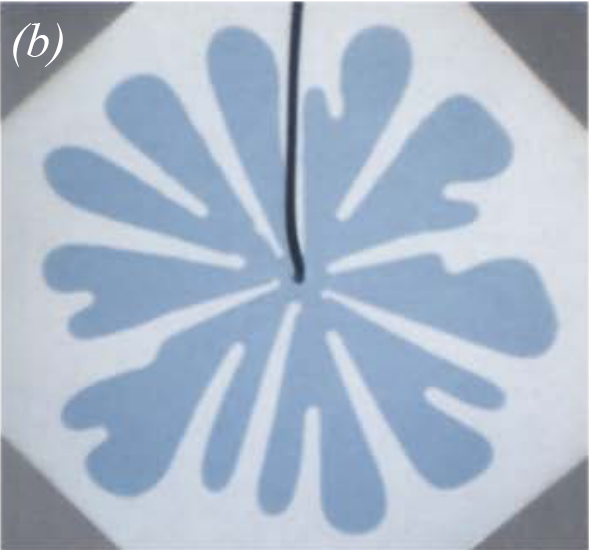}
	\includegraphics[width=0.28\linewidth]{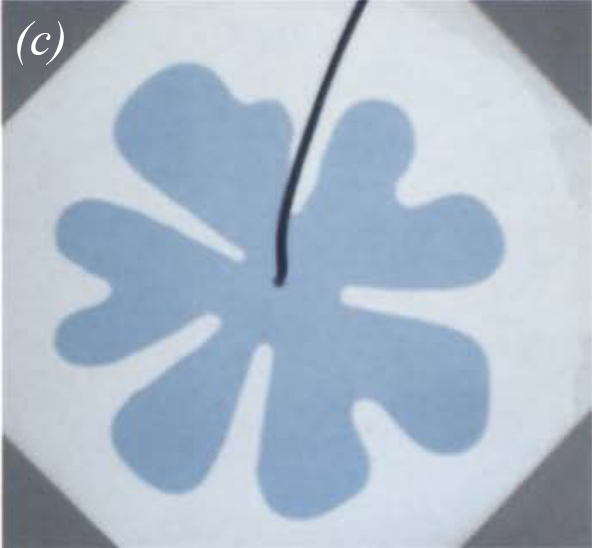}	\\	
	\includegraphics[width=0.28\linewidth]{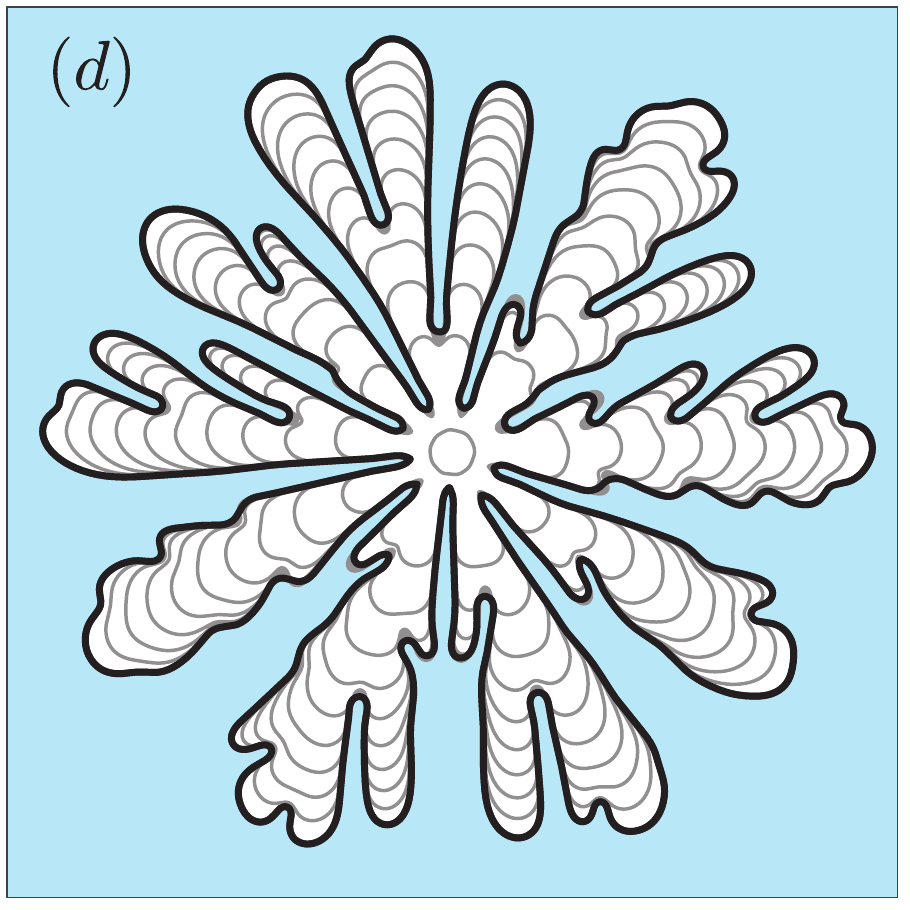}	
	\includegraphics[width=0.28\linewidth]{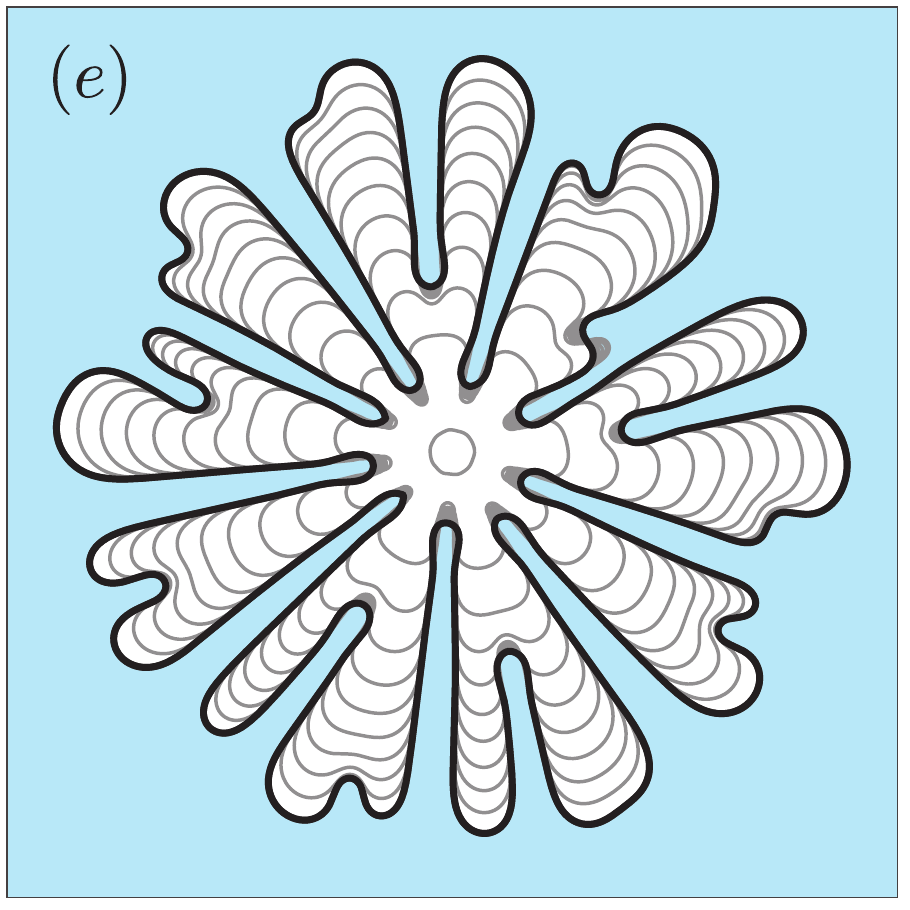}		
	\includegraphics[width=0.28\linewidth]{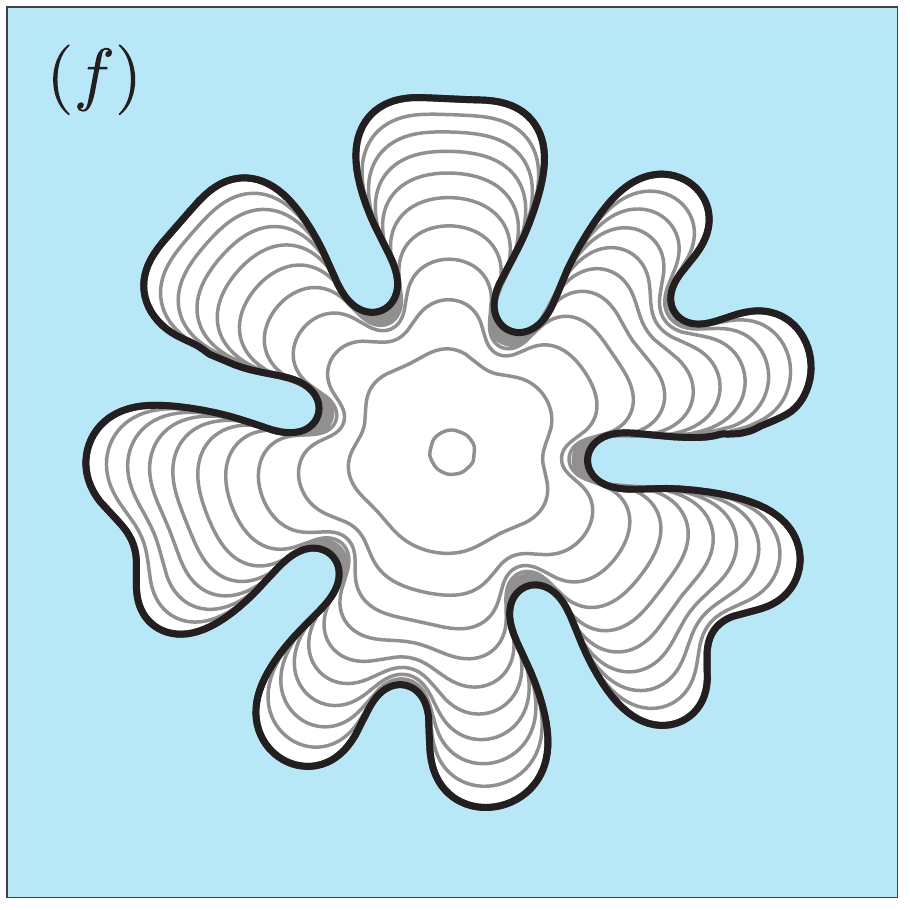}		
	\caption{$(a$-$c)$ Experimental results from \cite{Chen1987}, reproduced with permission from Springer Nature, comparing the development of viscous fingers for different injection rates and $(d$-$f)$ the corresponding numerical simulations. The gap thickness between plates is $7.5 \times 10^{-3}$ cm, and the injection rate and final times (left to right) are: $Q = 2.0 \times 10^{-3}$ mL/s and $t_f = 65$ s; $Q = 4.5 \times 10^{-4}$ mL/s and $t_f = 490$ s; and $Q = 1.4 \times 10^{-4}$ mL/s and $t_f = 1650$ s. Numerical solutions are plotted in time intervals of $t_f / 10$. Additionally, $\sigma = 20$ g/s$^2$, $\mu = 10.5$ g/(cm$\cdot$s), and $\omega = 0$ g/(s$^2 \cdot $mL). The initial condition is of the form \eqref{eq:InitialCondition} with $R_0 = 0.25$ cm. Simulations are performed on the domain $0 \le r \le 5$ and $0 \le \theta < 2 \pi$ using $750 \times 942$ equally spaced nodes.}
	\label{fig:Figure3}
\end{figure}

In figure \ref{fig:Figure3} we compare experimental results obtained by \cite{Chen1987} with our numerical simulations, and test some predictions made by linear stability analysis.  This figure illustrates the classic pattern formation in a standard Hele-Shaw configuration for three different injection rates in decreasing order.  For early times, we can apply \eqref{eq:UnstableMode} to predict the number of fingers that are produced.  Using figure \ref{fig:Figure3}$(e)$ as an example, equation \eqref{eq:UnstableMode} (with the appropriate parameter values) predicts that $n_{\max} \approx 9$ for small times, suggesting that 9 fingers should initially form.  A manual count of the onset of ``fingers'' in \ref{fig:Figure3}$(e)$ shows this prediction is consistent with the numerical simulation.

Another straightforward result from \eqref{eq:LSAclassic} is that, for fixed $\sigma$, $b$ and $\mu$, increasing the flow rate $Q$ results in a positive contribution to $\dot{\gamma}_n / \gamma_n$, which has a destabilising effect on each mode.  Further, we see from \eqref{eq:UnstableMode} that increasing $Q$ increases the most unstable wave number.  These observations are consistent with the experimental measurements performed by \cite{Chen1987} (and many others), which appear to show that increasing the injection rate results in larger wave numbers becoming more unstable, leading to branching and tip-splitting.  Of course, for later times, nonlinear effects become significant and linear stability analysis no longer provides an accurate description of the solution.   In this nonlinear regime, our numerical simulations are able to reproduce the main morphological features of these experiments for the different injection rates considered, as we can see by comparing images in each column of figure \ref{fig:Figure3}.  We view this comparison as a preliminary test of our numerical method.

\section{Reducing growth of viscous fingering pattern} \label{sec:MinisingInstabilities}

In this section, we investigate strategies for controlling viscous fingering when a prescribed amount of the inviscid fluid is injected over a finite period of time.  To begin, in \S~\ref{sec:TimeDependentInjection} and \ref{sec:TaperedPlates} we show how imposing a time-dependent injection rate or linearly tapering the gap between plates can suppress the development of viscous fingers.  As mentioned in the Introduction, both of these strategies have been previously considered; however, a comprehensive study comparing the effectiveness of these strategies to the standard configuration has not been conducted before now.  We extend this work in \S~\ref{sec:TaperedOptimal} by, for the first time, considering the strategy of imposing a time-dependent injection rate while the gap between the plates is tapered.  Our methodology involves deriving an injection rate that attempts to minimise the growth of the most unstable mode by adapting the ideas of \cite{Dias2012}.  Following these parts, in \S~\ref{sec:RotatingPlates}, we consider the effects on the bubble interface of rotating the Hele-Shaw cell while the inviscid fluid is injected.  For this geometry, we show for the first time how increasing the rotation rate of the  plates eventually act to stabilise the interface, while a careful choice of injection rate can accelerate this effect.

We consider two metrics for measuring how severe a fingering pattern is at the interface, namely the isoperimetric ratio
\begin{align}
	\mathcal{I}(t) = \frac{L^2}{4 \pi A}, \label{eq:IsoperimetricRatio}
\end{align}
where $L$ and $A$ are the arc length and area enclosed by the interface, and the ratio of the tip to base radii, which we refer to as the circularity ratio (the ``roundness''), defined as
\begin{align}
	\mathcal{C}(t) = \frac{R_{\mathrm{outer}}}{R_{\mathrm{inner}}},
	\label{eq:FingerLength}
\end{align}
where $R_{\mathrm{outer}}$ is the radius of the smallest circle (centred at the origin) that completely encloses the bubble and $R_{\mathrm{inner}}$ is the radius of the largest circle (centred at the origin) that contains only inviscid fluid.   Both $\mathcal{I}$ and $\mathcal{C}$ will be unity when the interface is circular and increase as the instabilities cause the interface to deform away from a circle.

In this section, the initial condition of the interface is
\begin{align} \label{eq:InitialCondition}
	s(\theta, 0) = R_0 \left( 1 + 0.01 \sum_{n=2}^{12} \cos \left( n \left(  \theta - 2 \pi \theta_n \right)  \right)  \right),
\end{align}
where $\theta_n$ is a uniformly random number between 0 and 1. Simulations are performed on the domain $0 \le r \le 7.5$ and $0 \le \theta < 2 \pi$ using $750 \times 628$ equally spaced nodes. For each parameter combination considered, 10 simulations are performed, and $\mathcal{I}$ and $\mathcal{C}$ are both averaged over these simulations. For simulations in this section we use $\sigma = 3$ g/s$^2$ and $\mu = 1.58$ g/(cm$\cdot$s).

\subsection{Time-dependent injection rate} \label{sec:TimeDependentInjection}

\begin{figure}
	\centering
	Parallel plates ($b = 0.2$ cm) with constant injection \\
	\includegraphics[width=0.19\linewidth]{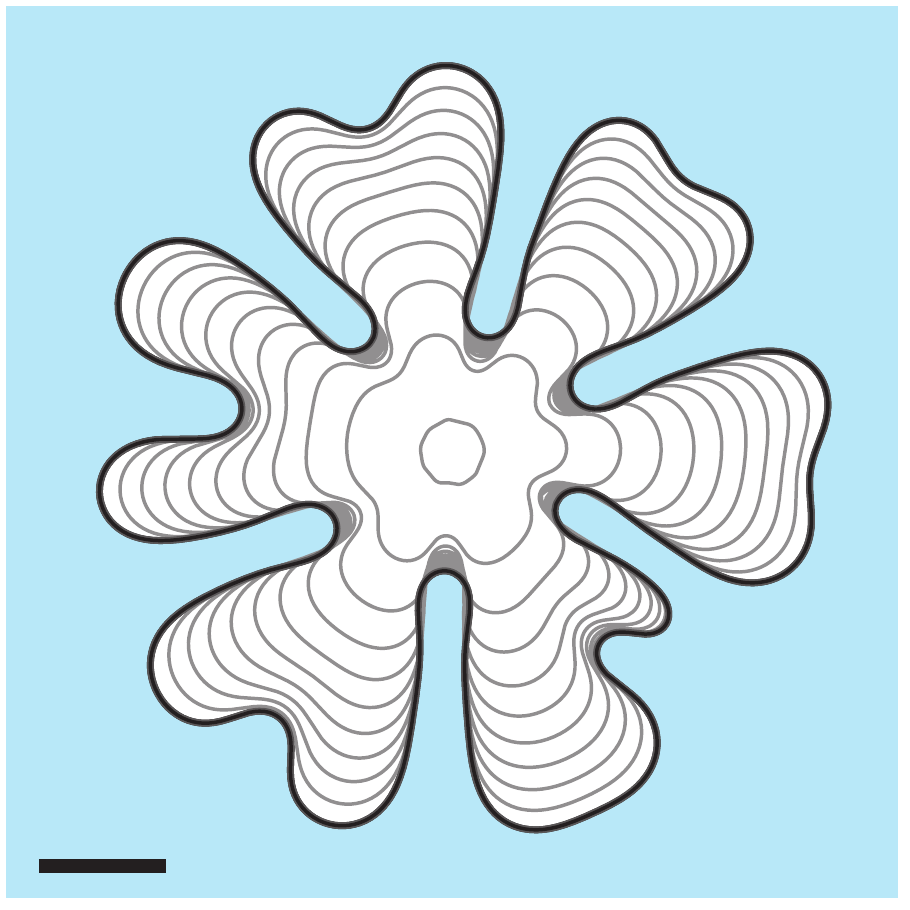} \hspace{0.1cm}
	\includegraphics[width=0.19\linewidth]{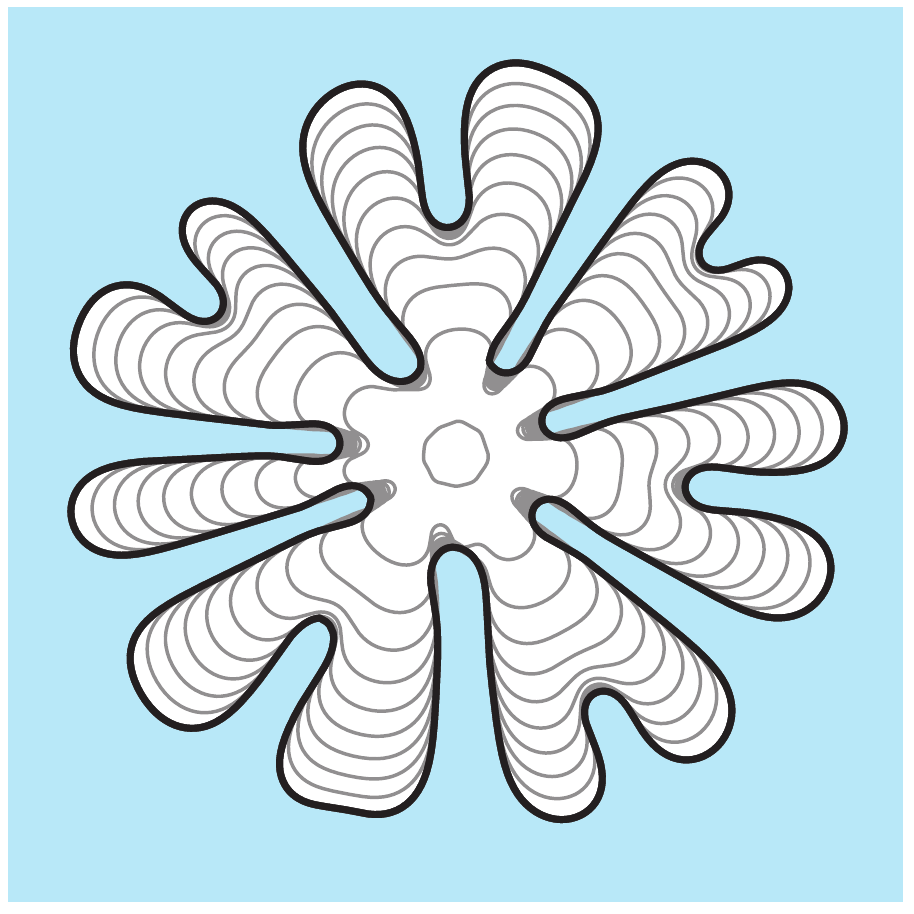}\hspace{0.1cm}
	\includegraphics[width=0.19\linewidth]{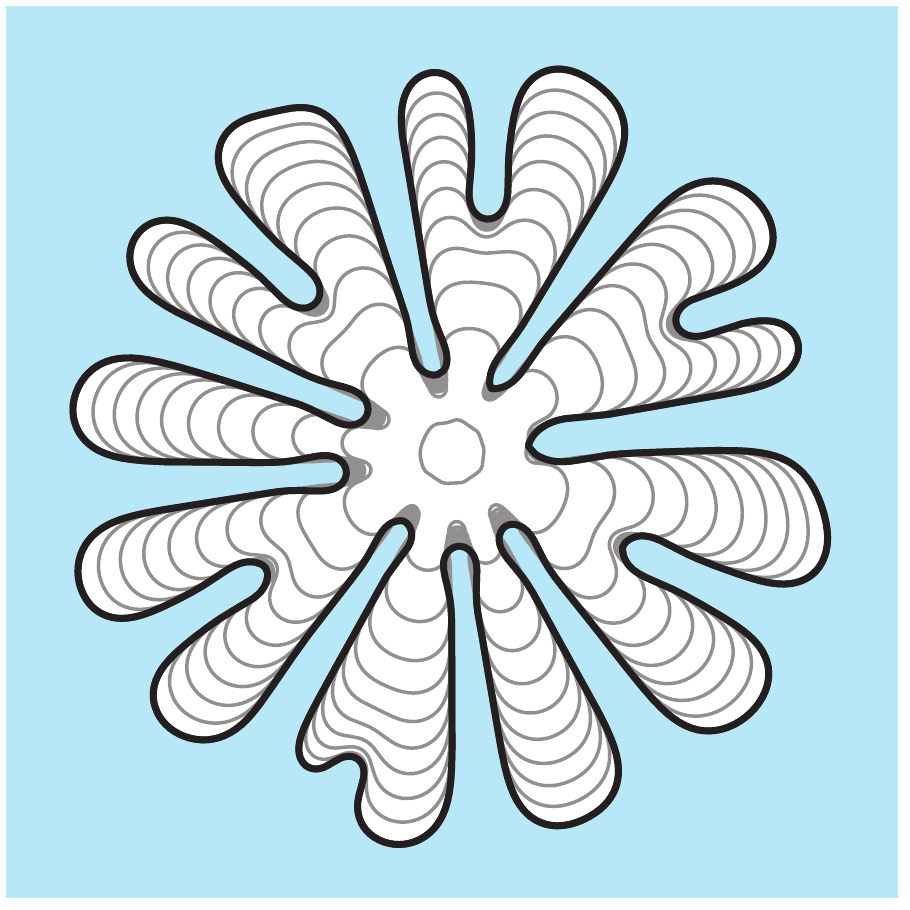}\hspace{0.1cm}
	\includegraphics[width=0.19\linewidth]{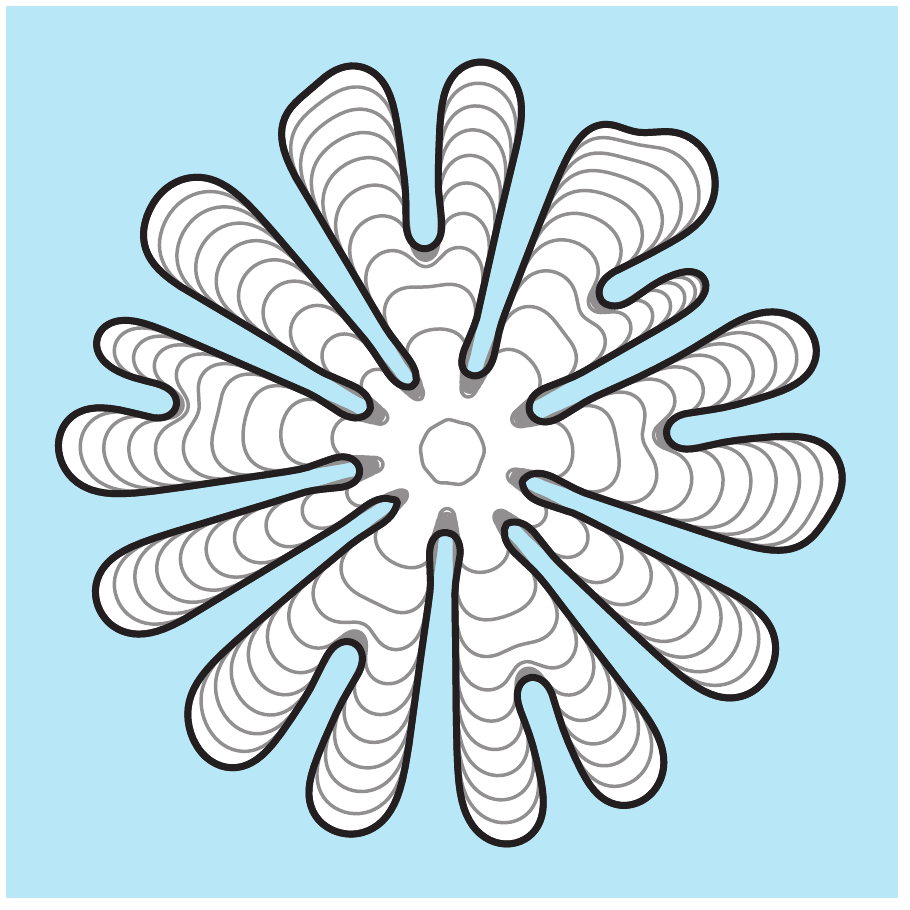}
	\\
	\vspace{0.15cm}
	Parallel plates ($b = 0.2$ cm) with linear injection\\	
	\includegraphics[width=0.19\linewidth]{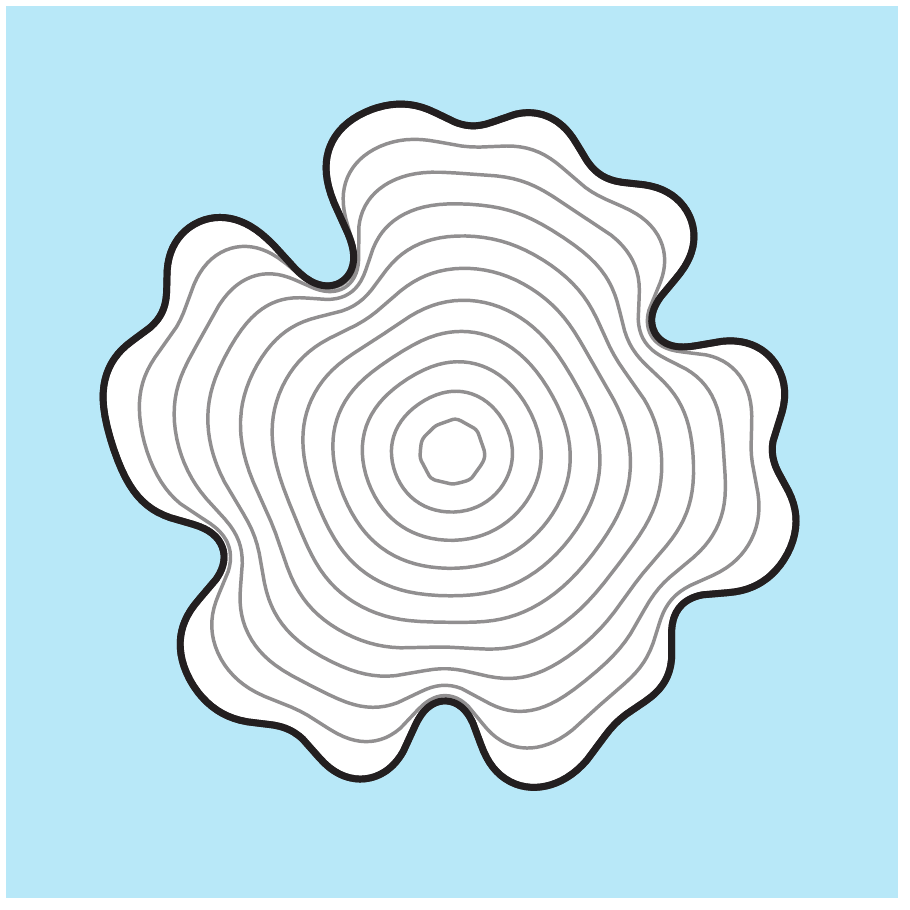} \hspace{0.1cm}
	\includegraphics[width=0.19\linewidth]{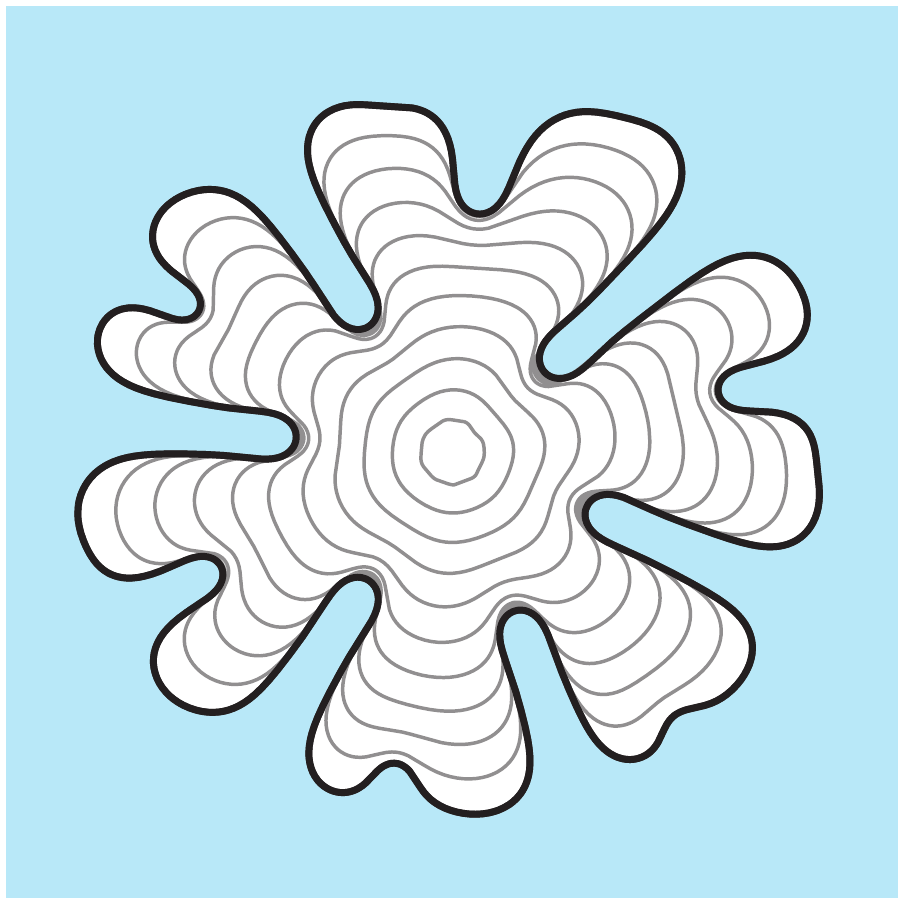} \hspace{0.1cm}
	\includegraphics[width=0.19\linewidth]{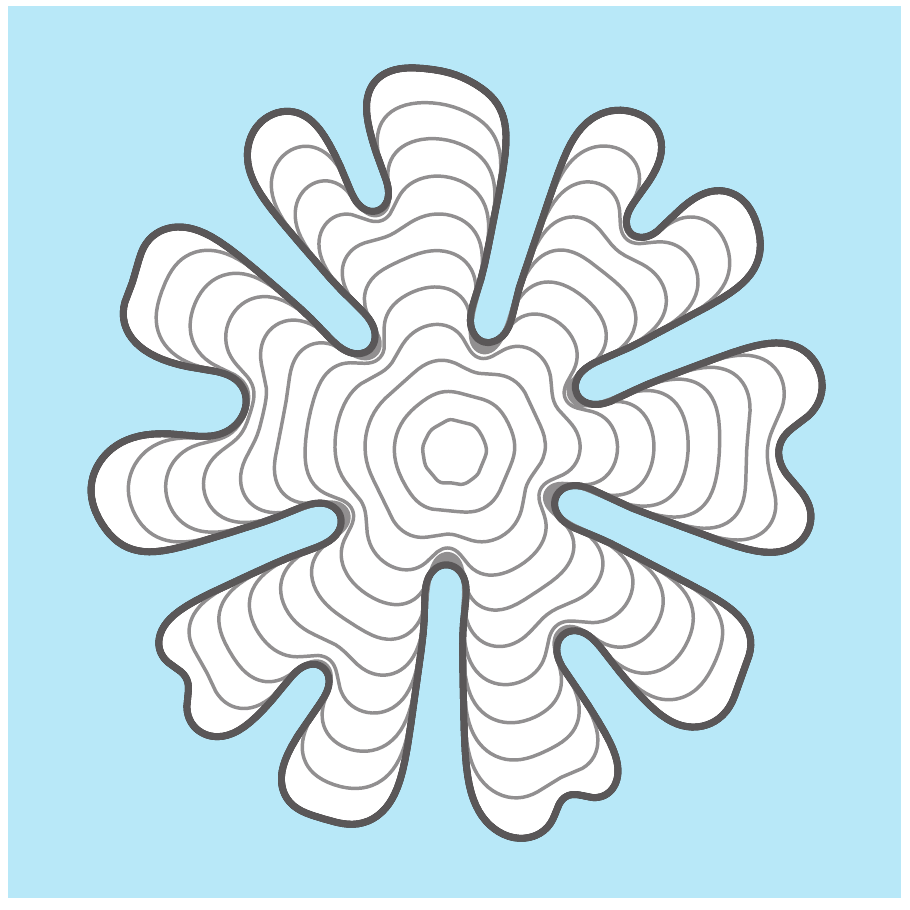} \hspace{0.1cm}
	\includegraphics[width=0.19\linewidth]{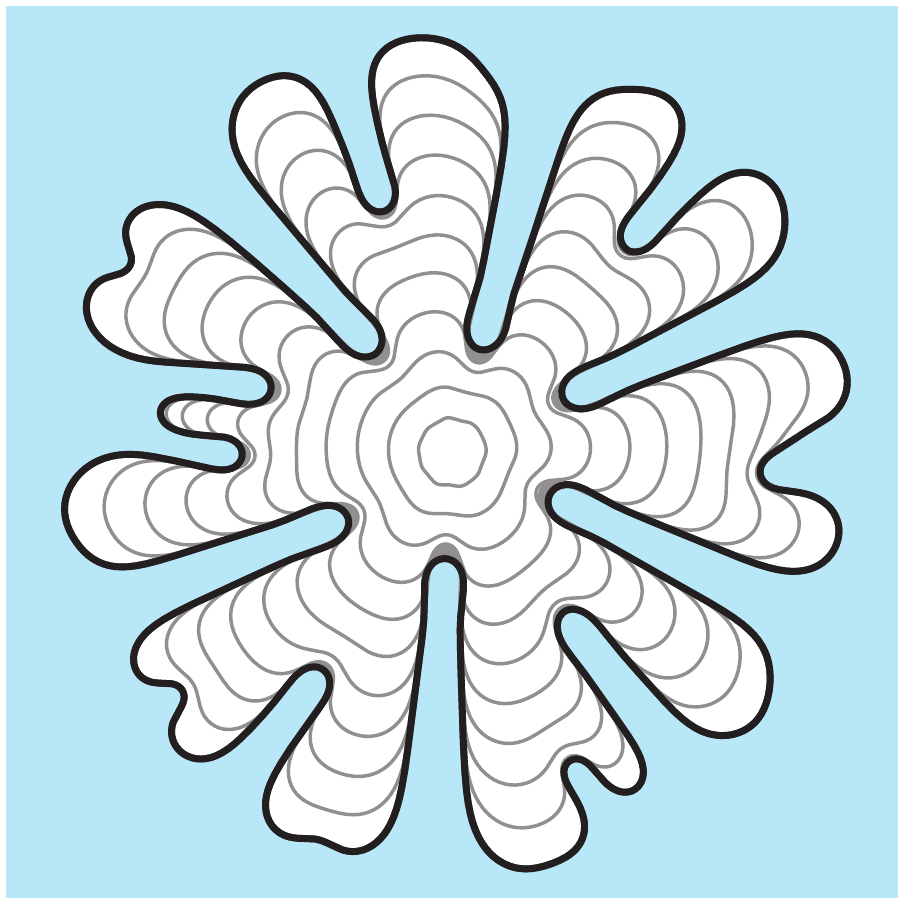}
	\\
	\vspace{0.15cm}
	Converging plates ($\alpha = 5.3 \times 10^{-2}$ $r_0 = 7$ cm $b_0 = 5\times 10^{-3}$ cm) with constant injection\\		
	\includegraphics[width=0.19\linewidth]{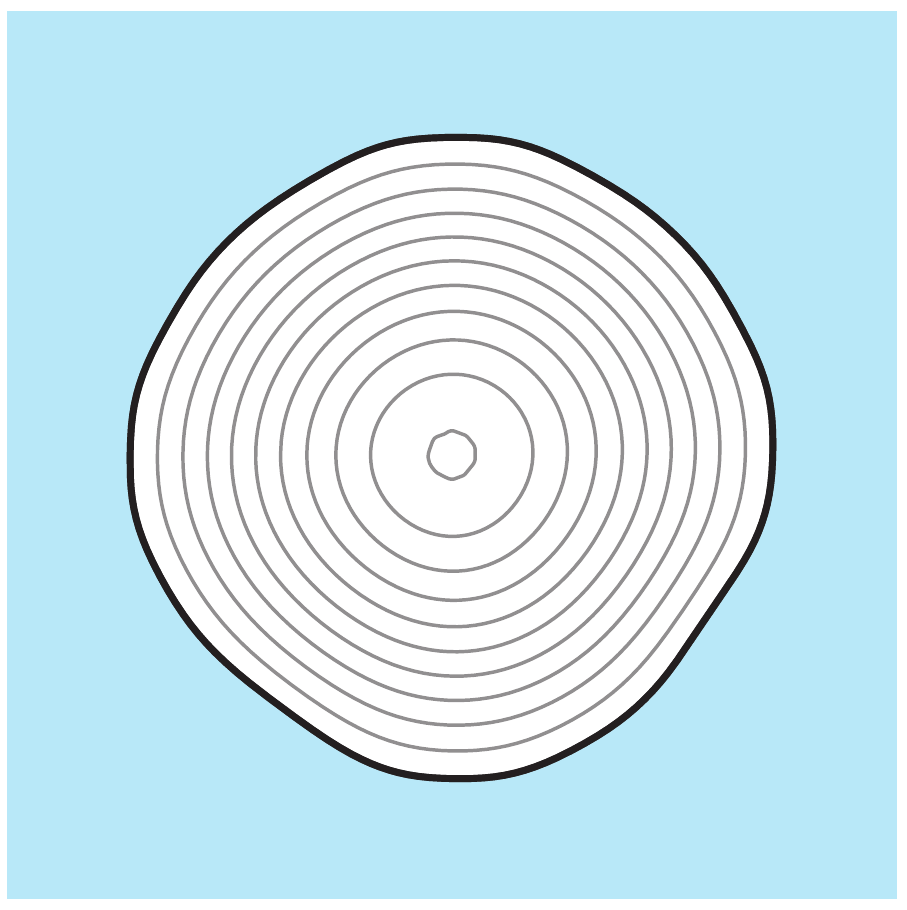} \hspace{0.1cm}
	\includegraphics[width=0.19\linewidth]{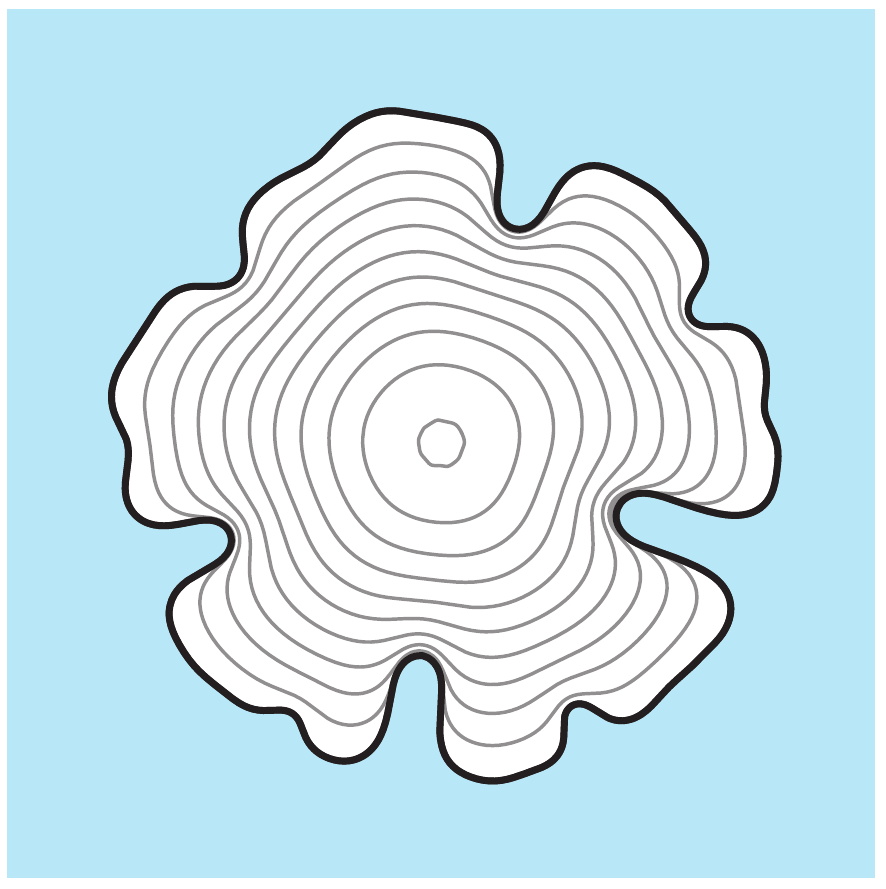} \hspace{0.1cm}
	\includegraphics[width=0.19\linewidth]{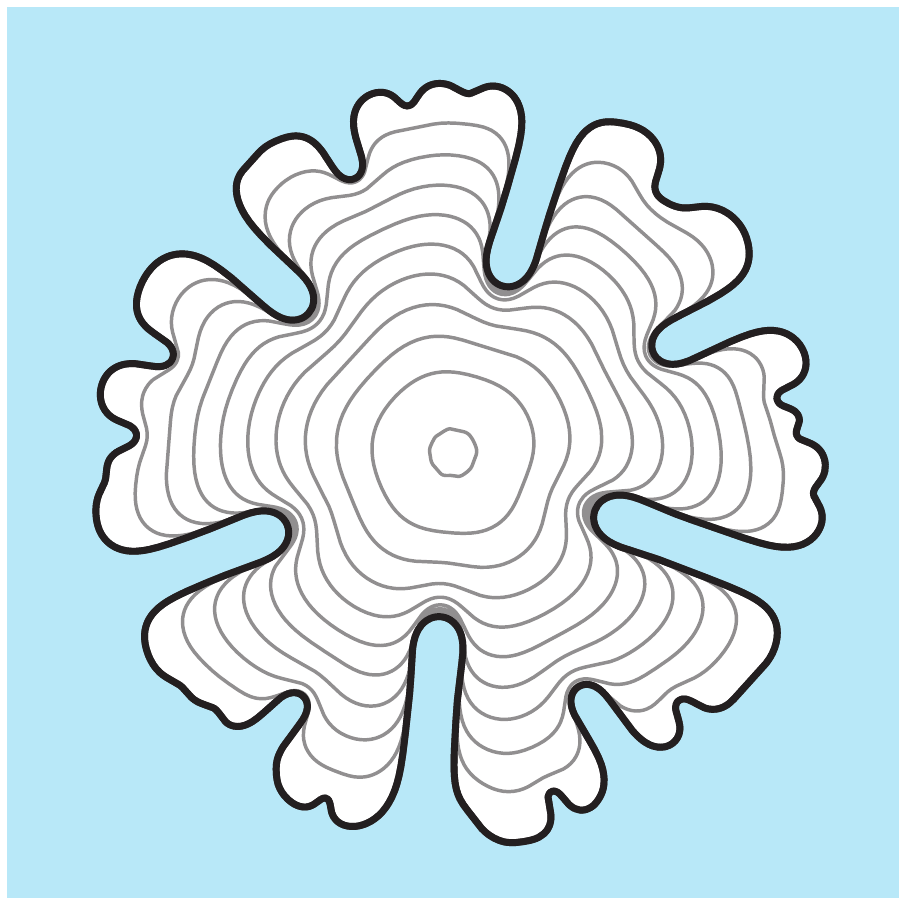} \hspace{0.1cm}
	\includegraphics[width=0.19\linewidth]{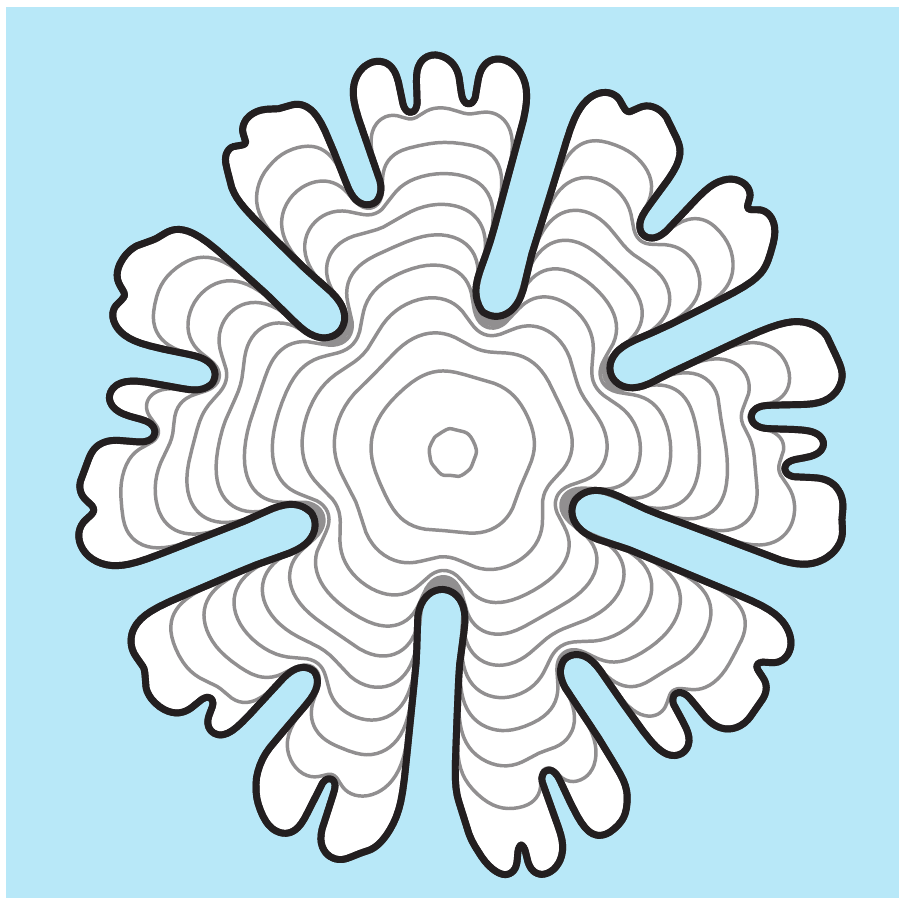}
	\\
	\vspace{0.15cm}
	Converging plates ($\alpha = 5.3 \times 10^{-2}$ $r_0 = 7$ cm $b_0 = 5\times 10^{-3}$ cm) with optimal injection \\		
	\includegraphics[width=0.19\linewidth]{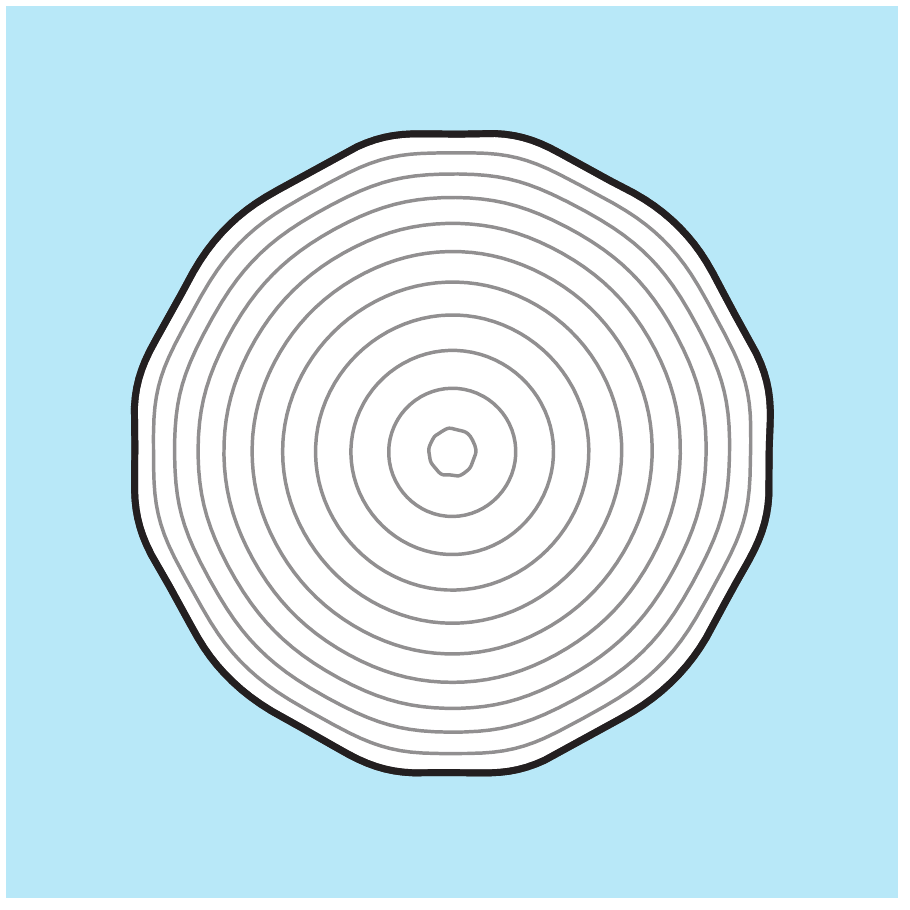} \hspace{0.1cm}
	\includegraphics[width=0.19\linewidth]{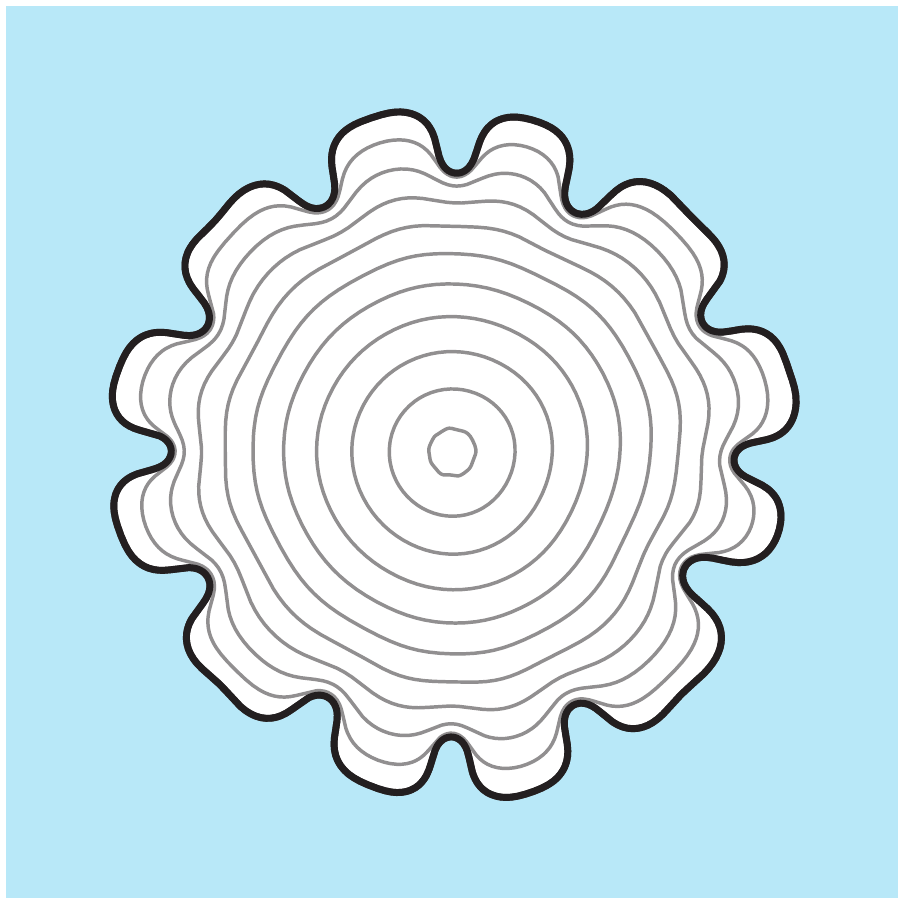} \hspace{0.1cm}
	\includegraphics[width=0.19\linewidth]{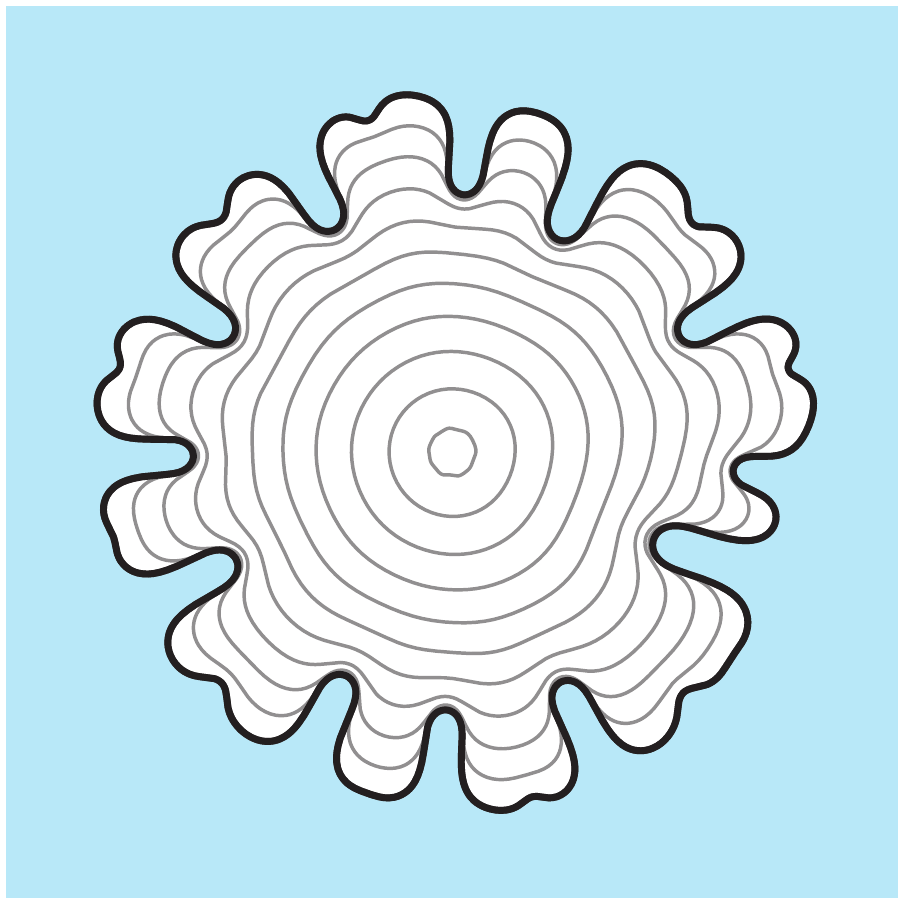} \hspace{0.1cm}
	\includegraphics[width=0.19\linewidth]{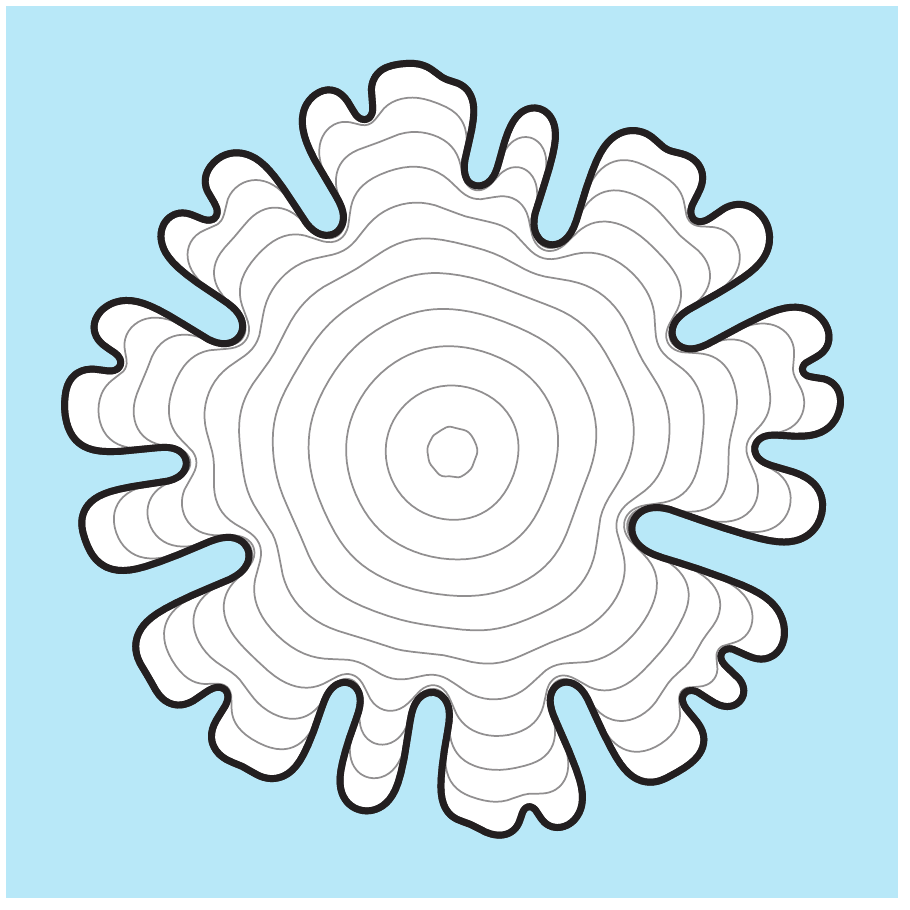}
	\\
	\vspace{0.15cm}
	Diverging plates, ($\alpha = -5.3 \times 10^{-2}$ $r_0 = 7$ cm $b_0 = 5\times 10^{-3}$ cm) with constant injection \\		
	\includegraphics[width=0.19\linewidth]{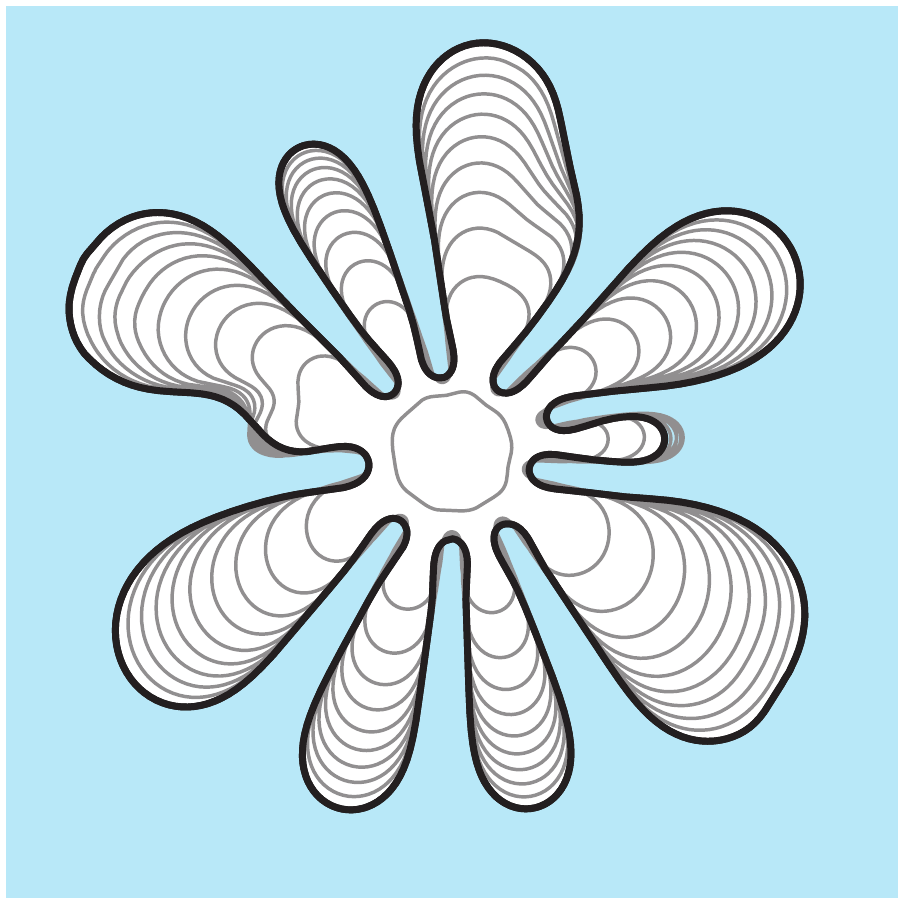} \hspace{0.1cm}
	\includegraphics[width=0.19\linewidth]{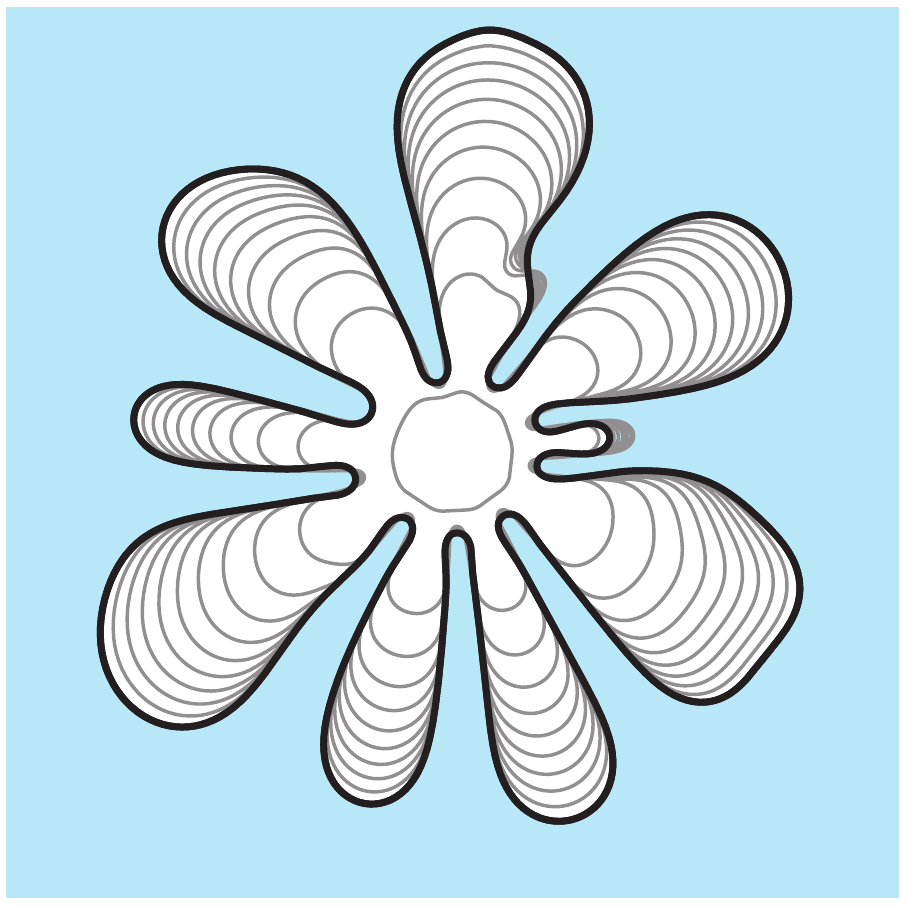} \hspace{0.1cm}
	\includegraphics[width=0.19\linewidth]{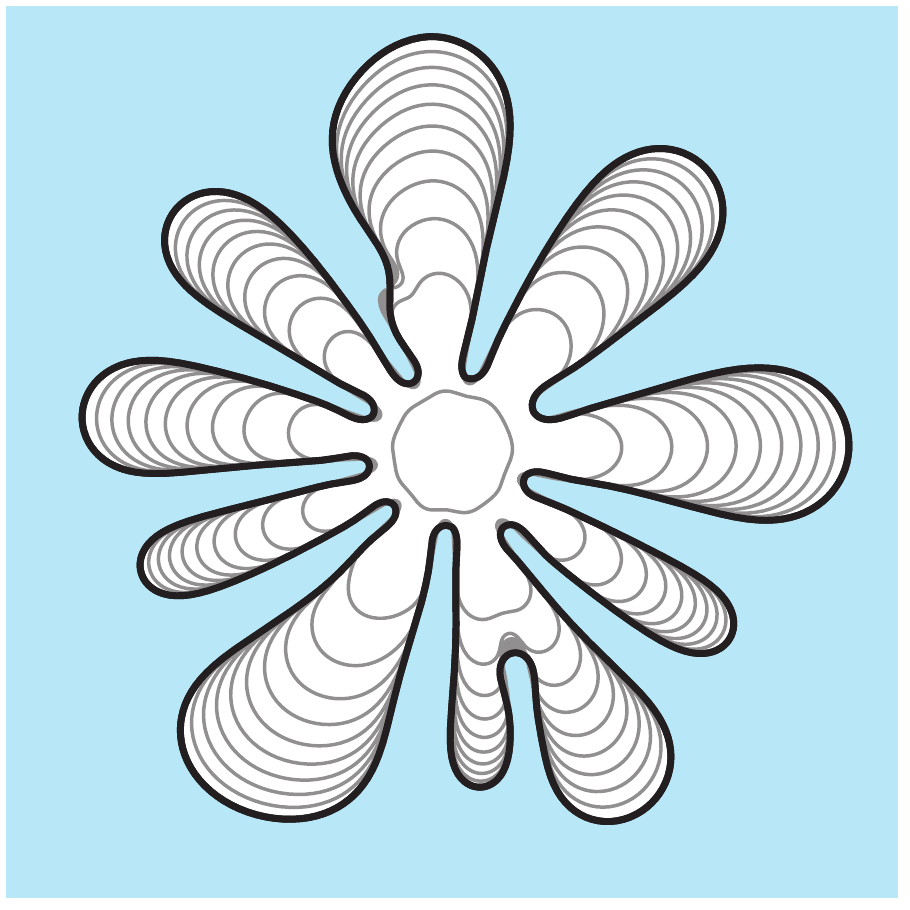} \hspace{0.1cm}
	\includegraphics[width=0.19\linewidth]{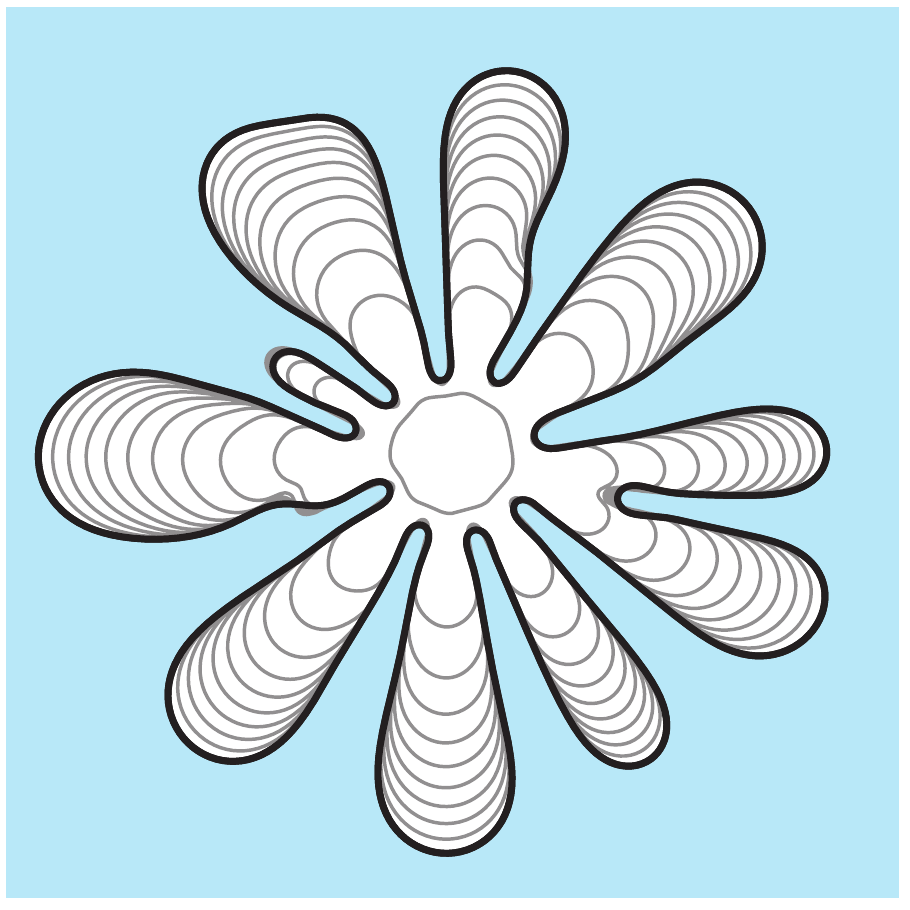}
	\\
	\vspace{0.15cm}
	Diverging plates ($\alpha = -5.3 \times 10^{-2}$ $r_0 = 7$ cm $b_0 = 5\times 10^{-3}$ cm) with optimal injection \\	
	\includegraphics[width=0.19\linewidth]{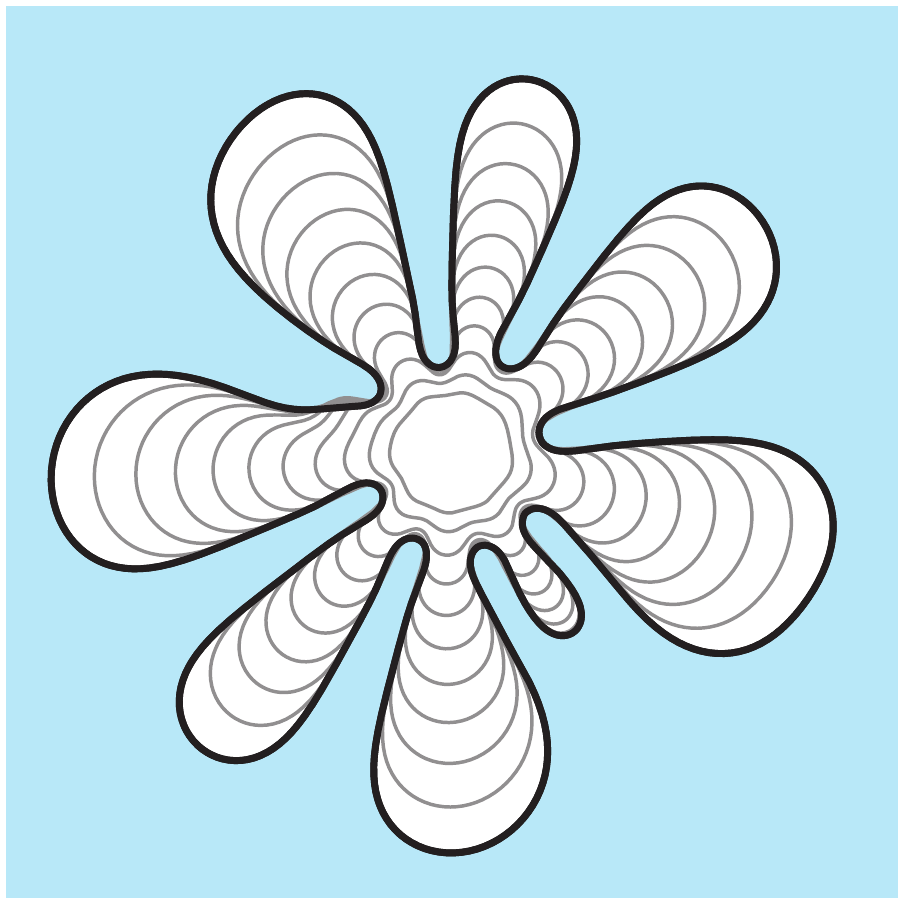} \hspace{0.1cm}
	\includegraphics[width=0.19\linewidth]{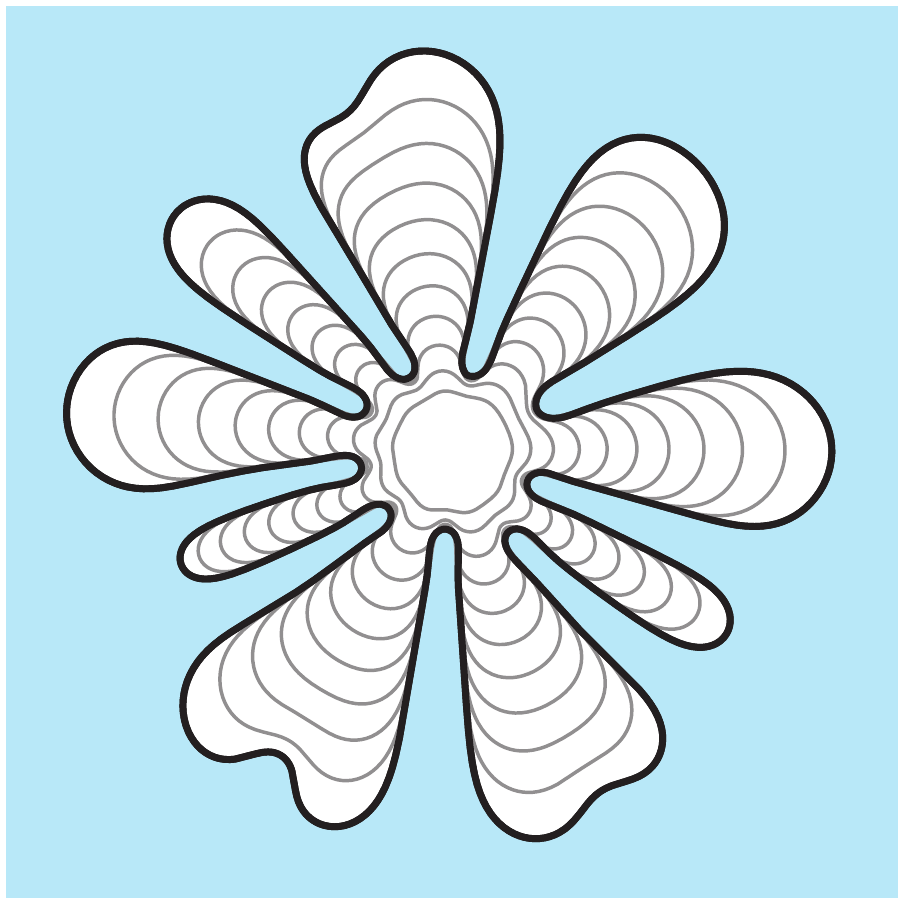} \hspace{0.1cm}
	\includegraphics[width=0.19\linewidth]{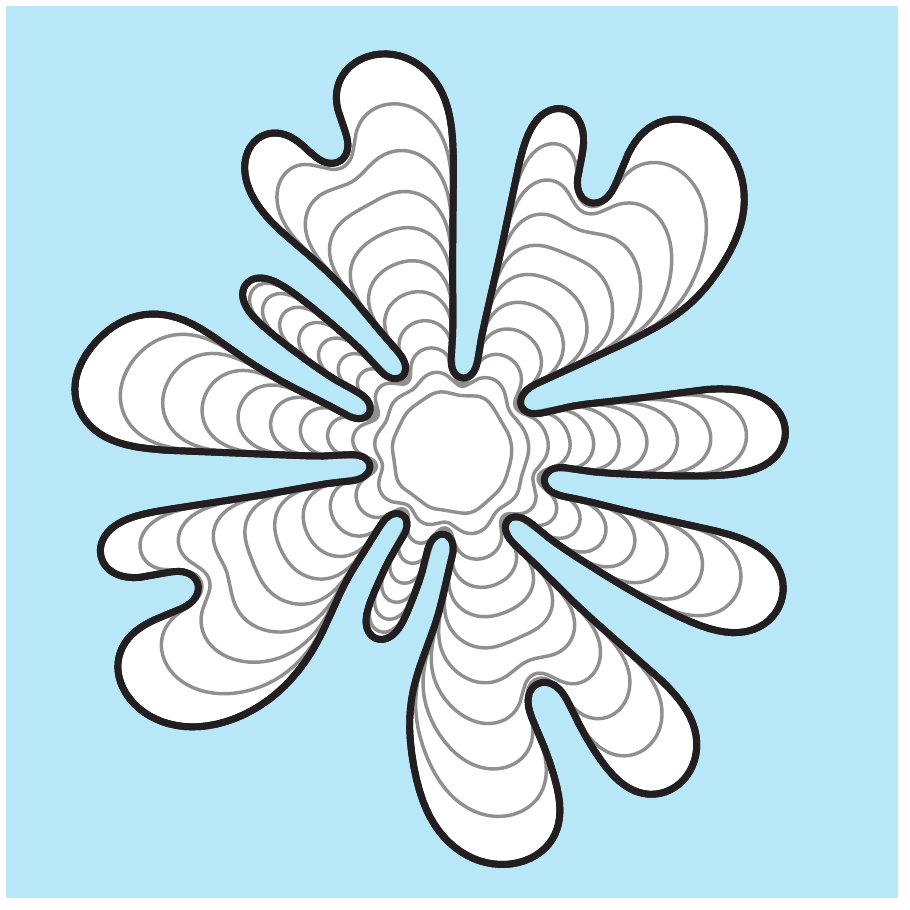} \hspace{0.1cm}
	\includegraphics[width=0.19\linewidth]{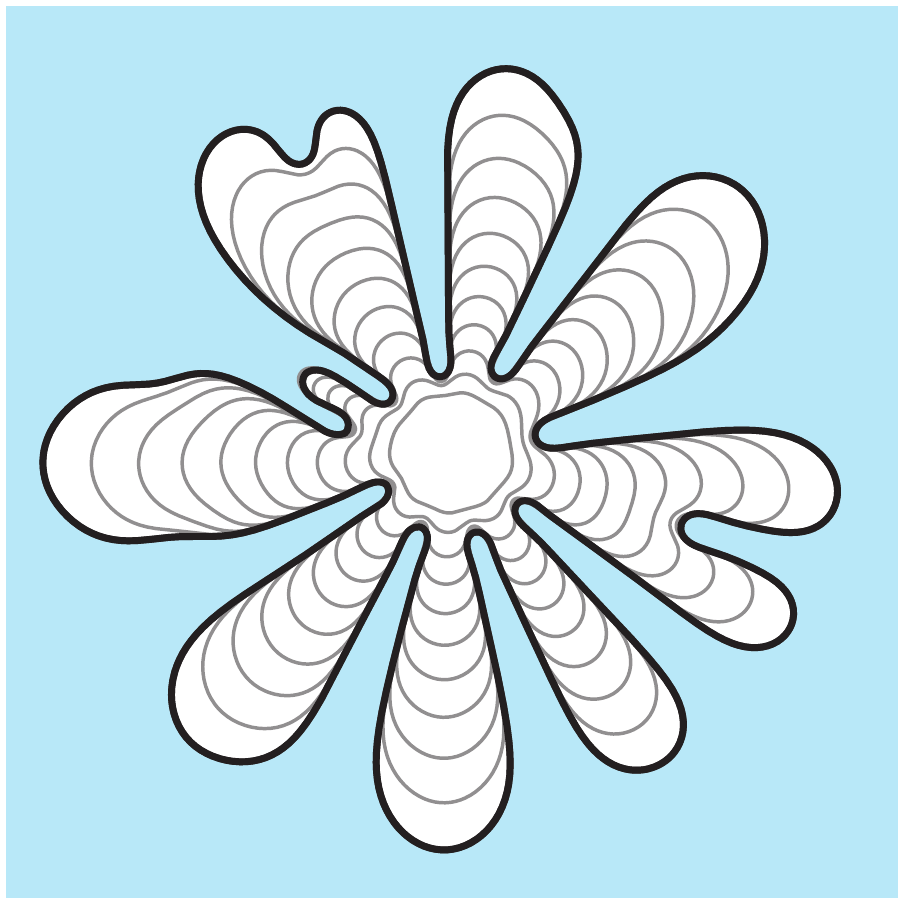}
		\caption{
			Comparison of numerical solutions with different injection schemes and plate gap thicknesses. For each column, the average injection rate is $\bar{Q} = 0.6$, 1.2, 1.6, and 2 mL/s and the final time is $t_f = 25.62$, 12.81, 9.61, and 7.78 s. Rows 1, 3, and 5 have constant injection rate $\bar{Q}$. For row 2, the injection rate is \eqref{eq:LinearInjection}, while for rows 4 and 6, the injection rate is determined from the solution to \eqref{eq:OptimalODE}. We use $R_0 = 0.5$ cm for rows 1 and 2, $R_0 = 0.37$ cm for rows 3 and 4, and $R_0 = 0.94$ cm for rows 5 and 6. Additionally, $R_f = 5$ cm for rows 2, 4, and 6. For all simulations, the initial volume of the bubble is approximately 0.157 mL, and the volume at $t = t_f$ is 15.7 mL. Profiles are plotted in time intervals of  $t_f / 10$. The scale bar represents a length of 2 cm.
		}
		\label{fig:Figure4}
	\end{figure}

The first strategy we consider is  proposed by \cite{Dias2012}, who, using linear stability analysis and optimal control theory, derived the optimal time-dependent injection rate when the plate gap thickness is uniform. By seeking solutions of the form \eqref{eq:LSA1} and \eqref{eq:LSA2}, Dias \textit{et al}.\ showed that the growth rate of the most unstable perturbations to the circular solution, $s_0$, when the inviscid fluid is injected over the time interval $0 \le t \le t_f$, are minimised when
\begin{align} \label{eq:LinearInjection}
	Q(t) = \frac{2 \pi b(R_f - R_0)}{t_f} \left( R_0 + \frac{R_f - R_0}{t_f} t \right),
\end{align}
where $s_0(0) = R_0$ and $s_0(t_f) = R_f$.  The average of \eqref{eq:LinearInjection} over this time period is
\begin{align} \label{eq:ConstantInection}
	\bar{Q} = \frac{\pi b(R_f^2 - R_0^2)}{t_f}.
\end{align}
We are interested in comparing results from the linear injection rate \eqref{eq:LinearInjection} with a constant injection rate where $Q=\bar{Q}$, so that in both cases the same amount of fluid is injected over the fixed time period. Using both experiments and numerical simulations, \cite{Dias2012} showed that \eqref{eq:LinearInjection} does suppress the growth of viscous fingers compared to \eqref{eq:ConstantInection}; however, only cases for which the injection rate is sufficiently low that viscous fingers were completely suppressed were considered. We extend this work by performing simulations over a much wider range of injection rates to better compare the development of viscous fingers between the injection rates of the forms \eqref{eq:LinearInjection} and \eqref{eq:ConstantInection}. Figure \ref{fig:Figure4} presents numerical solutions for the constant (top row) and linear (second row) injection rates.  The columns from left to right are for increasing values of $\bar{Q}$.  We observe that the linear injection rate appears to inhibit viscous fingering, and in particular, tip-splitting is delayed resulting in shorter fingers than the corresponding constant injection case.
	
\begin{figure}
	\centering
	\includegraphics[width=0.4\linewidth]{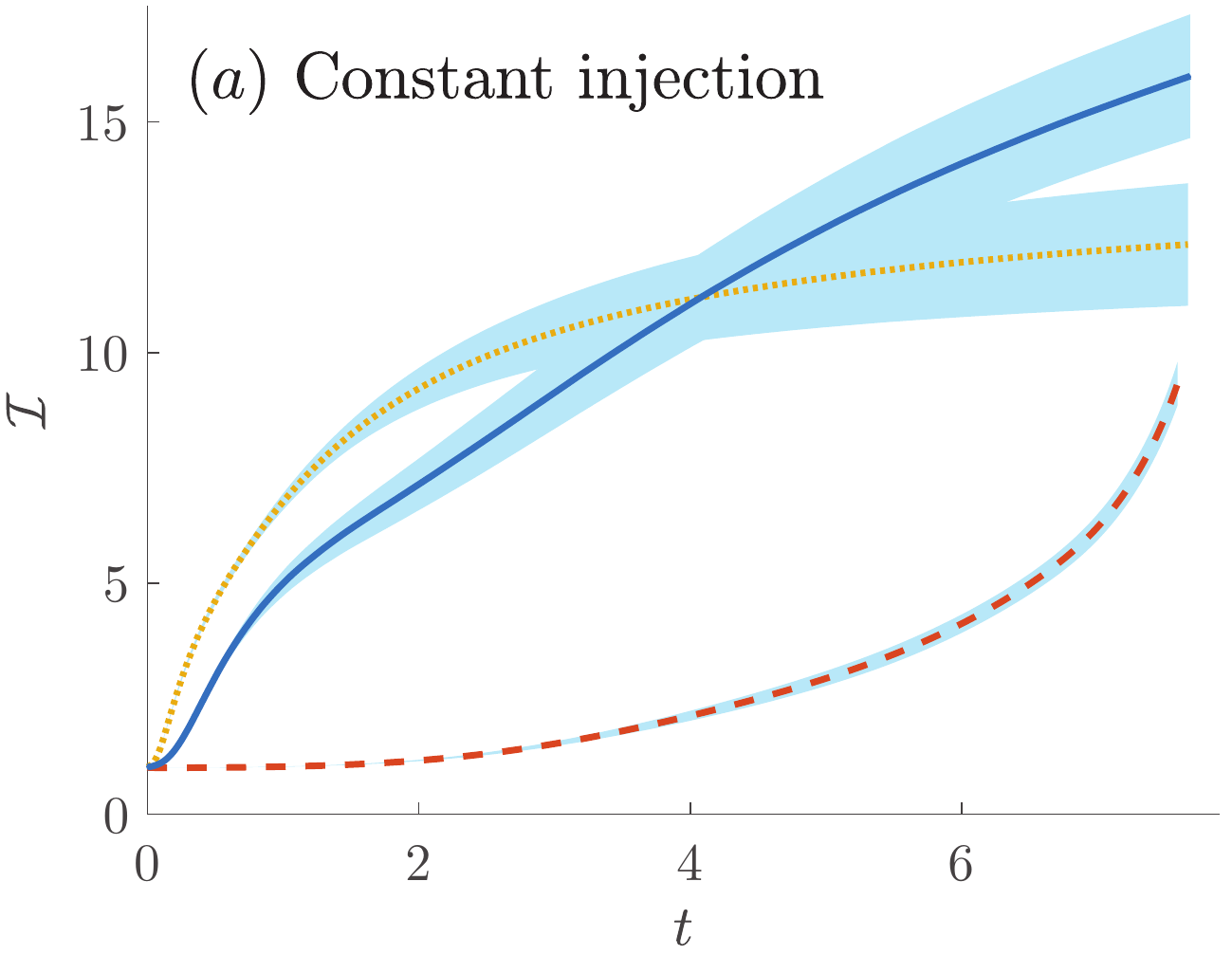}
	\includegraphics[width=0.4\linewidth]{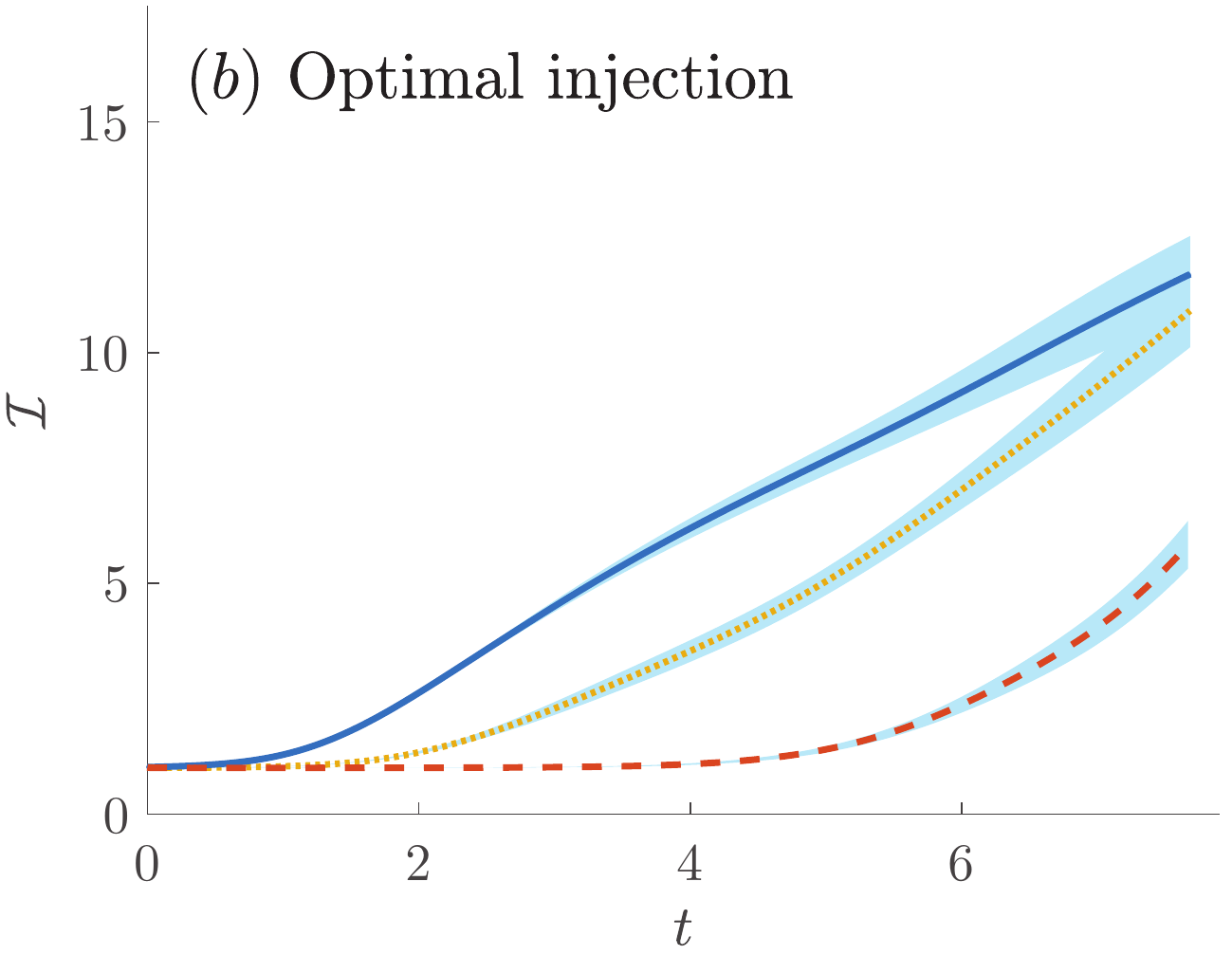}	
		\caption{Isoperimetric ratio for different plate tapering configurations with $(a)$ constant and $(b)$ time-dependent injection rates, where the average injection rate of all configurations is $\bar{Q} = 2$ mL/s and final time $t_f = 7.68$ s. Solid (blue) curve denotes the configuration with parallel plates $b = 0.2$ cm, dotted (yellow) curve is for plates tapered according to \eqref{eq:TaperedPlate} with $\alpha = -5.3 \times 10^{-2}$, $r_0 = 7$ cm, and $b_0 = 0.395$ cm, and dashed (red) curve is for tapered plate with $\alpha = 5.3 \times 10^{-2}$, $r_0 = 7$ cm, and $b_0 = 5 \times 10^{-3}$ cm. For $(b)$ the injection rate for the solid curve is given by \eqref{eq:LinearInjection} with $R_0 = 0.5$ cm and $R_f = 5$ cm, while for the dashed and dotted curves, the injection rate is determined from the solution to \eqref{eq:OptimalODE} with $R_0 = 0.37$ cm and $R_f = 5$ cm, and $R_0 = 0.94$ cm and $R_f = 5$ cm respectively. Shaded (blue) region represents one standard deviation above and below the mean.}
		\label{fig:IsoperimetricExample}
\end{figure}

	These are only visual observations.  Focusing on the representative case $\bar{Q} = 2$ mL/s (fourth column of figure~\ref{fig:Figure4}), a more quantitative measure is provided in figure \ref{fig:IsoperimetricExample}, where the isoperimetric ratio $\mathcal{I}$ is plotted against time for both injection schemes. Initially, the isoperimetric ratio of the linear injection rate case (solid blue curve in figure \ref{fig:IsoperimetricExample}$(b)$) grows much more slowly than the constant injection rate case (solid blue curve in figure \ref{fig:IsoperimetricExample}$(a)$), which is simply because the linear injection rate is lower than the constant injection rate for the first half of the simulation, and so the interface is less unstable.  For later times, the isoperimetric ratio of the linear injection rate case grows faster corresponding to times for which the linear injection rate is faster. Despite this switch in behaviour, the overall effect of the linear injection rate \eqref{eq:LinearInjection} is to noticeably reduce the isoperimetric ratio at the final time $t_f$ when compared to the constant injection rate.  These numerical results provide new quantitative evidence for how well \cite{Dias2012}'s `optimal' flow rate works in practice (we return to figure~\ref{fig:IsoperimetricExample} below).
	
\begin{figure}
	\centering
	\includegraphics[width=0.4\linewidth]{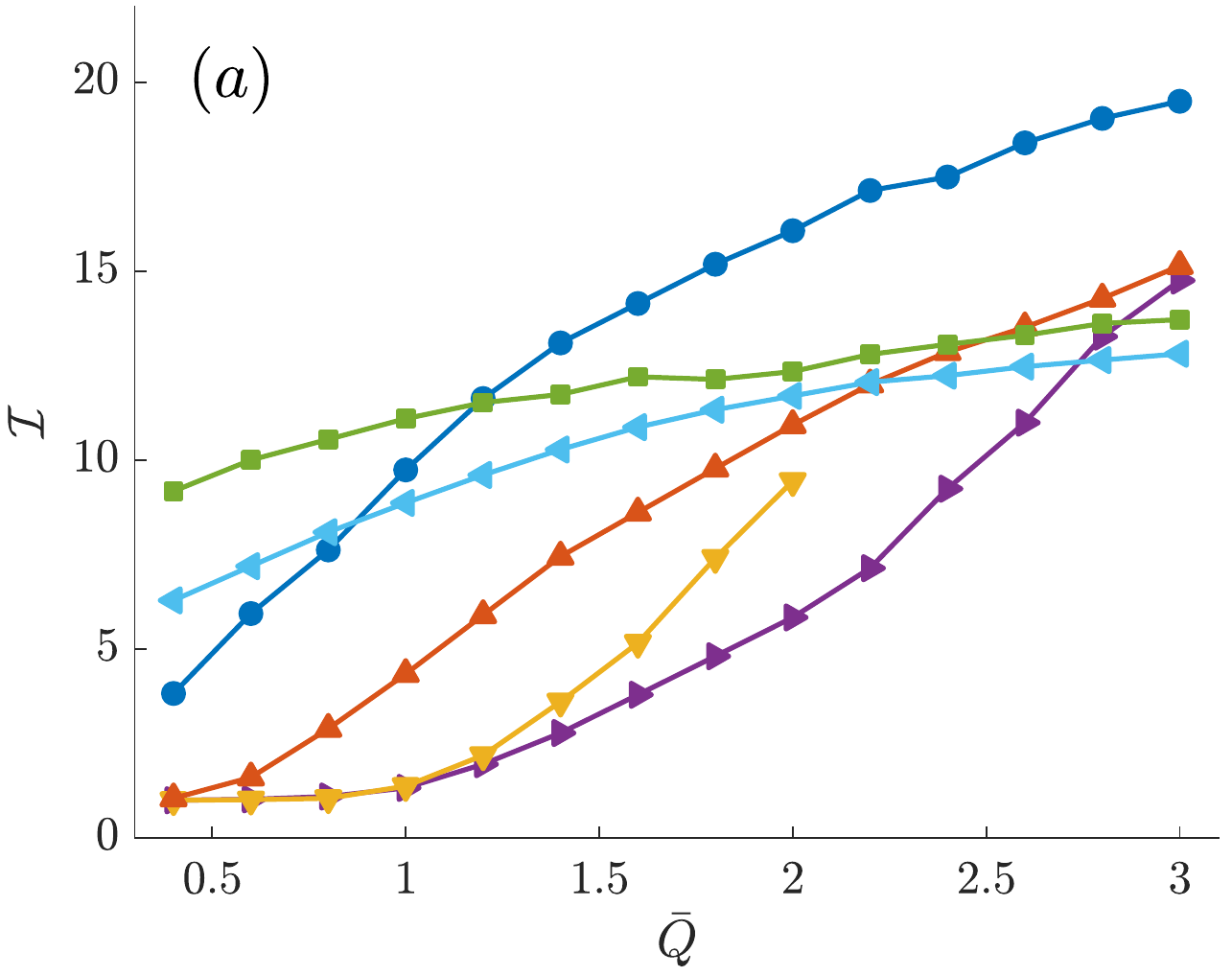}
	\includegraphics[width=0.4\linewidth]{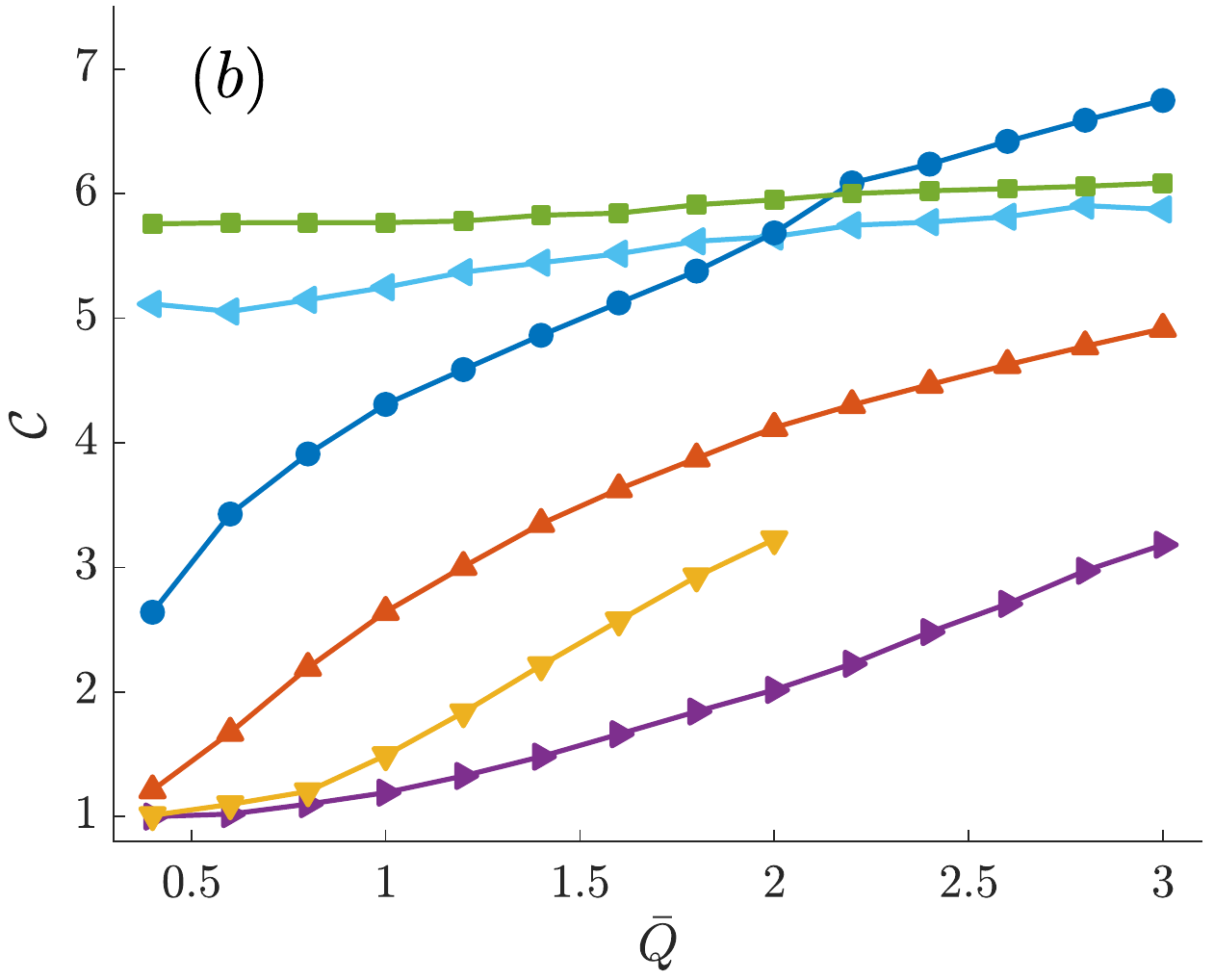}
	\caption{$(a)$ Isoperimetric \eqref{eq:IsoperimetricRatio} and $(b)$ circularity \eqref{eq:FingerLength} ratios at the final time $t = t_f$ for different injection schemes and plate tapering. The parameter $\bar{Q}$ (mL/s) is the average injection rate over a simulation, and the final time is $t_f = ((99 \pi / 20) \textrm{ mL}) / \bar{Q}$. Navy blue ($\bullet$) and red curves ($\blacktriangleup$) denote cases with parallel plates $b = 0.2$ cm, yellow ($\blacktriangledown$) and purple ($\blacktriangleright$) denotes cases with tapered plates where $\alpha = 5.3 \times 10^{-3}$, $r_0 = 7$ cm, and $b_0 = 5 \times 10^{-3}$ cm, and green ($\sqbullet$) and light blue ($\blacktriangleleft$) is tapered plates where $\alpha = -5.3 \times 10^{-3}$, $r_0 = 7$ cm, and $b_0 = 0.395$ cm. For $\bullet$, $\blacktriangledown$, and $\sqbullet$, the inviscid bubble is injected at a constant rate, $\bar{Q}$. For $\blacktriangleup$, injection rate is \eqref{eq:LinearInjection} with $R_0 = 0.5$ cm and $R_f = 5$ cm. For $\blacktriangleright$ and $\blacktriangleleft$, injection rate is computed from the solution to \eqref{eq:OptimalODE} with $R_0 = 0.37$ cm and $R_f = 5$ cm, and $R_0 = 0.94$ cm and $R_f = 5$ cm respectively.}
	\label{fig:Figure6}
\end{figure}

To investigate the robustness of the linear injection strategy as $\bar{Q}$ is varied, we compute both the isoperimetric ratio $\mathcal{I}$ and the circularity ratio $\mathcal{C}$ according to \eqref{eq:IsoperimetricRatio} and \eqref{eq:FingerLength} at $t = t_f$ for both the constant and linear injection rates, shown in figure \ref{fig:Figure6}. The results for the constant injection rates are denoted by (navy blue) $\bullet$, while the linear injection rate is indicated by (red) $\blacktriangleup$.  Recall that for each data point,  10 simulations are performed and the error bars indicate plus or minus one standard deviation.  Over the range of $\bar{Q}$ values considered, both of the measures $\mathcal{I}$ and $\mathcal{C}$ are considerably lower for the linear injection rate case compared to the constant injection rate.  Thus we conclude the linear injection scheme is successful in reducing the fingering pattern, regardless of $\bar{Q}$.
	
\subsection{Tapered Hele-Shaw geometry} \label{sec:TaperedPlates}

We now turn our attention to the configuration where the gap between the plates is linearly tapered in the direction of the flow such that
\begin{align} \label{eq:TaperedPlate}
	b(r) =
	\begin{cases}
		b_0 - \alpha (r - r_0) & \text{if } r \le r_0, \\
		b_0     & \text{if } r > r_0,
	\end{cases}
\end{align}
together with a constant injection rate.  The parameter $\alpha = (b(0) - b_0)/r_0$ controls the gradient of the taper. The influence of tapering the plates of a Hele-Shaw cell has been studied using linear stability analysis in both channel and radial geometry \citep{Al2012,Al2013b}. By considering $|\alpha| \ll 1$, \cite{Al2013b} derived an ordinary differential equation for $\gamma_n$
\begin{align} \label{eq:LSAtapered}
	\frac{\dot{\gamma}_n}{\gamma_n} = \frac{n-1}{s_0} \left( \dot{s}_0 - \frac{n(n+1)b(s_0)^2 \sigma}{12 \mu s_0^2} \right) - \left( \frac{n \sigma}{6 \mu s_0} - \frac{\dot{s}_0}{b(s_0)} \right) \alpha,
\end{align}
where $\dot{s}_0 = Q / 2 \pi b(s_0) s_0$, which suggests that the most unstable mode of perturbation is
\begin{align} \label{eq:nMaxTapererd}
	n_{\max} = \sqrt{\frac{1}{3} + \frac{4 \mu s_0^2 \dot{s}_0}{b(s_0)^2 \sigma} - \frac{2s_0^2 \alpha}{3 b(s_0)} }.
\end{align}
Of course, by setting $\alpha = 0$, \eqref{eq:LSAtapered} and \eqref{eq:nMaxTapererd} reduce to \eqref{eq:LSAclassic} and \eqref{eq:UnstableMode}. As noted by \cite{Al2013b}, diverging plates ($\alpha < 0$) introduces a negative offset to $\dot{\gamma}_n / \gamma_n$ in the form of $\dot{s}_0 \alpha / b(s_0)$, while for converging plates ($\alpha > 0$), this term acts to slow the growth of the most unstable mode of perturbation compared to the corresponding parallel plate case (analogous to the time-dependent injection rate considered in \S~\ref{sec:TimeDependentInjection}).  In the context of our objective of injecting a prescribed amount of fluid over a finite period of time, other comparisons between (\ref{eq:LSAtapered})-(\ref{eq:nMaxTapererd}) for the cases $\alpha<0$ and $\alpha>0$ are not straightforward as the speed of the interface, $\dot{s}_0$, is initially higher for the diverging case ($\alpha<0$) than it is for the converging case ($\alpha>0$), but subsequently switches over at some time to be lower, thus complicating the effects of each term in these equations.

\begin{figure}
	\centering
	\includegraphics[width=0.28\linewidth]{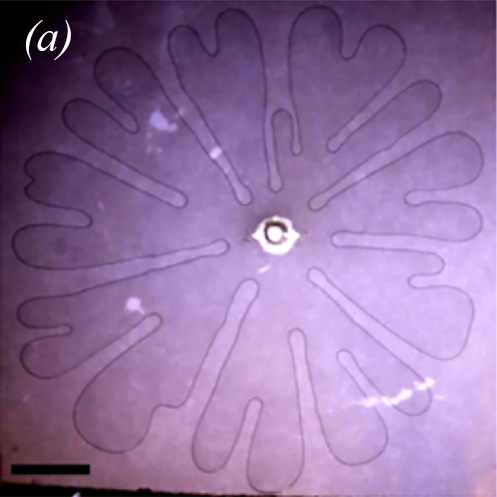}
	\includegraphics[width=0.28\linewidth]{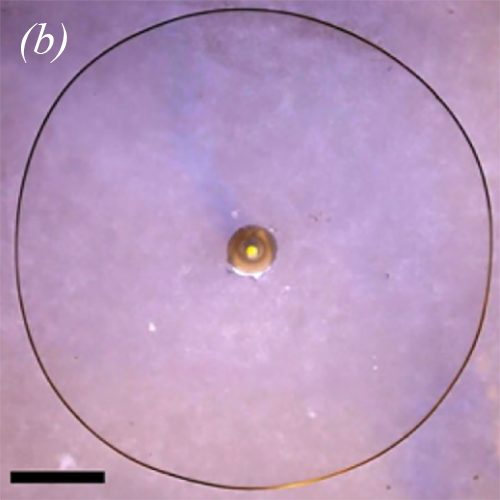}
	\includegraphics[width=0.28\linewidth]{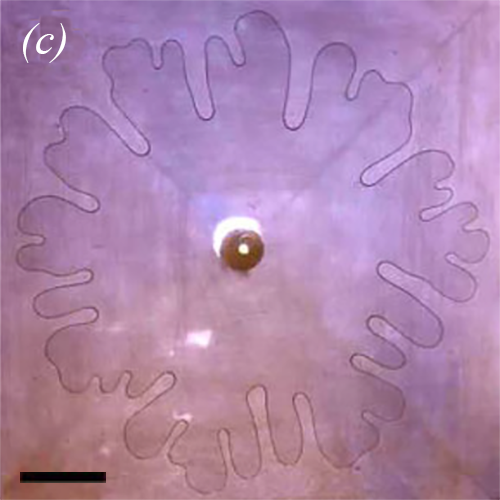}	
	
	\includegraphics[width=0.28\linewidth]{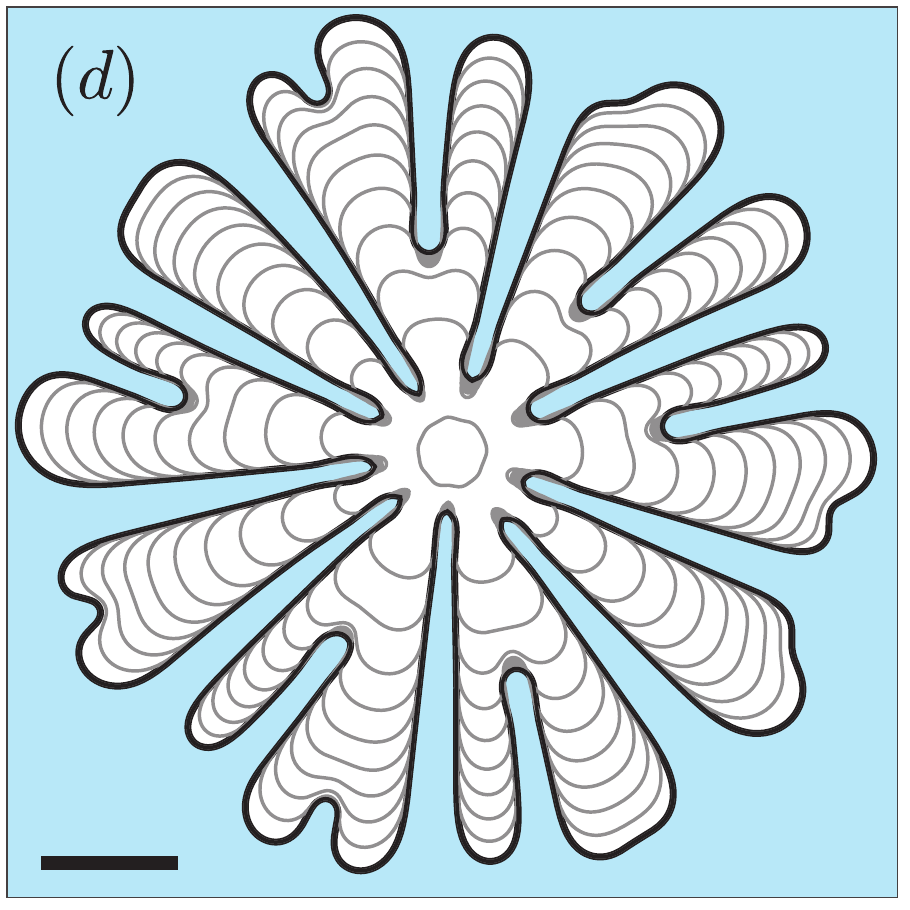}
	\includegraphics[width=0.28\linewidth]{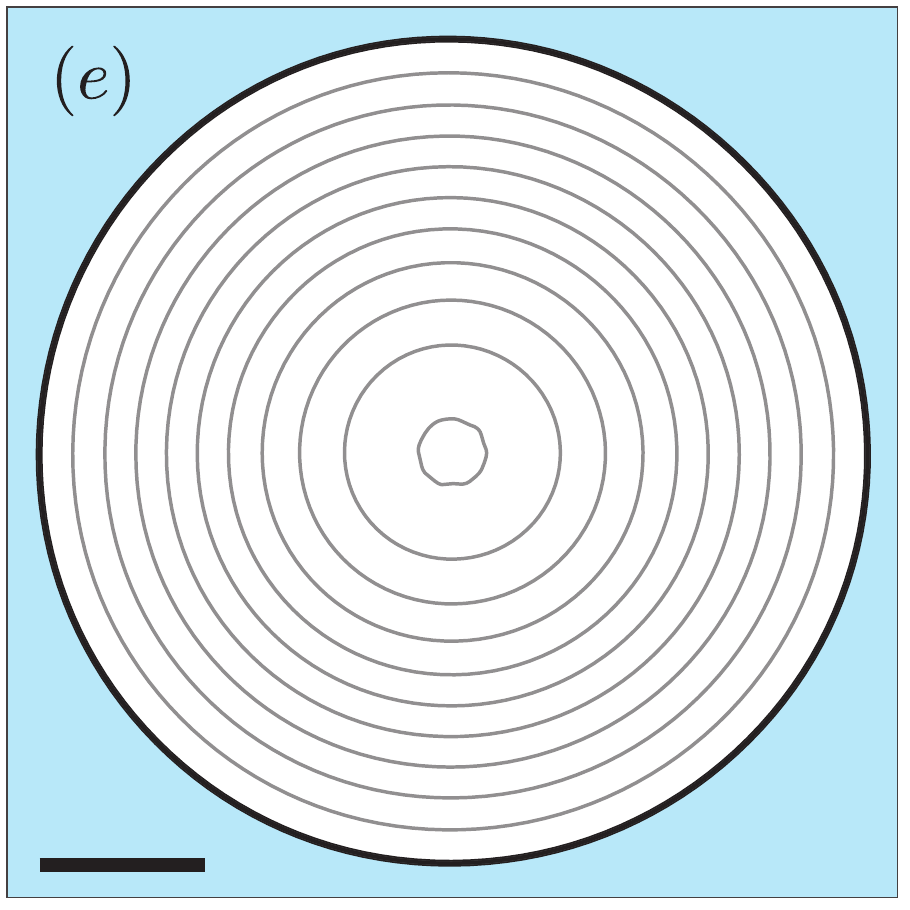}
	\includegraphics[width=0.28\linewidth]{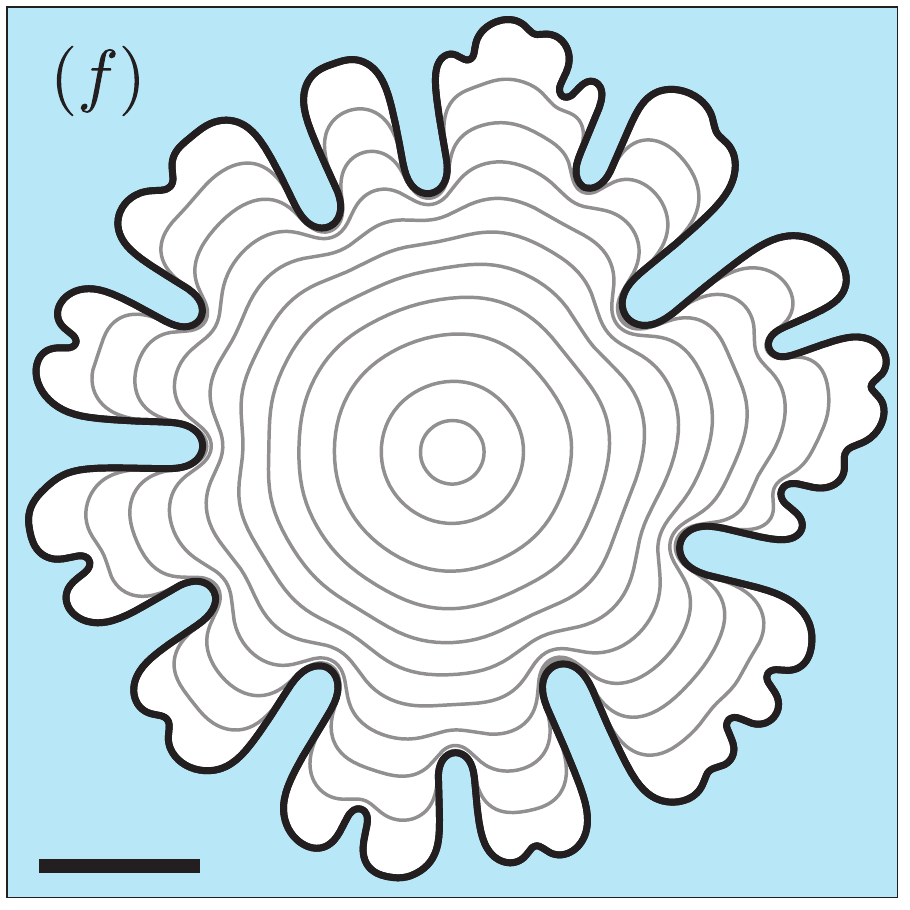}	
	\caption{$(a$-$c)$ Experimental results from \cite{Bongrand2018}, reproduced with permission  from the American Physical Society, comparing the viscous fingering pattern for different injection rates and plate configurations, and $(d$-$f)$ the corresponding numerical simulations.  The gap thickness is of the form \eqref{eq:TaperedPlate} with $(d)$ $Q = 2/3$ mL/s, $b_0 = 0.12$ cm, $\alpha = 0$, and $t_f = 14$ s, $(e)$ $Q = 2/3$ mL/s, $b_0 = 1.5 \times 10^{-2}$ cm, $\alpha = 6.67 \times 10^{-2}$, $r_0 = 7$ cm, and $t_f = 30$ s, and $(f)$ $Q = 11/6$ mL/s, $b_0 = 5 \times 10^{-2}$ cm, $\alpha = 4.75 \times 10^{-2}$, $r_0 = 7$ cm, and $t_f = 4.5$ s. The initial condition is of the form \eqref{eq:InitialCondition} with $R_0 = 0.5$ cm. The black scale bars represent a length of 2 cm.}
	\label{fig:Figure7}
\end{figure}

The tapered Hele-Shaw problem has recently been studied experimentally by \cite{Bongrand2018} for converging plates ($\alpha  > 0$), using weakly nonlinear analysis by \cite{Anjos2018} and numerically by \cite{Jackson2017} (who only considered the evolution of 8-fold symmetric bubbles). We extend these studies by providing insight into how effective tapering the plate gap is at reducing the development of viscous fingering by comparing simulations over a range of values of $\alpha$ and $Q$ to the corresponding parallel plate case. To confirm that our numerical solutions are consistent with the experimental results of \cite{Bongrand2018}, we compare simulations with these experiments for different values of $\alpha$, $b_0$, and $Q$, shown in figure \ref{fig:Figure7}.  For parallel plates ($\alpha = 0$) our simulations are able to reproduce the classic viscous fingering patterns observed experimentally (figure~\ref{fig:Figure7}$(a)$,$(d)$). When the plates are converging and the injection rate is sufficiently low, experimentally it is observed that the interface is stabilised, which is reproduced by our numerical simulations (figure~\ref{fig:Figure7}$(b)$,$(e)$). For a faster injection rate, the interface is unstable and develops fingers that appear slightly different to traditional viscous fingers (although the mechanism is presumably the same); our numerical solution is able to reproduce this morphology (figure~\ref{fig:Figure7}$(c)$,$(f)$).

Returning to our control objective that involves injecting the same volume of inviscid fluid over a fixed period of time, we compare numerical simulations in figure \ref{fig:Figure4} for different values of $\alpha$ over four different constant injection rates.  We see that tapering the plates in the direction of flow ($\alpha = 5.3 \times 10^{-2}$; third row of figure \ref{fig:Figure4}) delays tip-splitting and produces shorter fingers compared to the corresponding parallel plate configuration (first row of figure \ref{fig:Figure4}). Furthermore, for the lowest of the four injection rates, $\bar{Q} = 0.6$ mL/s (first column of figure \ref{fig:Figure4}), tapering the plates in this way completely suppresses the development of viscous fingers over the duration of the simulation. For faster injection rates, figure \ref{fig:Figure4} indicates that for times close to $t_f$, the interface develops numerous short fingers (\cite{Bongrand2018} refer to these as ``wavy'' fingers) which does not occur when $\alpha = 0$.  On the other hand, tapering the plates so they are diverging in the direction of flow has a qualitatively different effect.  Here ($\alpha = -5.3 \times 10^{-2}$; fifth row of figure \ref{fig:Figure4}), simulations indicate that the interface develops numerous long fingers and tip-splitting is reduced compared to the parallel plate case. Interestingly, increasing $\bar{Q}$ does not appear to significantly increase either the number or length of fingers that develop compared to the other configurations considered.

To better quantify how the interfacial instabilities develop when the plates are tapered over the duration of a simulation, we compare the isoperimetric ratio $\mathcal{I}$ for the tapered and parallel plate configurations for a particular flow rate, shown in figure \ref{fig:IsoperimetricExample}$(a)$. When the plates are converging (dashed red), the isoperimetric ratio initially grows much slower compared to when the plates are parallel (solid blue), while for later times, it increases at a faster rate. This behaviour can be explained by noting that, in order to inject the required volume of fluid over the time period, the interface must be slower in the tapered case for small times and faster for later times (analogous to the linear injection rate \eqref{eq:LinearInjection}). For the diverging case (dotted yellow), the opposite trend is observed; here, the isoperimetric ratio initially grows more rapidly than the parallel plate case as the velocity of the interface is initially higher.  However, as the bubble expands, we find that the growth of $\mathcal{I}$ slows for times leading up to $t_f$.  The leading order effects in \eqref{eq:LSAtapered} suggest the two mechanisms responsible for this reduction in the growth of $\mathcal{I}$ are the decrease in the interface's speed and an increase in the stabilising effect of surface tension.  Despite the differences in stabilising and destabilising effects for the converging and diverging cases, both result in a reduction in $\mathcal{I}$ at $t = t_f$.

Both the isoperimetric ratio $\mathcal{I}$ and circularity ratio $\mathcal{C}$ at the final time $t = t_f$ are shown in figure \ref{fig:Figure6} for various values of $\bar{Q}$.  For $\bar{Q} \le 2$ mL/s, tapering the plates such that they converge (yellow, $\blacktriangledown$) produces a more circular interface than the parallel plate case (navy blue, $\bullet$).  Furthermore, figure \ref{fig:Figure6}$(a)$ indicates that $\mathcal{I}$ may increase above the other configurations if $\bar{Q} > 2$ mL/s. This behaviour can by explained by noting that when the plates converge, the normal velocity of the interface can significantly increase for later times due to the reduction in gap between the plates, resulting in the `wavy' fingers observed in figure \ref{fig:Figure4}. However, simulations for faster injection rates indicate that the maximum radius of the interface can increase above 7 cm, which, according to \eqref{eq:TaperedPlate}, is where the plates are no longer tapered. While this increase in normal velocity results in a sharp increase in $\mathcal{I}$ as $\bar{Q}$ increases, it does not appear to significantly increase the length of the fingers as a corresponding sharp increase in $\mathcal{C}$ is not observed. For the diverging plate case (green, $\sqbullet$), both the isoperimetric and circularity ratios are larger than that of the parallel plate case when the injection rate is slow. However, as $\bar{Q}$ is increased, both of these quantities become smaller compared to the parallel case.  Further, over the range of values of $\bar{Q}$ considered, there is relatively little variation in $\mathcal{I}$ and $\mathcal{C}$ compared to the other configurations.	Thus our results indicate that compared to the corresponding parallel configuration, we can produce a more circular interface for both slower and faster injection rates by imposing linearly converging and diverging plates, respectively.

\begin{figure}
	\centering
	\includegraphics[width=0.5\linewidth]{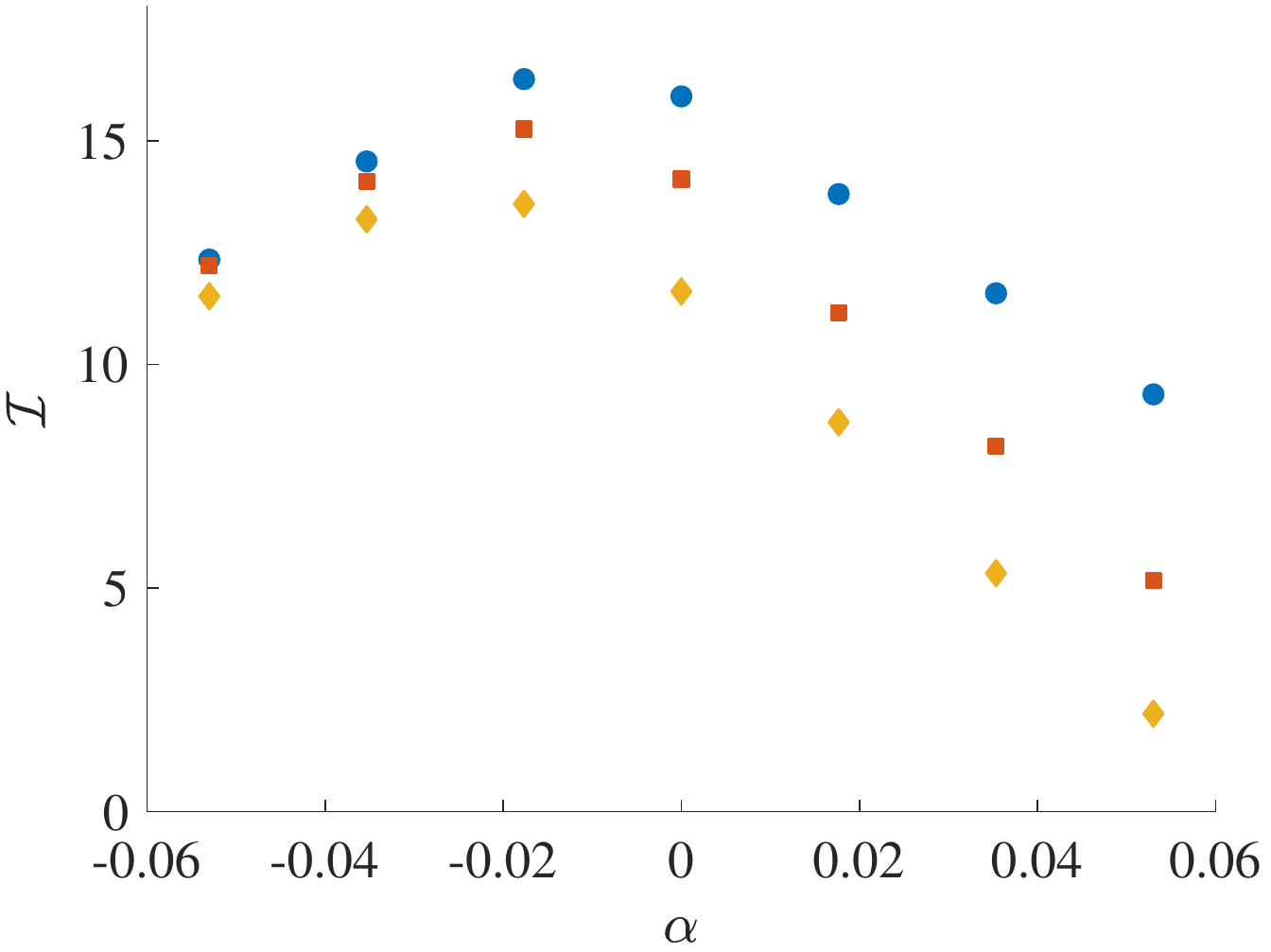}
		\caption{Isoperimetric ratio, $\mathcal{I}$, at final time $t_f$ as a function of the gradient of the taper, $\alpha$, for injection rate $Q = 2$ (blue, $\bullet$), 1.6 (red ,$\sqbullet$), and 1.2 (yellow, $\blackdiamond$) mL/s with $t_f = 7.78$, 9.61, and 12.81 s. For all simulations, $r_0 = 7$ cm, and $b_0$ and $R_0$ are chosen such that for each simulation, the volume of the inviscid bubble is 0.157 mL at $t = 0$ and 15.7 mL at $t = t_f$. The value of $\mathcal{I}$ is computed by performing 10 numerical simulations with initial condition \eqref{eq:InitialCondition}, and averaging $\mathcal{I}$ between the simulations.}
		\label{fig:isoperimetricalpha}
\end{figure}

To clarify the influence of $\alpha$ on the development of viscous fingering, we compute the isoperimetric ratio from numerical simulations for values of $\alpha$ between $-5.3 \times 10^{-2}$ and $5.3 \times 10^{-2}$ with $Q = $ 2 (blue, $\bullet$), 1.6 (red ,$\sqbullet$), and 1.2 (yellow, $\blackdiamond$) mL/s, and show the results in Figure \ref{fig:isoperimetricalpha}.  This figure indicates that for each value of $Q$ considered, $\mathcal{I}$ is a non-monotonic function of $\alpha$, and has a maximum at $\alpha \approx -0.02$. Furthermore, for $Q = 1.6$ and $2$ mL/s, $\alpha = -5.3 \times 10^{-2}$ and $5.3 \times 10^{-2}$ result in a reduction of $\mathcal{I}$ compared to $\alpha = 0$ (that is, tapering either way reduces the fingering pattern). However, choosing $\alpha = 5.3 \times 10^{-2}$ results in the smallest value of $\mathcal{I}$ for each of the injection rates considered (for these injection rates, converging plates has a greater effect of reducing the fingering pattern than diverging plates).
		
\subsection{Tapered plates with time-dependent injection} \label{sec:TaperedOptimal}

In \S~\ref{sec:TimeDependentInjection} and \S~\ref{sec:TaperedPlates}, we demonstrate that employing either the linearly increasing injection rate \eqref{eq:LinearInjection} (in a parallel plate configuration), or tapering the gap between the plates according to \eqref{eq:TaperedPlate} (with a constant injection rate) results in a less unstable interface than the corresponding constant injection case with parallel plates.   We now go further by, in the spirit of \cite{Dias2012}, deriving an optimal injection rate that attempts to minimise viscous fingering when the gap thickness is of the form \eqref{eq:TaperedPlate}.  We also perform numerical simulations to investigate the effectiveness of this configuration.  In contrast to the strategies discussed in \S~\ref{sec:TimeDependentInjection} and \S~\ref{sec:TaperedPlates}, the combination of imposing a time-dependent injection rate when the plates are tapered has not previously been considered.
	
\begin{figure}
	\centering
	\includegraphics[width=0.50\linewidth]{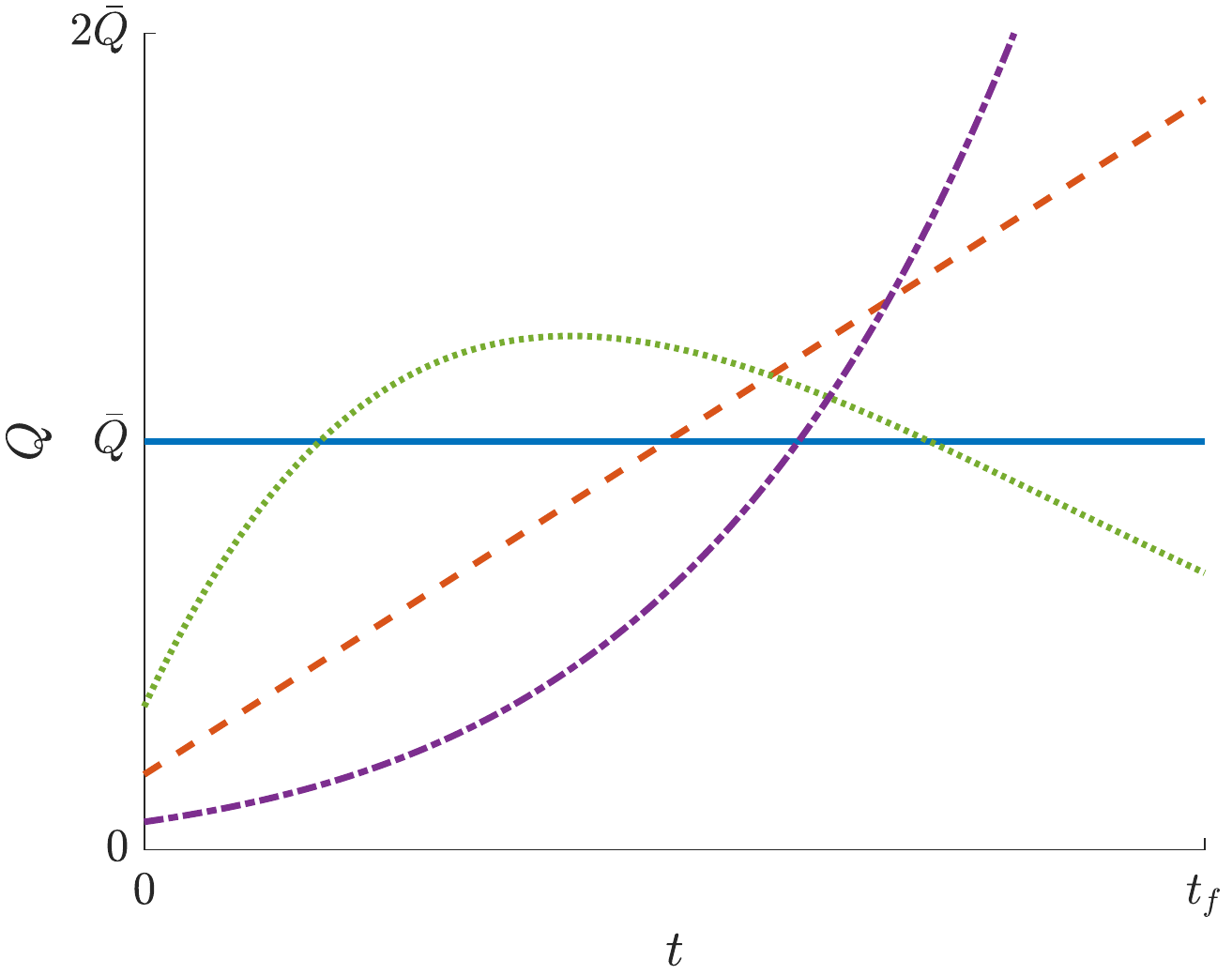}
		\caption{Sketch of different injection rates as a function of time. Solid (blue) curve is constant injection rate ($Q = 1.8$ mL/s) with parallel plates ($b = 0.2$ cm). Dashed (red) curve is optimal injection rate (\eqref{eq:LinearInjection} where $R_0 = 0.5$ cm) with parallel plates ($b = 0.2$ cm). Dotted (green) curve is optimal injection rate computed from the solution to \eqref{eq:OptimalODE} with $\alpha = 5.3 \times 10^{-2}$, $b_0 = 5 \times 10^{-3}$ cm, $r_0 = 7$ cm, and $R_0 = 0.37$ cm. Dash-dotted (purple) curve is optimal injection rate computed from the solution to \eqref{eq:OptimalODE} with $\alpha = -5.3 \times 10^{-2}$, $b_0 = 0.395$ cm, $r_0 = 7$ cm, and $R_0 = 0.94$ cm. The total amount of inviscid fluid injected over the time interval $[0,\hspace{0.5em} t_f]$ is the same for all injection rates. Additional parameters are $t_f = 8.54$ s and $R_f = 5$ cm.}
		\label{fig:Figure5}
\end{figure}

To derive the relevant optimal injection rate, we assume that $12 \mu \dot{s}_0 s_0^2 / \sigma b^2 \gg 1$ such that the most unstable mode of perturbation can be approximated by
\begin{align} \label{eq:nMaxTapered}
	n_{\max} \approx \sqrt{ \frac{4 \mu \dot{s}_0 s_0^2}{\sigma b(s_0)^2} - \frac{2s_0^2 \alpha}{3b(s_0)}}.
\end{align}
Furthermore, \eqref{eq:LSAtapered} evaluated at $n = n_{\max}$ reduces to
\begin{align} \label{eq:LSAtapered2}
	\lambda(s_0, \dot{s}_0) = \frac{\dot{\gamma}_{n_{\max}}}{\gamma_{n_{\max}}} \approx \frac{4}{3} \sqrt{\frac{\mu}{b(s_0)^2 \sigma}} \left( \dot{s}_0 - \frac{\sigma \alpha}{6 \mu} \right)^{3/2} - \frac{\dot{s}_0}{s_0} + \frac{\dot{s}_0 \alpha}{b(s_0)}.
\end{align}
The idea presented by \cite{Dias2012} is to determine an injection rate that minimises the integral
\begin{align}
	\int_{0}^{t_f} \lambda \hspace{0.2em} \textrm{d} t,
\end{align}
which is found from the solution to the Euler-Lagrange equation
\begin{align} \label{eq:EulerLagrange}
	\frac{\textrm{d}}{\textrm{d} t} \left( \frac{\partial \lambda}{\partial \dot{s}_0} \right) =\frac{\partial \lambda}{\partial s_0} .
\end{align}
By substituting \eqref{eq:LSAtapered2} into \eqref{eq:EulerLagrange}, we arrive at the second order nonlinear differential equation
\begin{align} \label{eq:OptimalODE}
	b(s_0) \ddot{s}_0 + \frac{2 \alpha \dot{s}_0^2}{3} + \frac{\alpha^2 \sigma \dot{s}_0}{9 \mu} = \frac{\alpha^3 \sigma^2}{27 \mu^2},
\end{align}
with boundary conditions $s_0(0) = R_0$ and $s_0(t_f) = R_f$. When $\alpha = 0$, \eqref{eq:OptimalODE} reduces to $\ddot{s} = 0$, the same equation derived by \cite{Dias2012} as expected. We solve \eqref{eq:OptimalODE} numerically and compute the optimal injection rate as $Q(t) = 2 \pi \dot{s}_0 s_0 b(s_0)$.

The result of this computation is presented in figure \ref{fig:Figure5}, where we compare the time-dependent injection rate for the tapered and parallel plate cases. This figure illustrates that the optimal flow rate for the tapered geometry when the plates converge in the direction of the flow (dotted green curve) is non-monotone in such a way that it is lower than the corresponding constant (solid blue curve) rate both when $t = 0$ and $t = t_f$. This choice of injection rate acts to slow the speed of the interface for small time (analogous to the linear injection rate \eqref{eq:LinearInjection}), while for later times it acts to prevent the rapid increase in speed that occurs when the interface reaches the region in which the gap between the plates is smaller.  For the diverging plates, figure \ref{fig:Figure5} shows that the optimal injection (dash-dotted purple curve) rate is monotonically increasing such that the normal velocity of the interface is slowed down when the gap between plates is smallest in exchange for a faster injection rate for later times when the gap becomes larger.

To illustrate the effect of implementing our new optimal injection rate when the plates are tapered, we include numerical results in rows four and six of figure \ref{fig:Figure4}.  To ensure the comparison with the previous three configurations is fair, we compute the solutions for the same average flow rates (we use $\bar{Q} = 0.6$, $1.2$, $1.6$, and $2$ mL/s in columns 1-5, respectively) over the same period of time.  When imposing a time-dependent injection rate while the plates linearly converge in the direction of the flow (row four), the interface appears to remain stable until near the very end of the simulation, where numerous stubby fingers form over the final time interval. It is interesting to note these fingers appear significantly shorter than the fingers that develop for the other configurations, as we discuss below. In comparison, when the plates diverge in the direction of the flow, row 6 of figure \ref{fig:Figure4} indicates that the optimal injection rate appears to have less impact on the morphology of the interface, both in regards to the number and length of fingers that develop.

For the representative case $\bar{Q} = 2$ mL/s (third column of figure~\ref{fig:Figure4}), the isoperimetric ratio as a function of time is shown in figure \ref{fig:IsoperimetricExample}(b).   Compared to the corresponding case where the injection rate is constant shown in figure~\ref{fig:IsoperimetricExample}$(a)$, we see that the optimal injection rate acts to reduce the growth rate of $\mathcal{I}$ for small time for all three values of $\alpha$. We see that when the plates are converging and the injection rate is optimal (dashed red), the interface remains essentially circular for almost all of the simulation. For late times, the isoperimetric ratio sharply increases as the interface appears to `switch' from stable to unstable and short fingers (observed in the fourth row of figure \ref{fig:Figure4}) begin to develop. When compared to the corresponding constant injection case, imposing a time-dependent injection rate decreases $\mathcal{I}(t_f)$. For diverging plates (yellow dotted), we observe that the isoperimetric ratio appears to grow slowly over the first half of the simulation, and faster over the second half, which is the opposite behaviour to the constant injection case. While the time-dependent injection rate has resulted in a decrease in $\mathcal{I}(t_f)$, this decrease is less substantial compared to the converging plate configuration.

In figure \ref{fig:Figure6}, we compare the isoperimetric ratio $\mathcal{I}$ and the circularity ratio $\mathcal{C}$ at $t = t_f$ as a function of $\bar{Q}$ for both the constant and optimal injection rates with $\alpha < 0$ (light blue, $\blacktriangleleft$) and $\alpha > 0$ (purple, $\blacktriangleright$). As was observed for the constant injection rate case (yellow, $\blacktriangledown$) discussed in \S~\ref{sec:TaperedPlates}, when the plates converge, $\mathcal{I}$ is less than the other configurations for $\bar{Q} < 2.6$ mL/s, and rapidly increases as $\bar{Q}$ becomes large. This increase in $\mathcal{I}$ corresponds to the large number of fingers that develop for late times, as seen in fourth row of figure \ref{fig:Figure4}. While the number of fingers significantly increases for large values of $\bar{Q}$, the lengths of these fingers are short compared to the diverging case, and this is reflected in the value of $\mathcal{C}$, which is the lowest over the range of $\bar{Q}$ considered for each of the configurations. As we noted for the constant injection rate case in \S~\ref{sec:TaperedPlates}, we find that injection rates above $3$ mL/s result in the maximum radius increasing above 7 cm (which is where the plates are no longer tapered). For the diverging case, as was noted by comparing the fifth and sixth rows of figure~\ref{fig:Figure4}, imposing the time-dependent injection rate does not appear to significantly impact either the number or length of fingers, and thus we see a relatively small reduction in $\mathcal{I}$ and $\mathcal{C}$. We conclude that when compared to the corresponding constant injection case, imposing a carefully chosen injection rate does result in a more circular interface for both converging and diverging plates depending on the choice of $\bar{Q}$, but the interface can exhibit very different morphological features dependent on $\alpha$. We discuss this issue further in \S~\ref{sec:Conclusion}.

\subsection{Rotating plates} \label{sec:RotatingPlates}
	
\begin{figure}
	\centering
	Constant injection \\
	\includegraphics[width=0.19\linewidth]{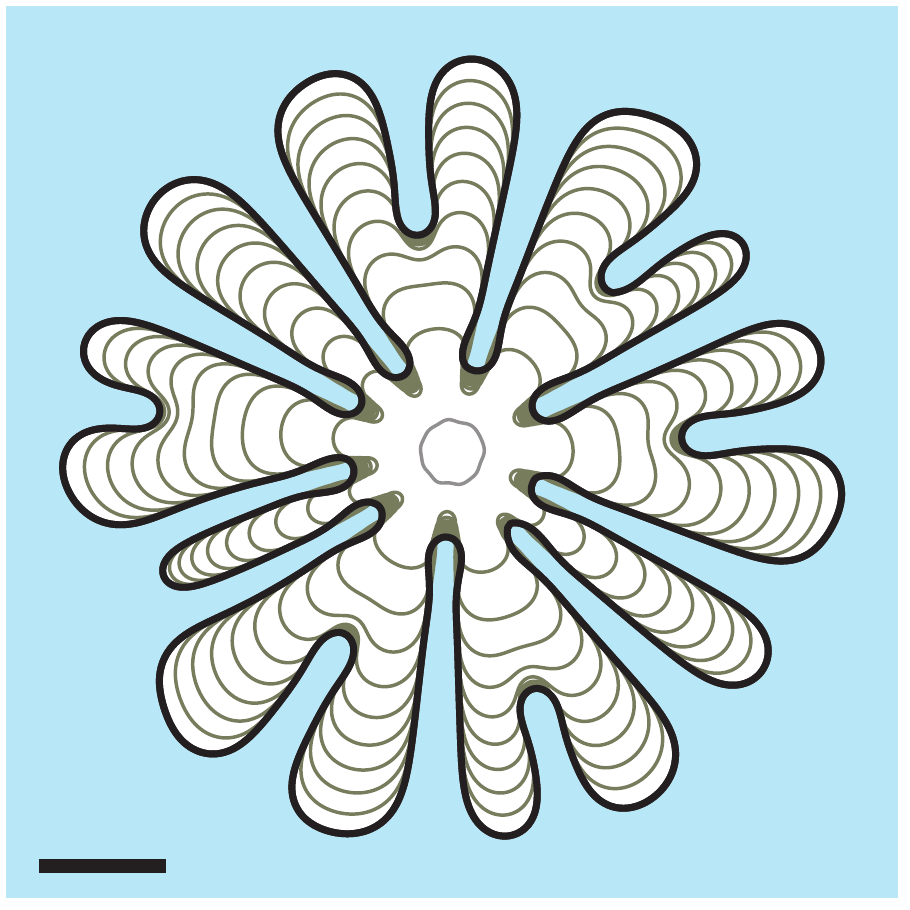}
	\includegraphics[width=0.19\linewidth]{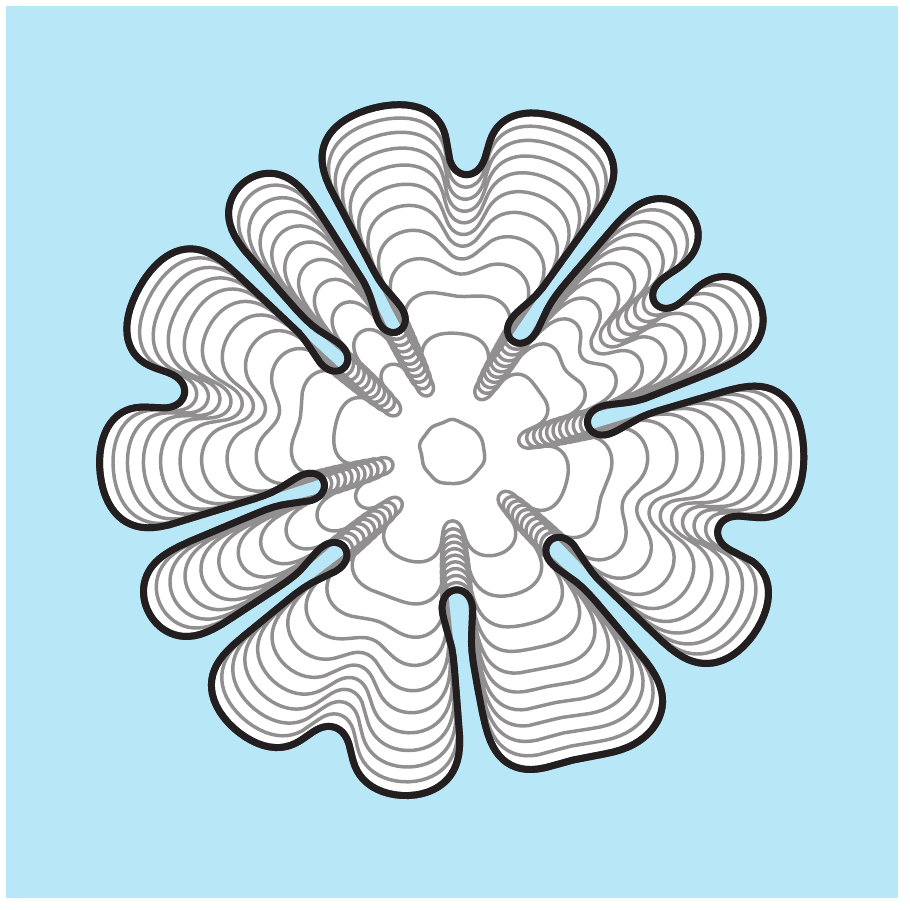}
	\includegraphics[width=0.19\linewidth]{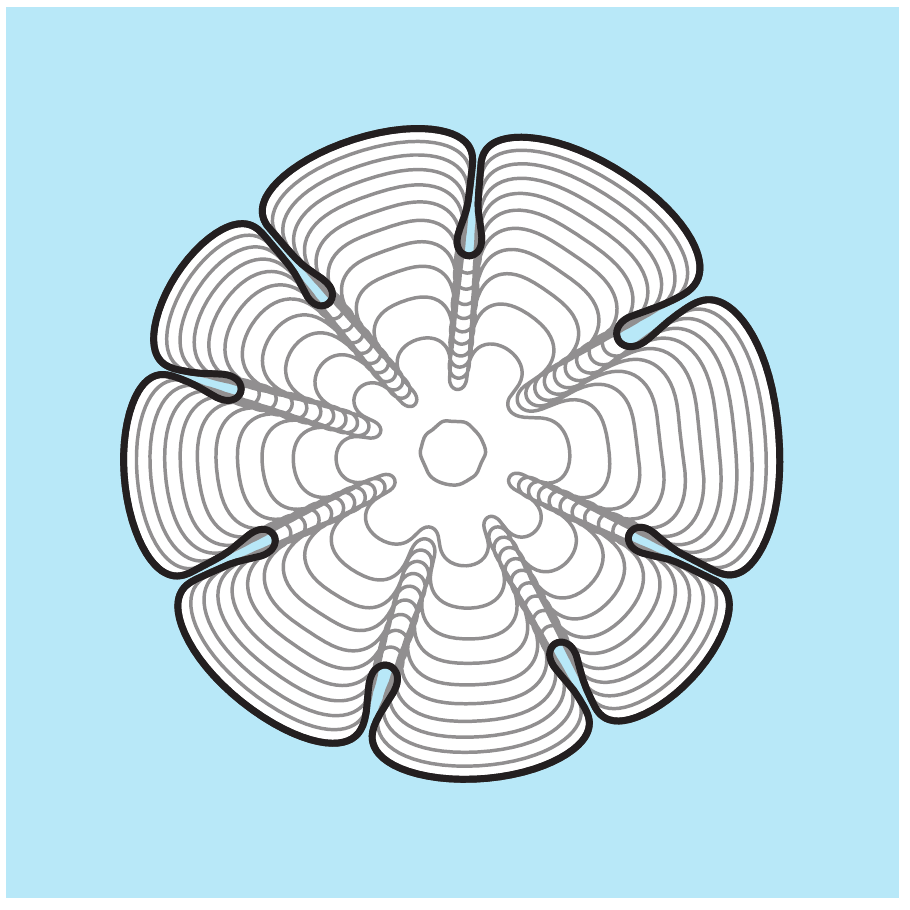}
	\includegraphics[width=0.19\linewidth]{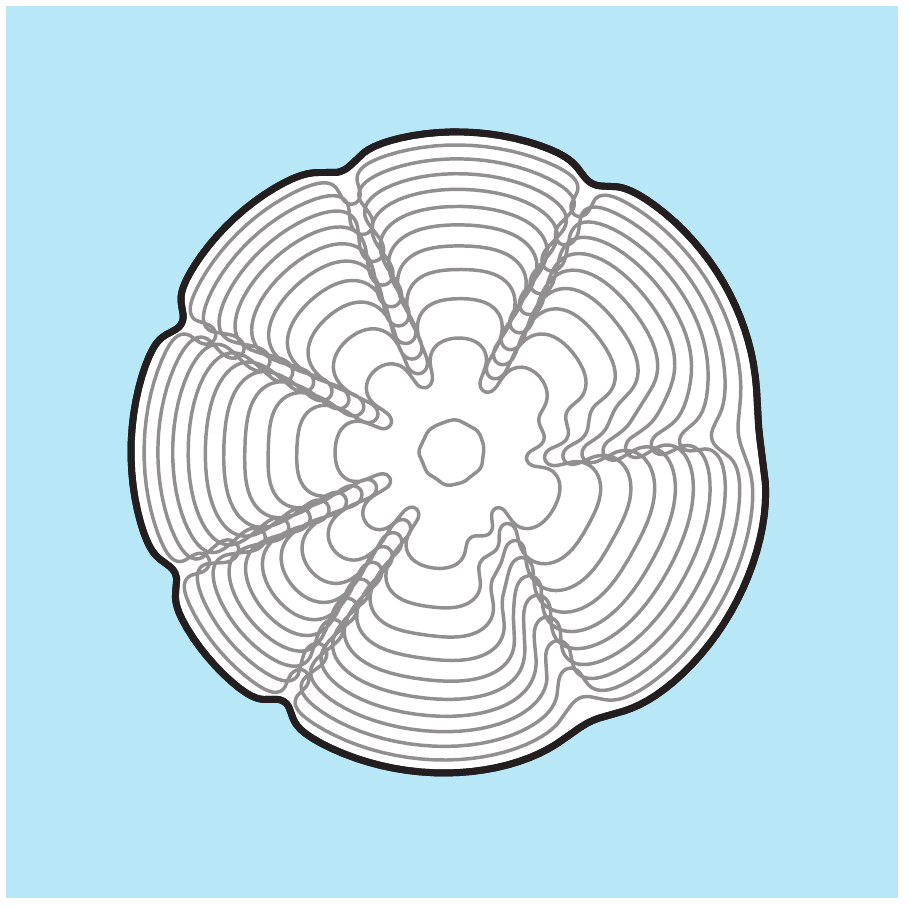}			
	\includegraphics[width=0.19\linewidth]{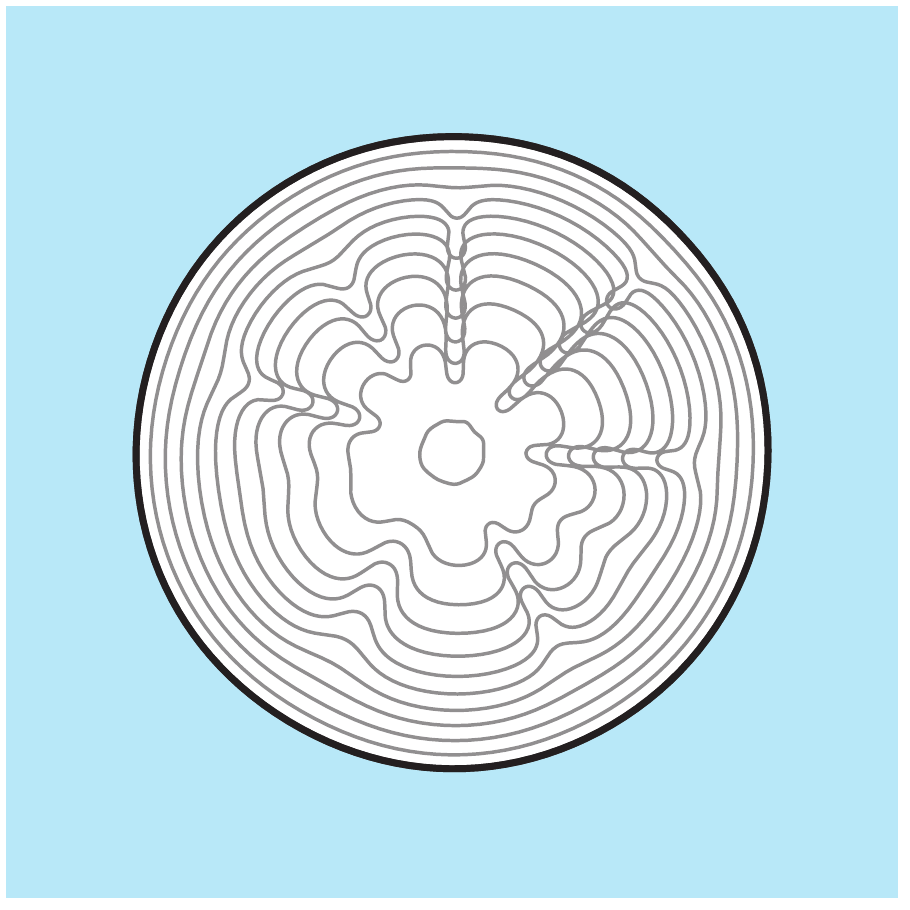} \\	
	\vspace{0.2cm}
	Optimal injection \\
	\includegraphics[width=0.19\linewidth]{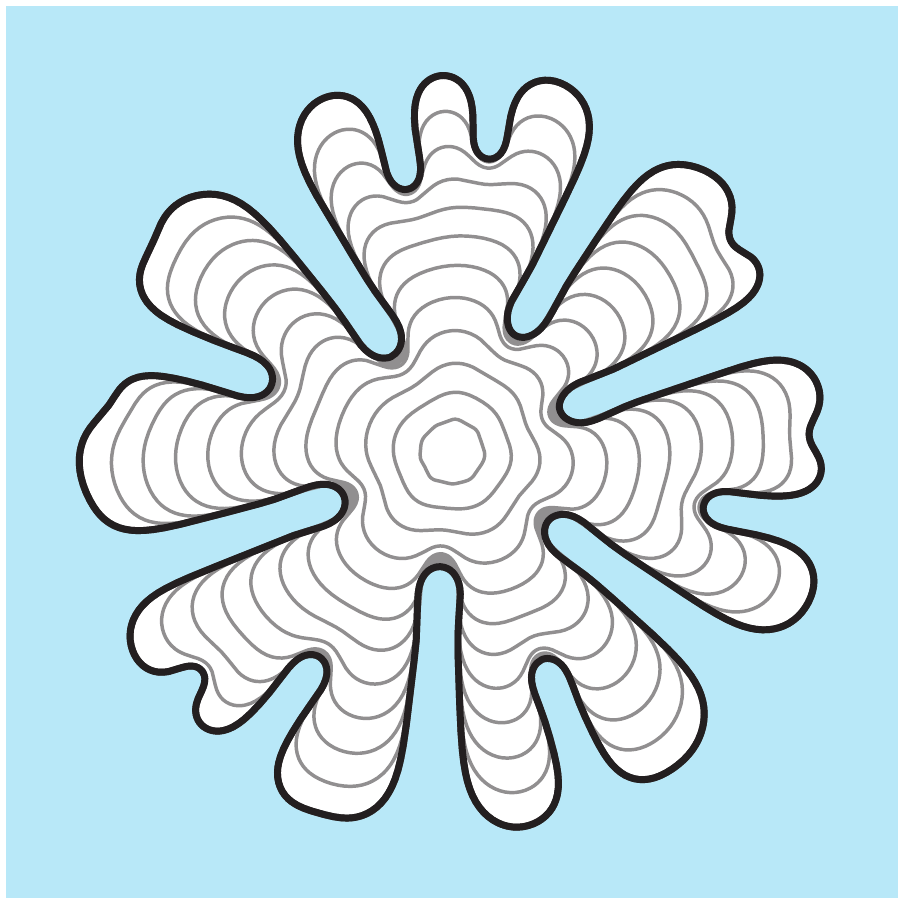}
	\includegraphics[width=0.19\linewidth]{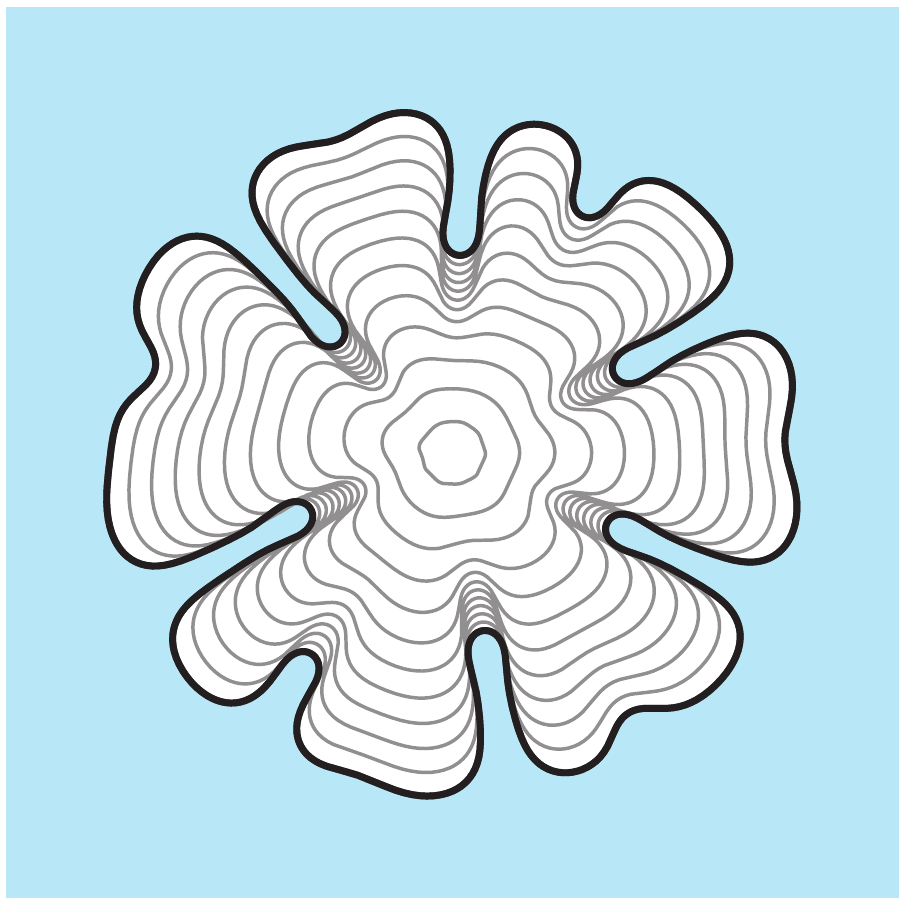}
	\includegraphics[width=0.19\linewidth]{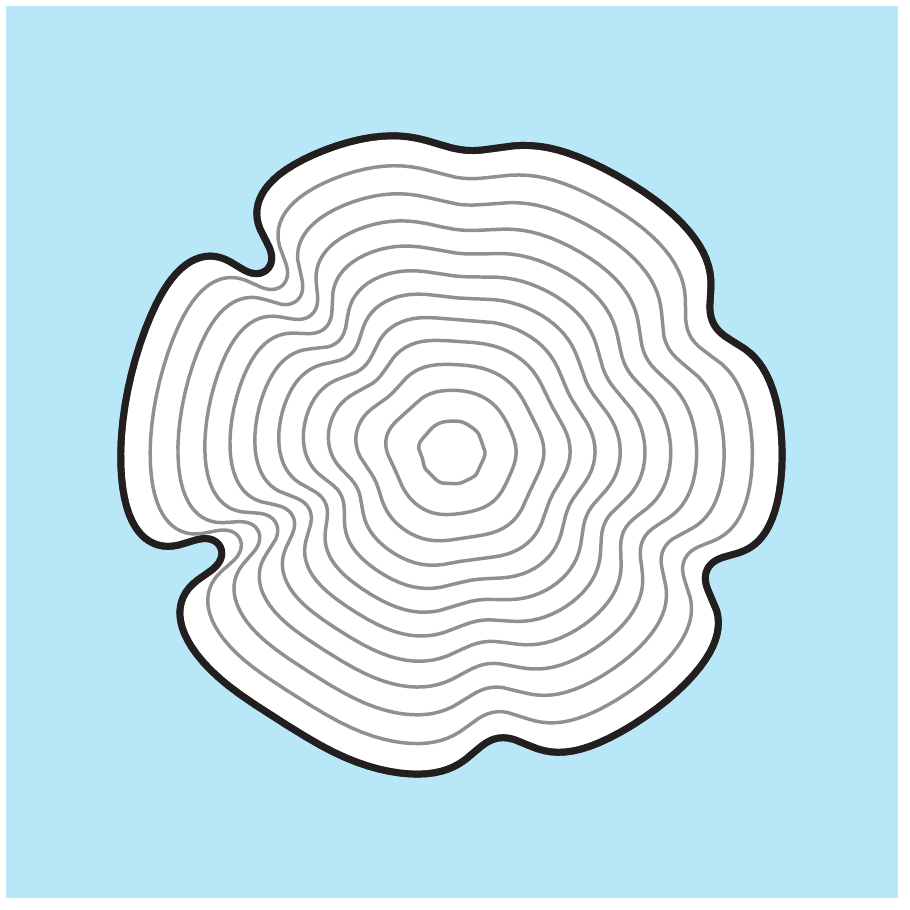}
	\includegraphics[width=0.19\linewidth]{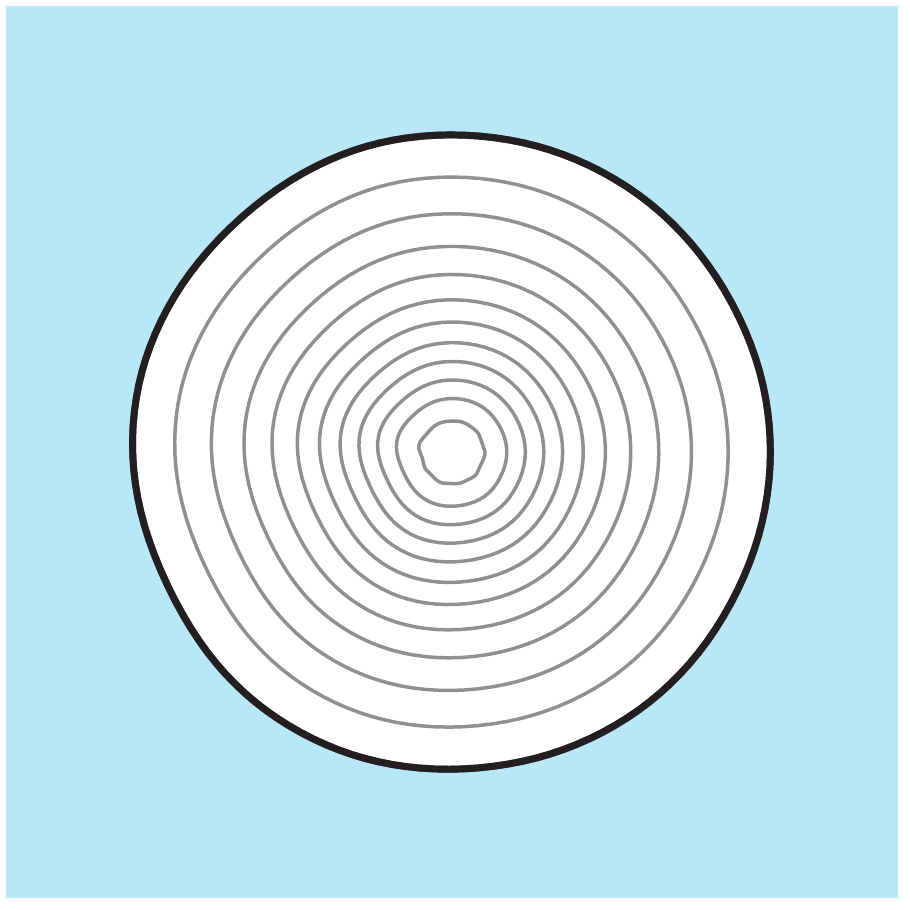}			
	\includegraphics[width=0.19\linewidth]{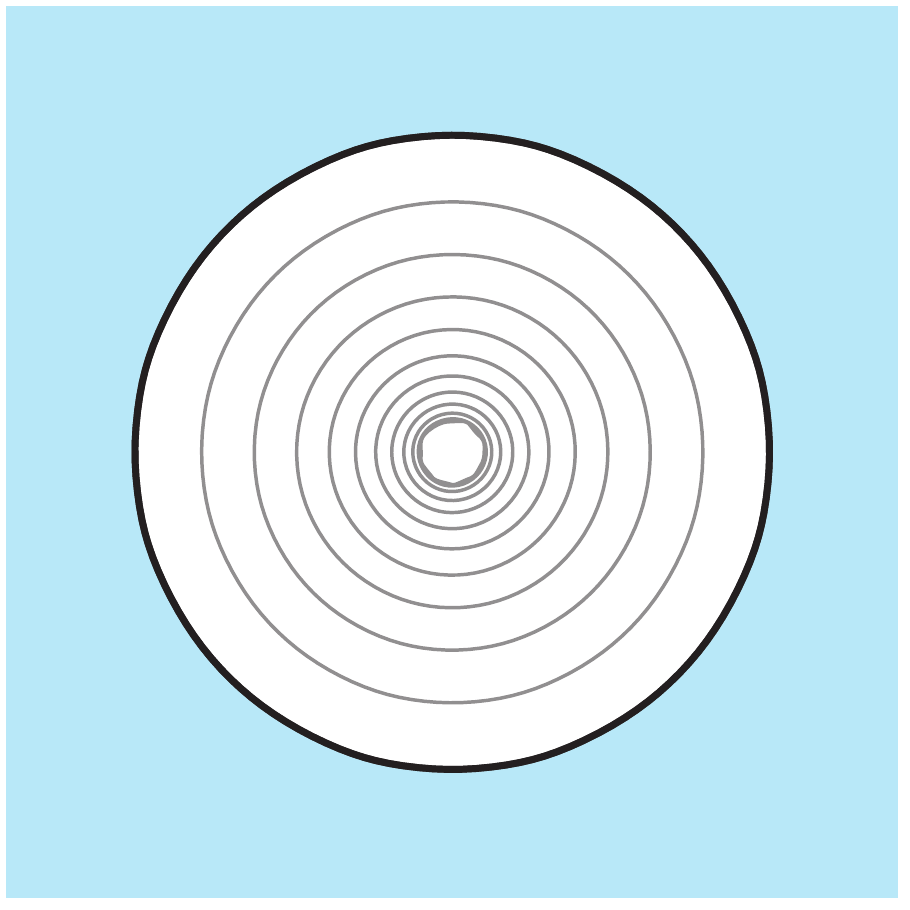} 	
		\caption{Numerical solution to \eqref{eq:Model1}-\eqref{eq:Model6} with (left to right) $\omega = 0$, 10, 20, 30, and 40 g/(s$^2\cdot$mL). Top row is for rotating plates with constant injection rate $Q = 1.6$ mL/s, and second row is for injection of the form \eqref{eq:RotatingOptimal}. The initial condition is of the form \eqref{eq:InitialCondition} with $R_0 = 0.5$ cm and $R_f = 5$ cm. For all simulations, $t_f = 9.61$ and $b = 0.2$ cm. The scale bar represents a length of 2 cm.}
		\label{fig:Omega}
\end{figure}

In this subsection, we now turn our attention to the case for which the gap between the plates is constant and the Hele-Shaw cell is rotated while the inviscid bubble is injected.  By seeking solutions to \eqref{eq:Model1}-\eqref{eq:Model6} of the form \eqref{eq:LSA1} and \eqref{eq:LSA2} when $b$ is a constant and $\omega > 0$, we find that
\begin{align} \label{eq:LSArotating}
	\frac{\dot{\gamma}_n}{\gamma_n} = \frac{(n-1)}{s_0} \left(  \frac{Q}{2 \pi b s_0} - \frac{n(n+1)b^2\sigma}{12\mu s_0^2} \right) - \frac{b^2 n \omega}{6 \mu},
\end{align}
and
\begin{align} \label{eq:nMaxRotating}
	n_{\max} = \sqrt{\frac{1}{3} \left( 1 + \frac{6 Q \mu s_0}{b^3 \pi \sigma} - \frac{2 \omega s_0^3}{\sigma} \right) }.
\end{align}
Thus, the centrifugal force acts as a stabilising term as it contributes a negative offset to $\dot{\gamma}_n / \gamma_n$ and decreases the most unstable mode of perturbation.
Furthermore, by noting that
\begin{align}
	\gamma_n(s_0) = \gamma_n(0) s_0^{n-1} \exp\left( \frac{b^3 n \pi ((n^2 - 1)\sigma - s_0^3 \omega)}{6 Q \mu s_0} \right),
\end{align}
it follows that when $n \ge 2$, $Q>0$, and $\omega > 0$, $\gamma_n(s_0) / s_0 \to 0$ as $s_0 \to \infty$. By comparison, when $\omega = 0$, $\gamma_n(s_0) / s_0 \to \infty$ as $s_0 \to \infty$. This suggests that when the plates are rotating, there exists a critical radius where all modes of perturbation will be stable and the interface will become circular. Interestingly, this result is similar to case where the plates are stationary and the bubble is contracting ($\omega = 0$ and $Q <0$), as it can be shown that for $n \ge 2$, $\gamma_n(s_0)/s_0 \to 0$ as $s_0 \to 0^+$ \citep{Dallaston2013}.

\begin{figure}
	\centering
	\includegraphics[width=0.4\linewidth]{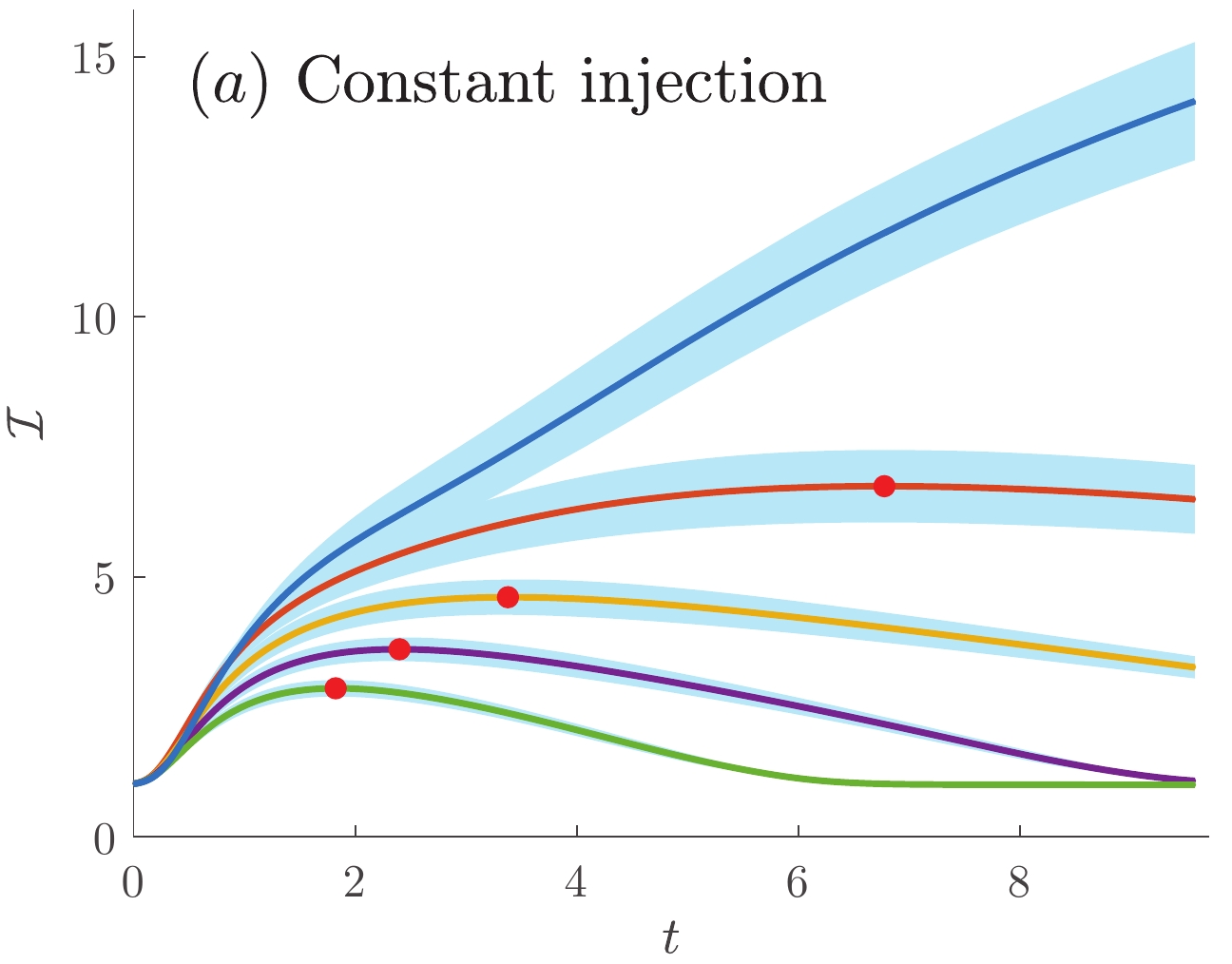}
	\includegraphics[width=0.4\linewidth]{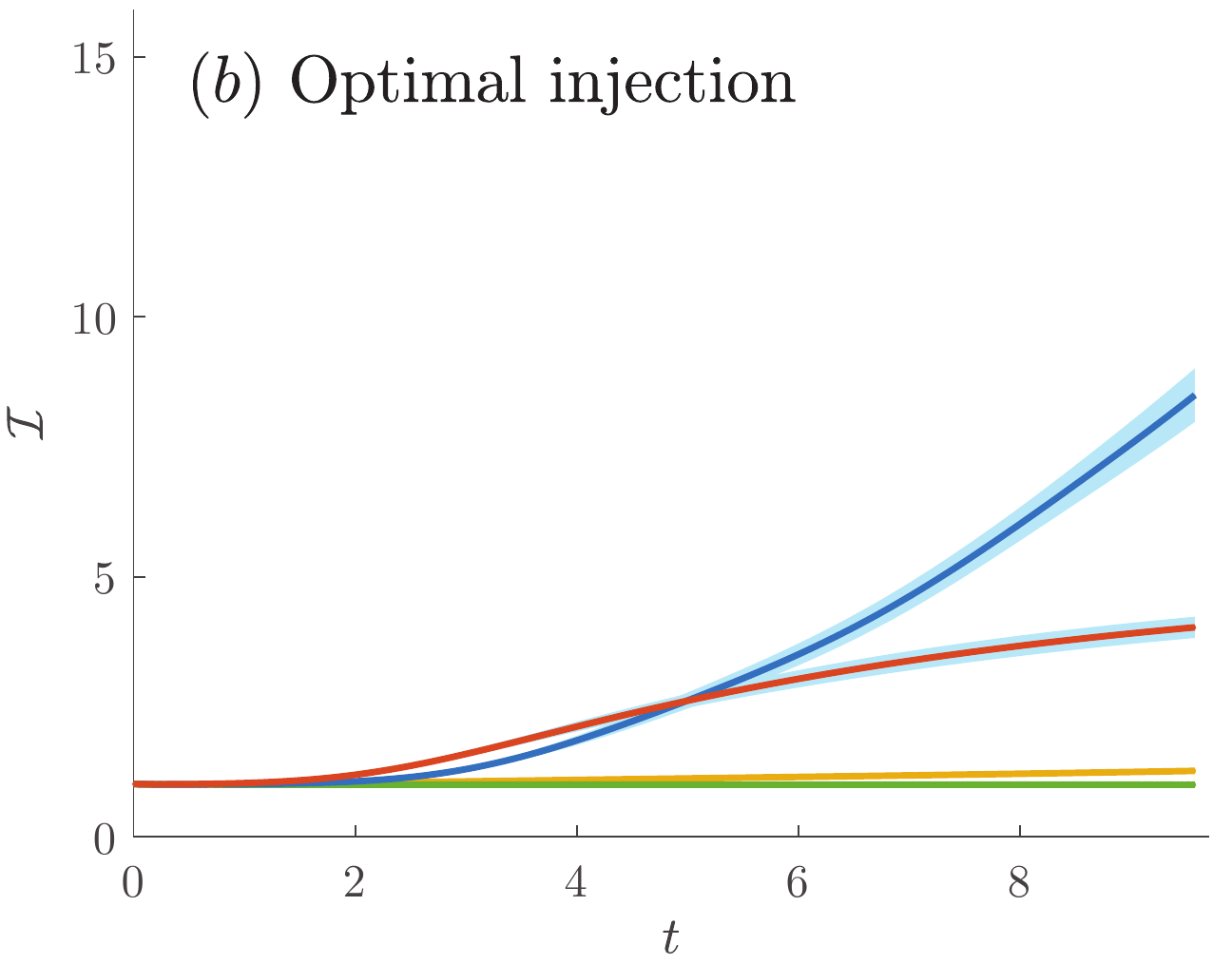}	
		\caption{Isoperimetric ratio of the numerical solution to \eqref{eq:Model1}-\eqref{eq:Model6} (top to bottom) $\omega = 0$, 10, 20, 30, and 40 g/(s$^2\cdot$mL) for $(a)$ constant and $(b)$ time-dependent injection rate \eqref{eq:RotatingOptimal}. For all simulations, the same amount of is injected over time period where $t_f = 9.61$ and the average injection rate is $\bar{Q} = 1.6$ mL/s. The initial condition is of the form \eqref{eq:InitialCondition} with $R_0 = 0.5$ cm and $R_f = 5$ cm. Shaded (blue) region represents one standard deviation above and below the mean. For all simulations, $t_f = 9.61$ and $b = 0.2$ cm.}
		\label{fig:RotatingIsoperimetric}
\end{figure}

To illustrate the nonlinear behaviour of solutions to \eqref{eq:Model1}-\eqref{eq:Model6} when $\omega > 0$, we perform numerical simulations for different values of $\omega$, shown in the first row of  figure \ref{fig:Omega}. For $\omega = 10$ g/(s$^2\cdot$mL) (second column), we see that both the number and length of fingers that develop is less than that for the case in which the plates are stationary (first column). For larger values of $\omega$, we find that fingers initially develop; however, as time increases, the base of these fingers appear to be `pulled' towards the finger tips, and in the case of $\omega = 40$ g/(s$^2\cdot$mL) (fifth column), the interface appears to be essentially circular at $t = t_f$. We compute the corresponding isoperimetric ratio of these simulations, shown in figure \ref{fig:RotatingIsoperimetric}$(a)$. For cases where $\omega > 0$, we find that while $\mathcal{I}$ initially increases as it does for when $\omega = 0$, there exists a turning point (denoted by red dots) after which $\mathcal{I}$ monotonically decreases. This is consistent with the behaviour predicted by linear stability analysis, and thus our results suggest that when $\omega > 0$, the interface will become circular after a sufficient amount of time has passed.
	
The explanation for why the centrifugal force causes the interface to become circular for sufficiently large times relates to the dynamic boundary condition \eqref{eq:Model4}. This equation indicates that as the interface expands and fingers develop, the centrifugal force creates a pressure differential between the base and tip of the fingers. From the kinematic boundary condition \eqref{eq:Model5}, we see that this pressure differential acts to increase the normal velocity of the interface at the base of the finger. By comparison, when $\omega = 0$ this pressure differential is absent and the normal velocity of the base fingers tends to be slower than that of the tips (see row one figure \ref{fig:Figure4} for example). Furthermore, the normal velocity of the base of the finger increases linearly in $r$, suggesting that as the interface grows, the effect of the centrifugal force becomes stronger. Thus, while surface tension can be thought of as `penalising' regions where curvature is high, the centrifugal term acts to penalise longer fingers.

\begin{figure}
	\centering
	\includegraphics[width=0.5\linewidth]{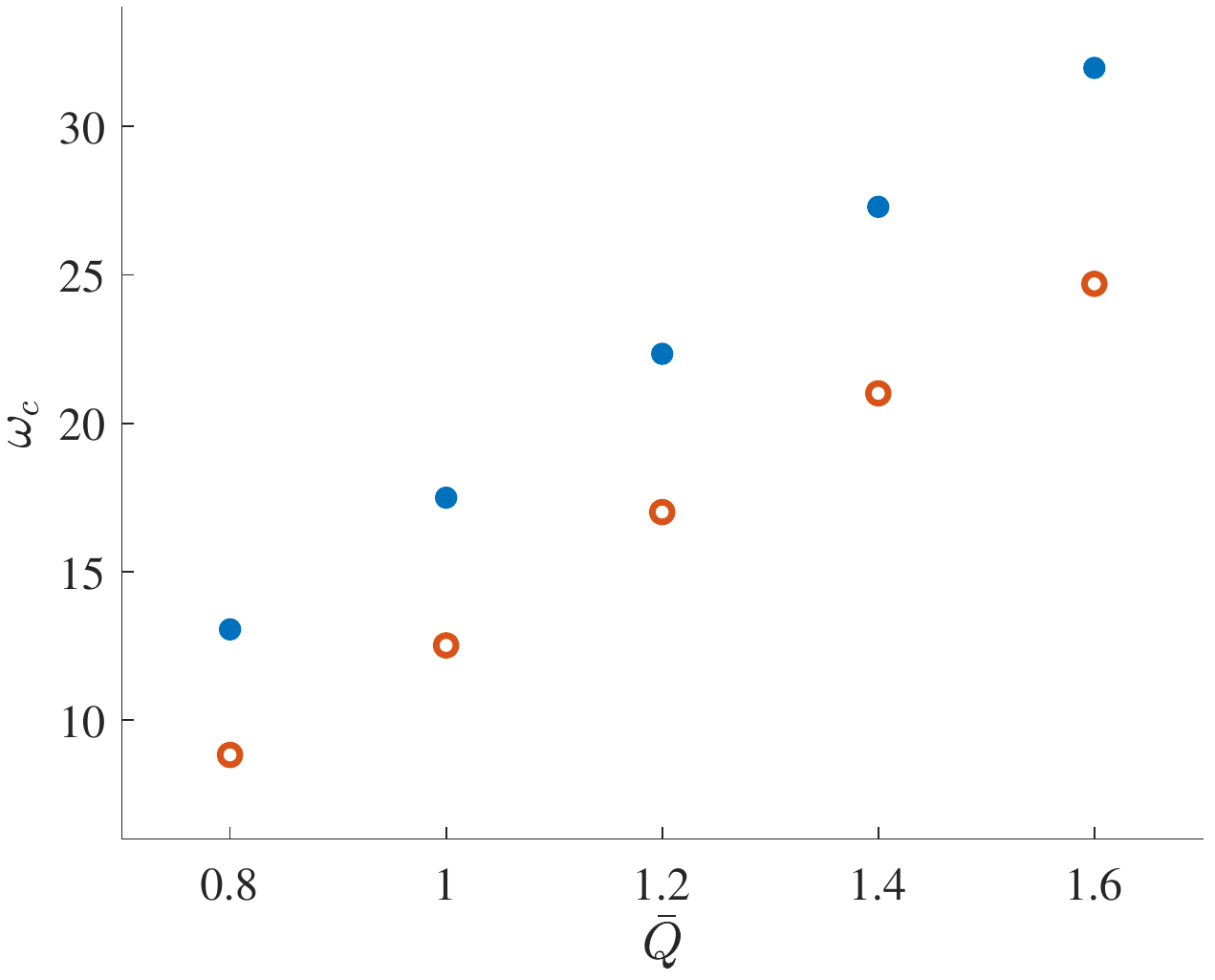}
		\caption{Stability diagram denoting the critical centrifugal parameter, $\omega_c$, such that $\mathcal{I}(t_f) = 1.01$ for constant ($\bullet$, blue) and time-dependent ($\circ$, red) injection rates. The parameter $\bar{Q}$ (mL/s) is the average injection rate over a simulation and the final time is  $t_f = ((99 \pi / 20) \textrm{ mL}) / \bar{Q}$. The initial condition is of the form \eqref{eq:InitialCondition} with $R_0 = 0.5$ cm. For the time-dependent injection cases, $Q$ is \eqref{eq:RotatingOptimal} with $R_f = 5$ cm. Additionally, $b = 0.2$ cm.}
		\label{fig:StabilityDiagram}
\end{figure}

Regarding our objective of reducing the growth of viscous fingers when a prescribed amount of fluid is injected over a finite period of time, figure \ref{fig:RotatingIsoperimetric}$(a)$ suggests that for a particular $Q$, the interface will be essentially circular at $t = t_f$ for sufficiently large $\omega$. Thus, we wish to determine the minimum value of $\omega$ that ensures the interface will be circular at $t = t_f$. We perform a parameter sweep of $\omega$ for a particular value of $Q$, and determine the critical value of the centrifugal parameter, $\omega_c$, as the value of $\omega$ where $\mathcal{I}(t_f) = 1.01$, denoted in figure \ref{fig:StabilityDiagram} by $\bullet$ (blue). For each value of $Q$ considered, we are able to approximate $\omega_c$ such that the interface is circular at the end of the simulation when $\omega > \omega_c$. Of course, when $\omega < \omega_c$, we expect the interface to become circular if the simulations were run for a longer period of time. It is interesting to note that for the tapered plate configuration discussed in \S~\ref{sec:TaperedPlates}, $\alpha$ can be chosen such that $\mathcal{I}(t_f)$ will be less than the corresponding parallel plate configuration (see figure \ref{fig:isoperimetricalpha} for example). However, our results suggest that there does not exist an analogous critical taper angle for every $Q$ such that the interface will be completely stabilised over the duration of a simulation.

Returning to time-dependent injection rates, as was discussed in \S~\ref{sec:TimeDependentInjection} and \S~\ref{sec:TaperedOptimal} we can derive the optimal rate for the situation in which a prescribed amount of inviscid fluid is injected over a finite period of time. Following \cite{Dias2012} and the methodology presented in \S~\ref{sec:TaperedOptimal}, we find $s_0$ satisfies a linear second-order differential equation with constant coefficients, which reduces to $\ddot{s}_0=0$ when $\omega=0$.  We can easily solve this equation exactly and compute the flow rate via $Q=2\pi b s_0\dot{s}_0$ to give
\begin{align} \label{eq:RotatingOptimal}
	Q(t) = \frac{b^3 \pi \omega}{3 \mu}  \left(C_2 \mathrm{e}^{\frac{b^2 \omega}{6 \mu} t} - 2C_1\mathrm{e}^{-\frac{b^2 \omega}{3 \mu} t} \right)  \left( C_1 \mathrm{e}^{-\frac{b^2 \omega}{3 \mu} t} +C_2  \mathrm{e}^{\frac{b^2 \omega}{6 \mu} t} \right),
\end{align}
where
\begin{align}
	C_1 = \mathrm{e}^{\frac{b^2 t_f \omega}{3 \mu}} \left(R_0\mathrm{e}^{\frac{b^2 t_f \omega}{6 \mu}} - R_f \right) \left(\mathrm{e}^{\frac{b^2 \omega t_f}{2 \mu}} - 1\right)^{-1},
	\quad C_2 = R_0 - C_1.
\end{align}
To illustrate the effect of implementing this injection rate, we compare fully nonlinear simulations of our Hele-Shaw problem using \eqref{eq:RotatingOptimal} (second row of figure \ref{fig:Omega}) with the corresponding constant injection case (first row). Imposing \eqref{eq:RotatingOptimal} has the effect of both reducing the number and length of viscous fingers that form, and the interface appears to be completely stabilised for $\omega = 30$ and 40 g/(s$^2\cdot$mL). Regarding the isoperimetric ratio (figure \ref{fig:RotatingIsoperimetric}$(b)$), we find that, similar to the parallel and tapered plate configurations discussed in \S~\ref{sec:TimeDependentInjection}-\ref{sec:TaperedOptimal}, the optimal injection rate initially reduces the growth of $\mathcal{I}$ compared to the corresponding constant injection rate (figure \ref{fig:RotatingIsoperimetric}$(a)$), in exchange for a faster growth rate for times leading up to $t_f$. We also note that $\mathcal{I}$ is now a monotonically increasing function in time, and the turning point observed when $Q$ is constant is absent. As expected, implementing \eqref{eq:RotatingOptimal} reduces $\mathcal{I}(t_f)$ compared to the corresponding constant injection case, and in particular, for $\omega = 30$ and 40 g/(s$^2\cdot$mL), the interface is stabilised over the entire duration of the simulation. Finally, we are able to determine the critical centrifugal parameter, $\omega_c$, (denoted as $\circ$ in figure \ref{fig:StabilityDiagram}) and, as expected, implementing \eqref{eq:RotatingOptimal} results in a reduction of $\omega_c$ compared to the corresponding constant injection case (denoted with $\bullet$).
	
\section{Controlling the number of fingers} \label{sec:NumberOFingers}
	
\begin{figure}
	\centering
	\includegraphics[width=0.22\linewidth]{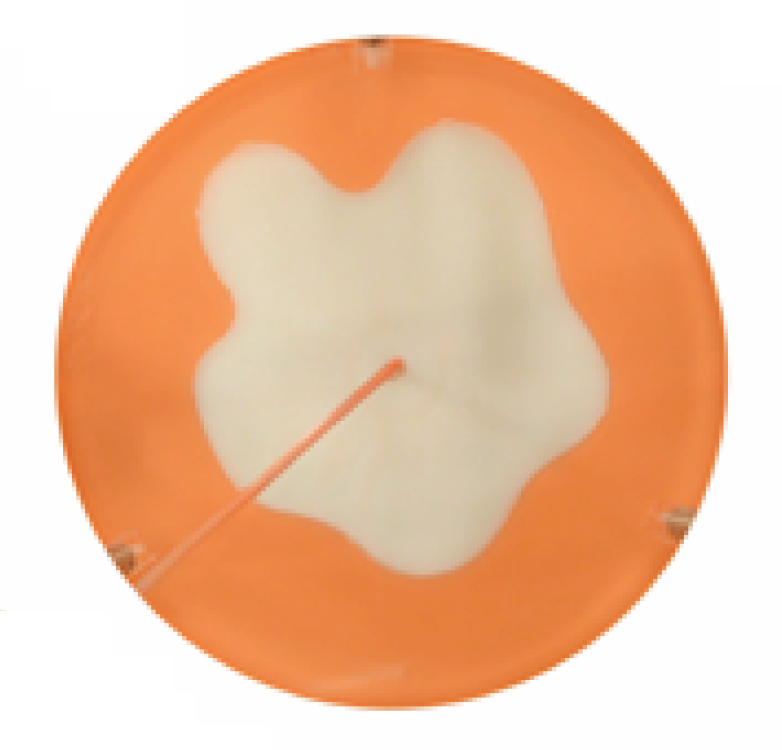}
	\includegraphics[width=0.22\linewidth]{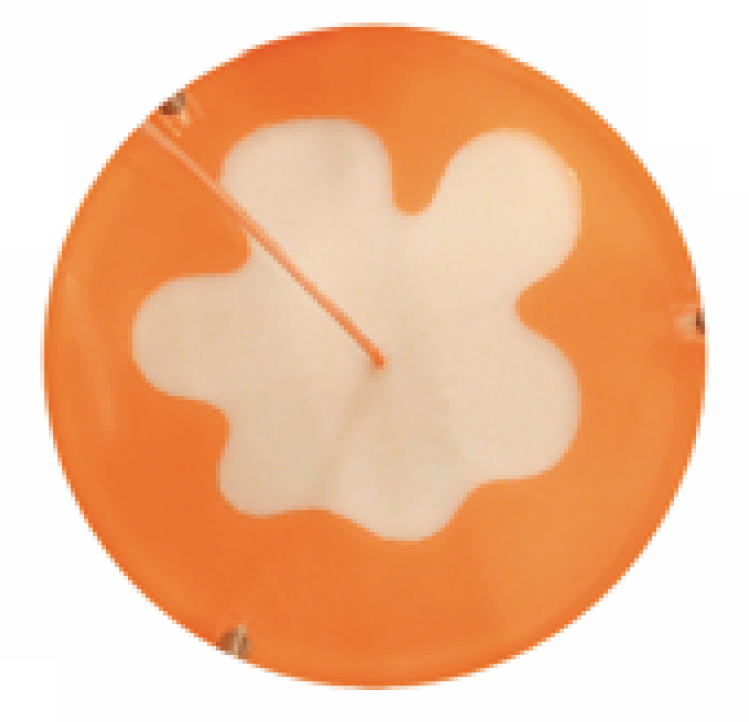}
	\includegraphics[width=0.22\linewidth]{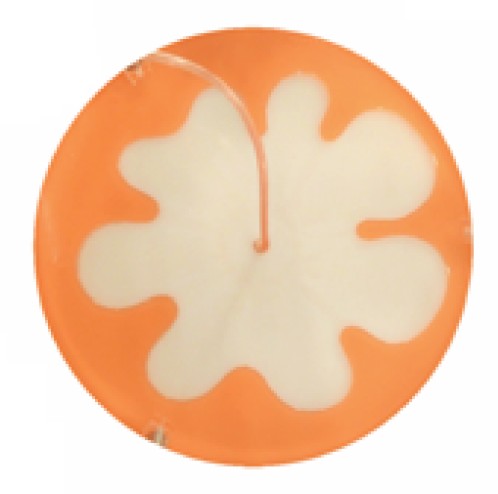}
	\includegraphics[width=0.22\linewidth]{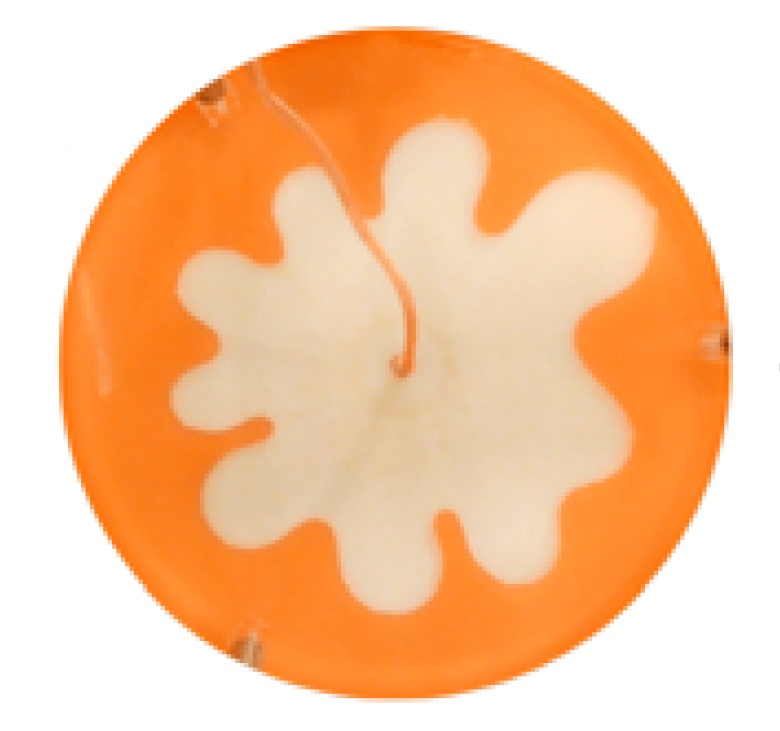} \\
	\includegraphics[width=0.22\linewidth]{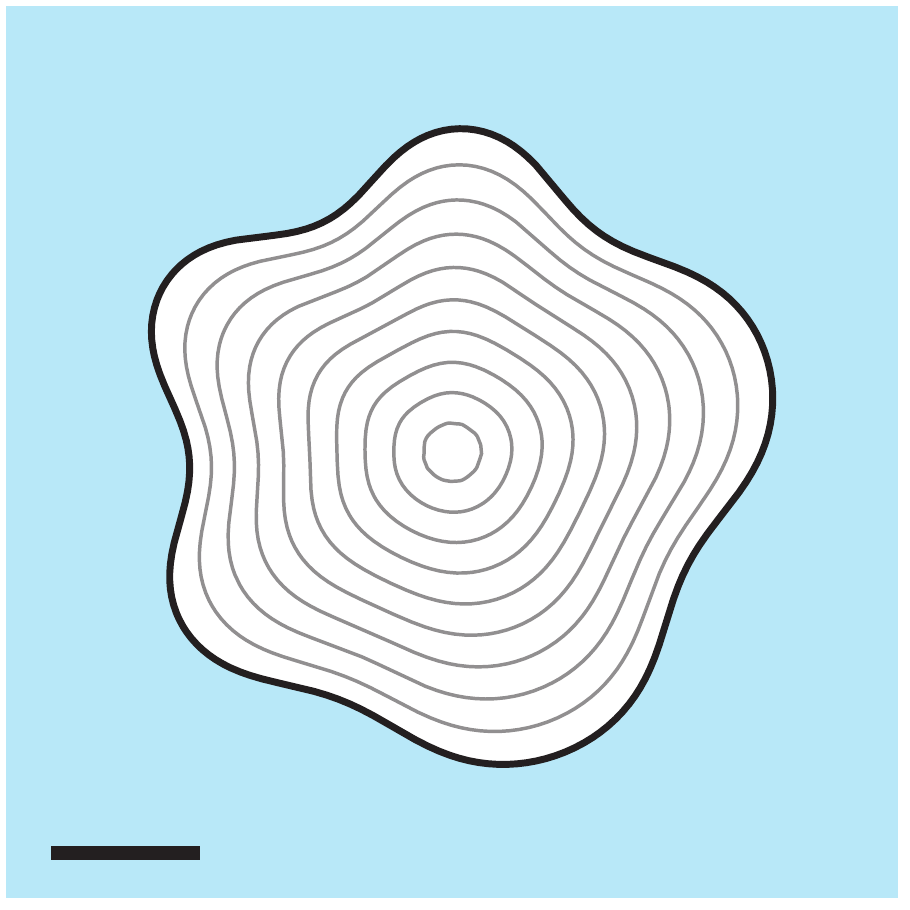}
	\includegraphics[width=0.22\linewidth]{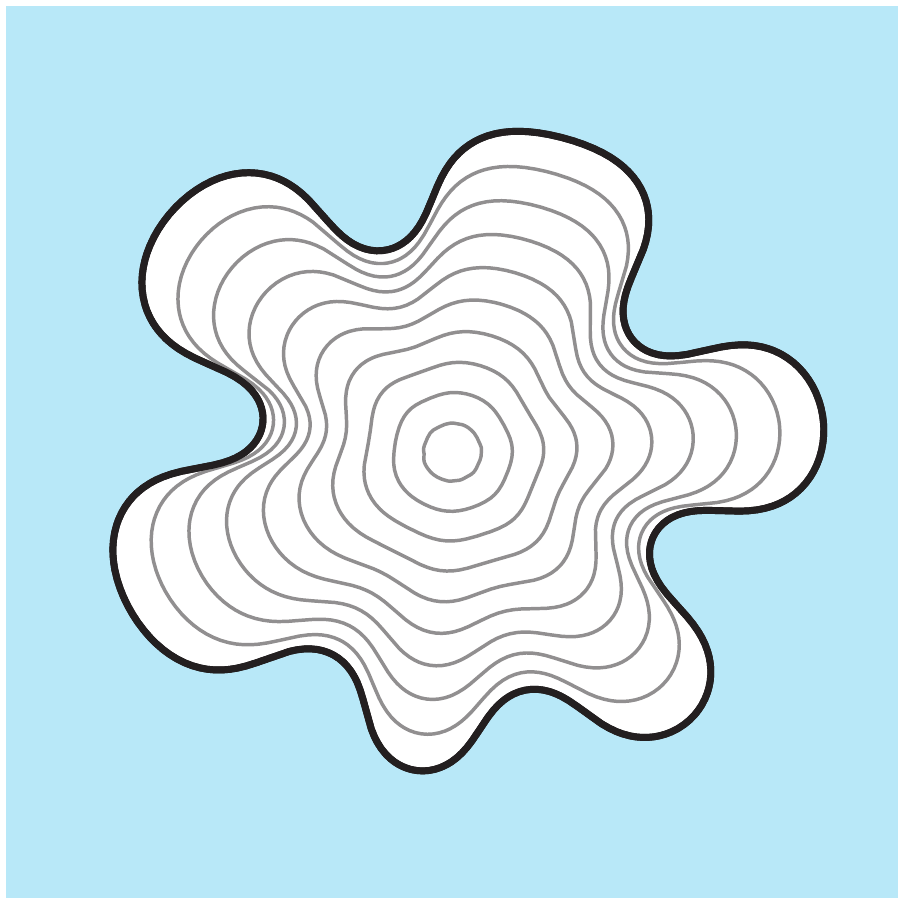}
	\includegraphics[width=0.22\linewidth]{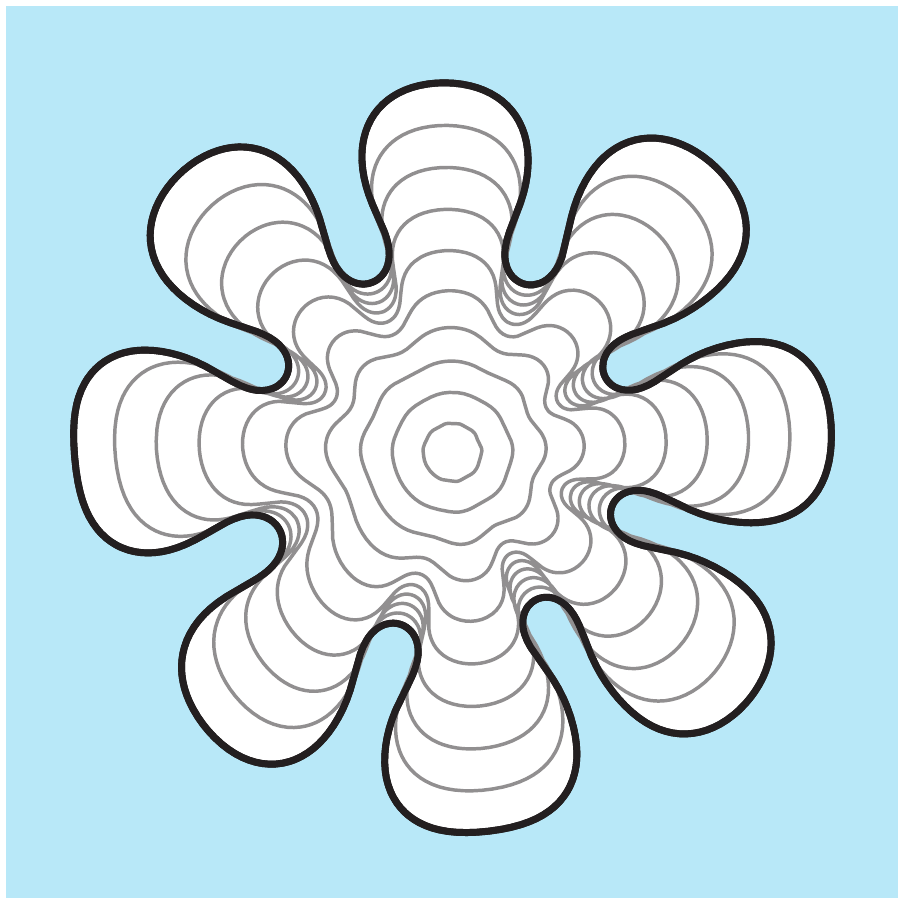}
	\includegraphics[width=0.22\linewidth]{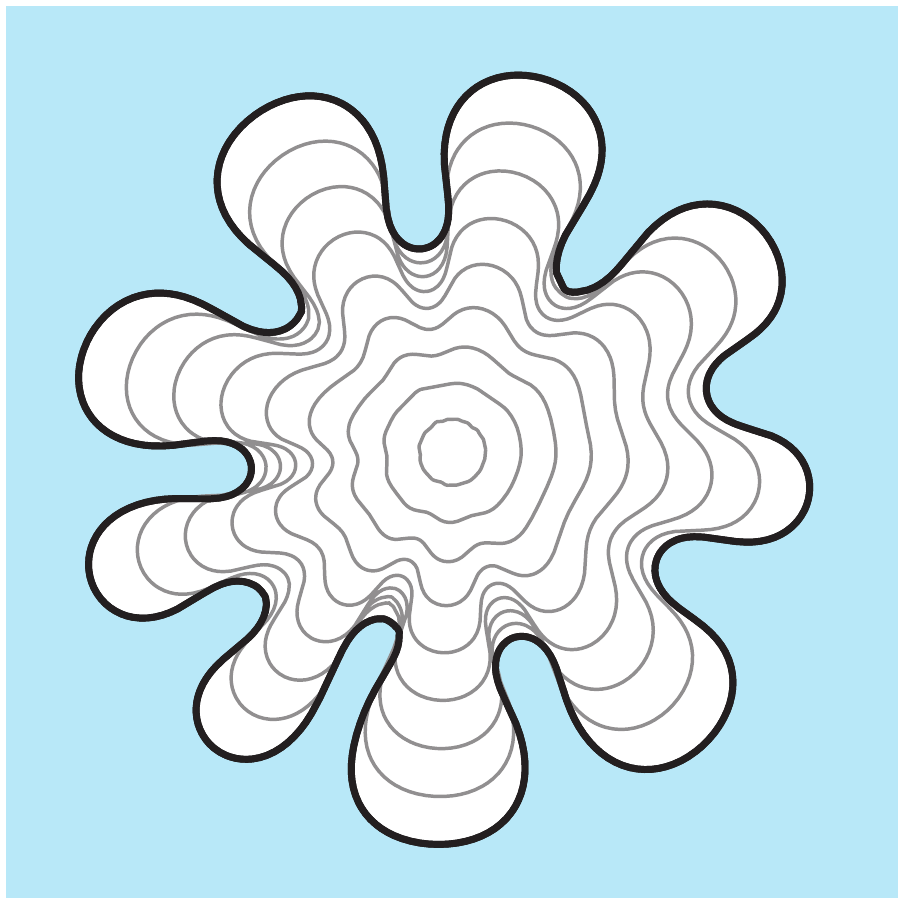}
		\caption{Row 1 is experimental results from \cite{Zheng2015} where the plates are separated according to $b = b_1 t^{1/7}$, each produced for a different value of the control parameter $\hat{J}$, reproduced with permission from the American Physical Society. Row 2 is the corresponding numerical simulations with parameters $\mu = 0.95$ Pa s, $\sigma = 2.1$ g/s$^2$, $b_1 = 8.23 \times 10^{-2}$ cm/s$^{1/7}$, and injection rate (left to right) $Q = 0.09$, $0.12$, $0.17$, and $0.18$ mL/s. The initial condition for all simulations is \eqref{eq:nFingersIC}. The scale bar denotes a length of $5$ cm.}
		\label{fig:ZhengComparison}
\end{figure}
	
\begin{figure}
	\centering
	\begin{tabular}{c|ccccc}
		\hline
		& $J = 107$  & $J = 146$  & $J = 191$  & $J = 242$  \\
		\hline
		\rotatebox{90}{$\alpha_b = 1/7$, $\alpha_Q = 0$}
		&
		\includegraphics[width=0.22\linewidth]{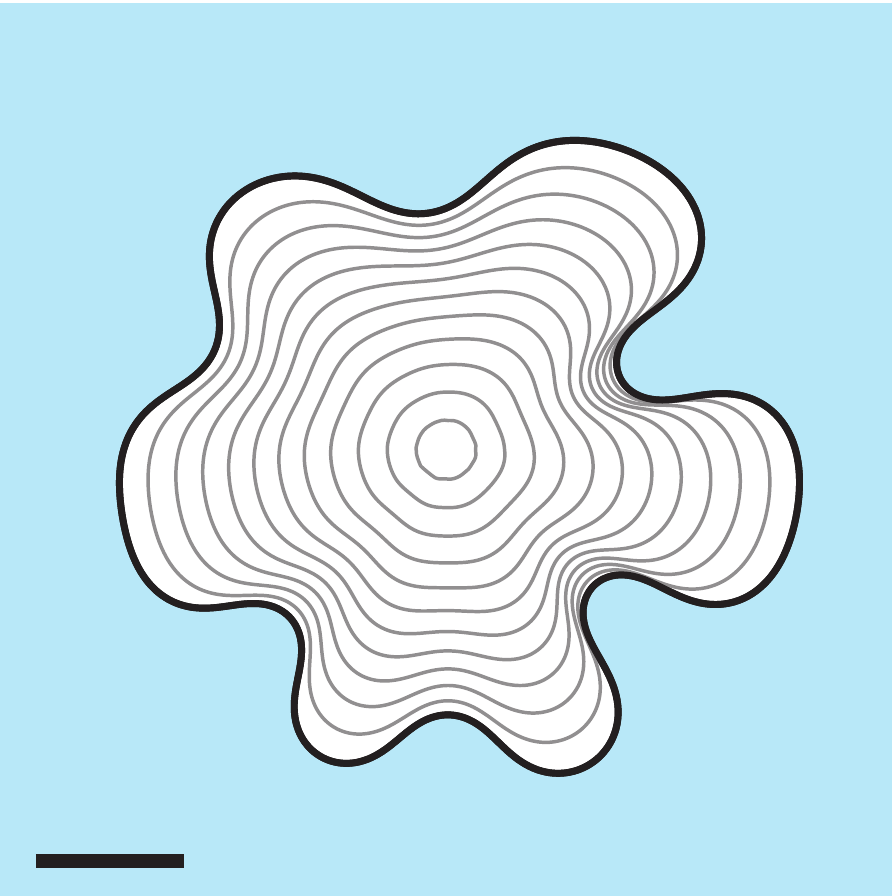}  &
		\includegraphics[width=0.22\linewidth]{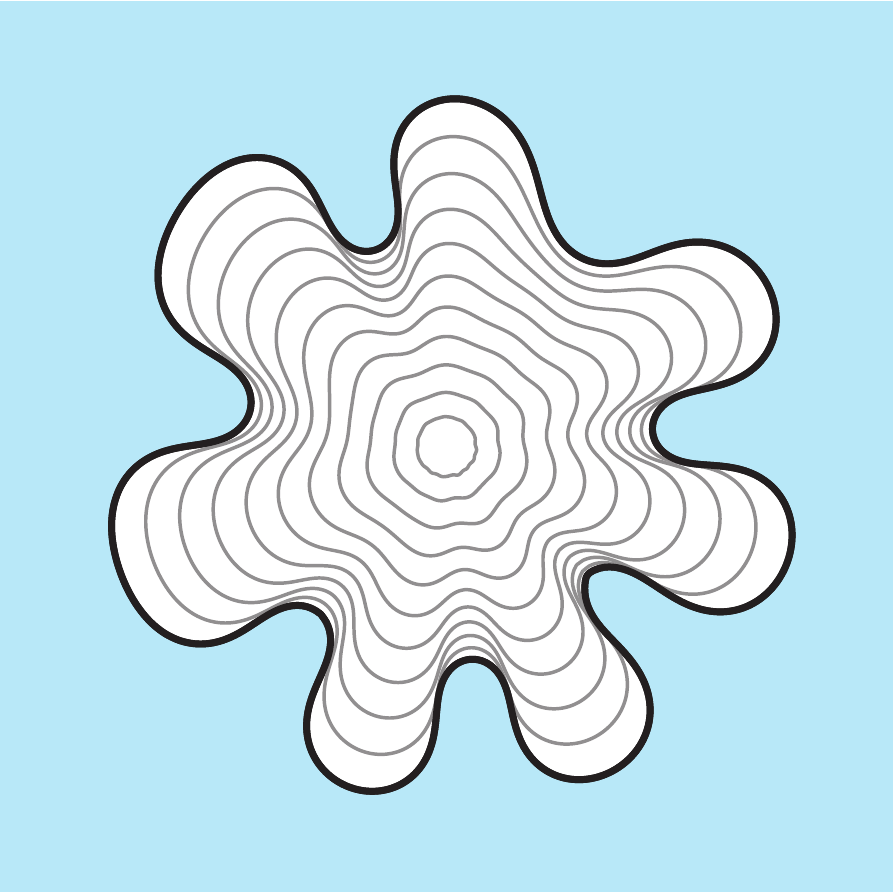} &
		\includegraphics[width=0.22\linewidth]{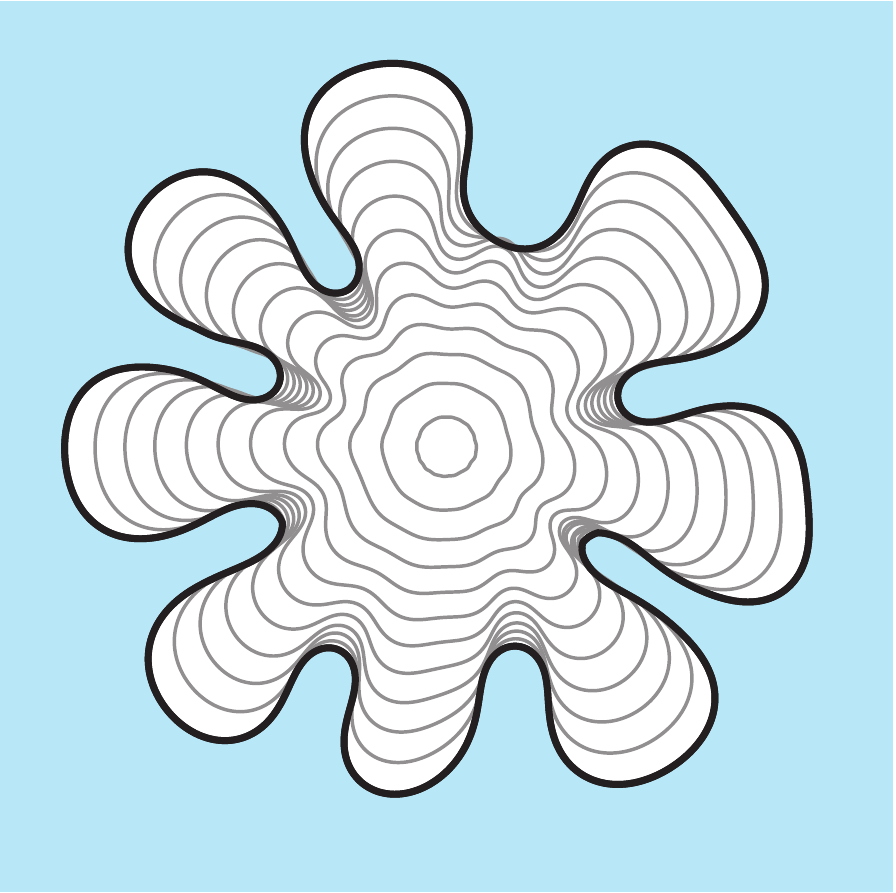} &
		\includegraphics[width=0.22\linewidth]{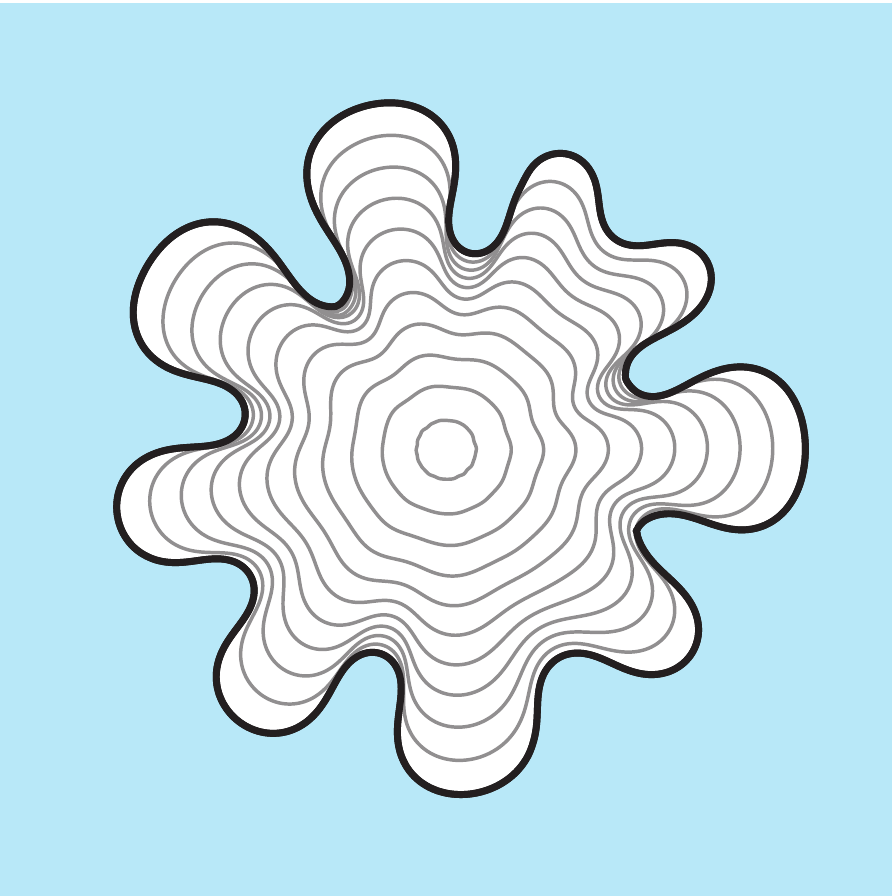}
		\\
		\hline
		\rotatebox{90}{$\alpha_b = 0$, $\alpha_Q = -1/3$}
		&
		\includegraphics[width=0.22\linewidth]{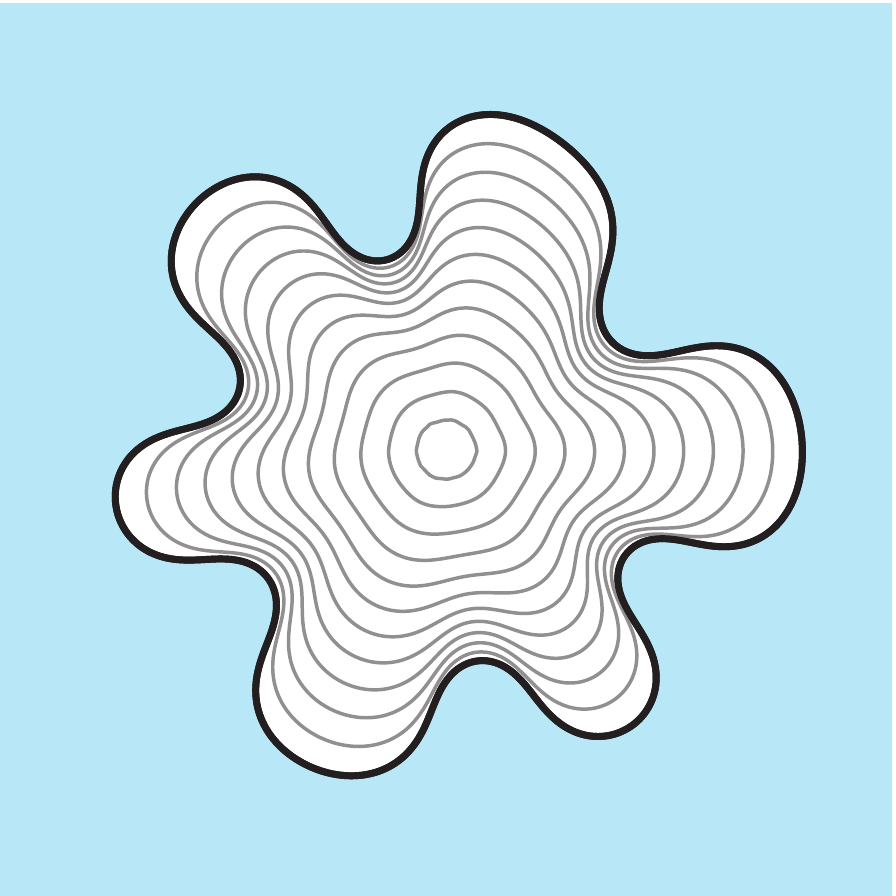} &
		\includegraphics[width=0.22\linewidth]{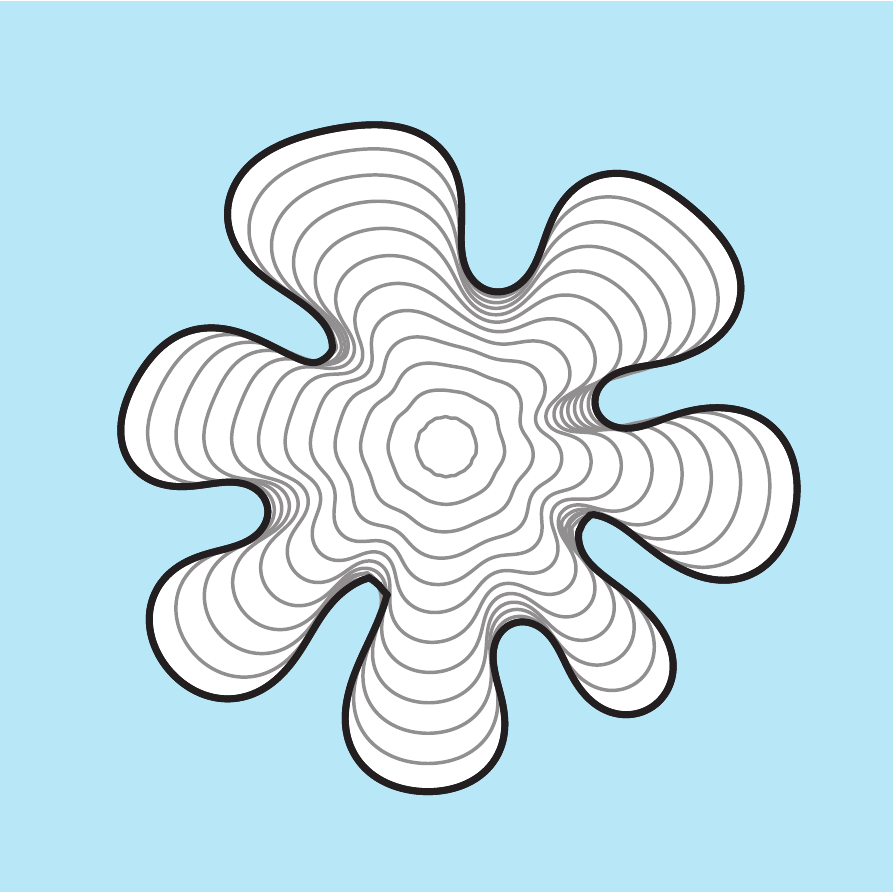} &
		\includegraphics[width=0.22\linewidth]{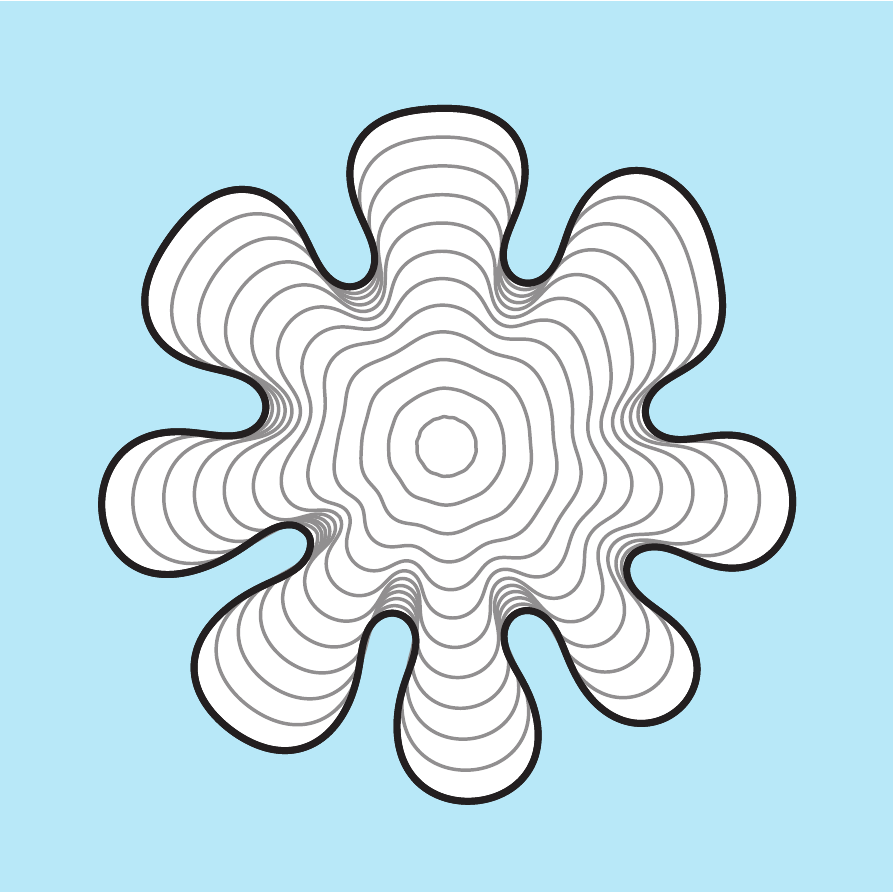} &
		\includegraphics[width=0.22\linewidth]{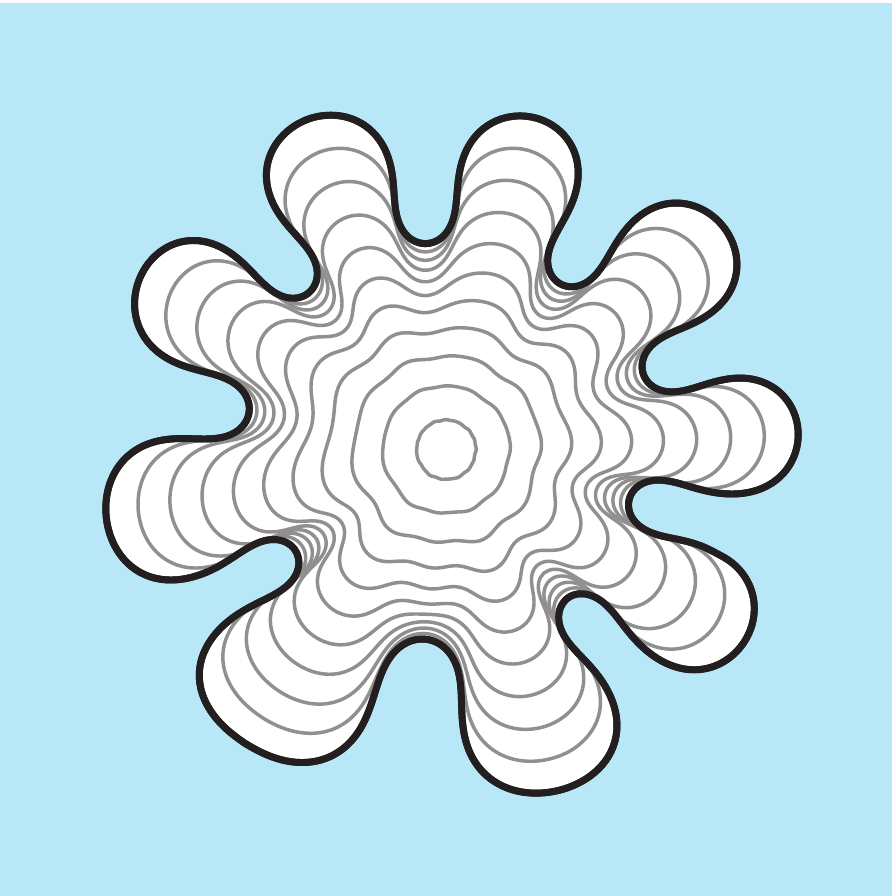}
		\\
		\hline
		\rotatebox{90}{$\alpha_b = 1/14$, $\alpha_Q = -1/6$}
		& 	
		\includegraphics[width=0.22\linewidth]{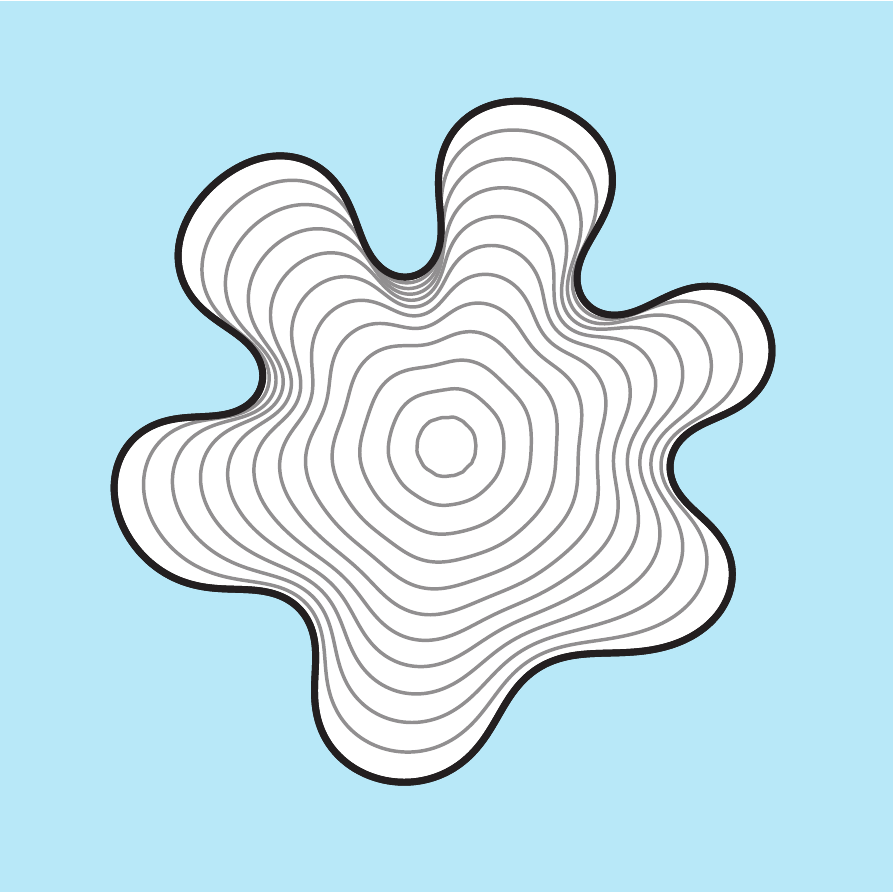} &
		\includegraphics[width=0.22\linewidth]{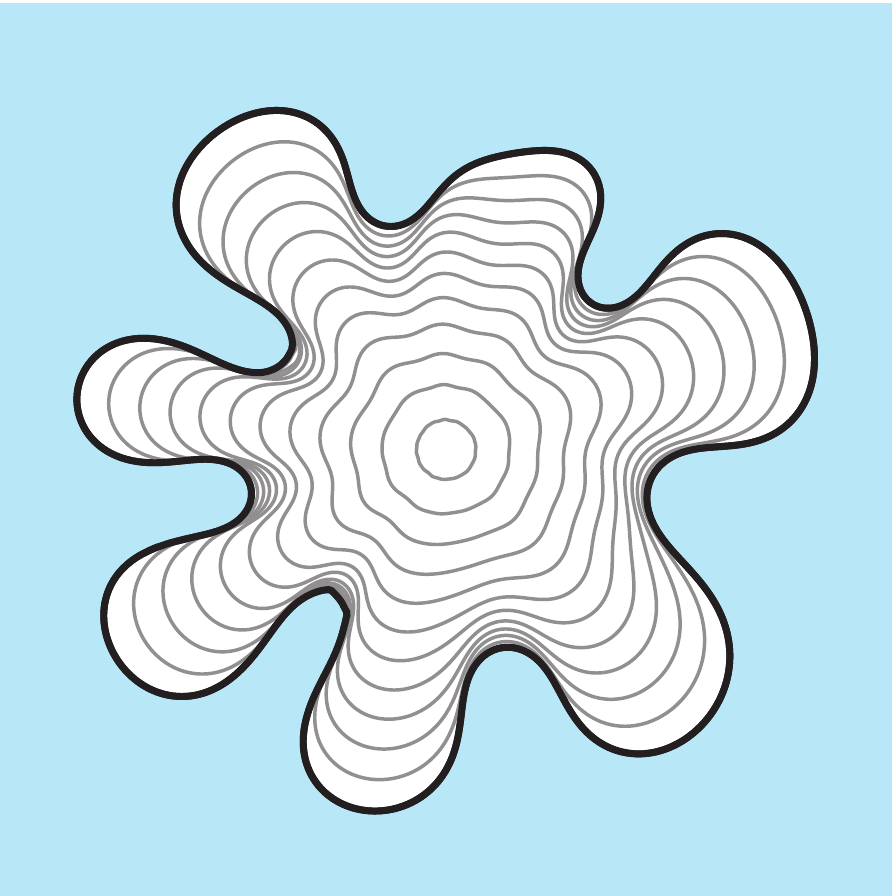} &
		\includegraphics[width=0.22\linewidth]{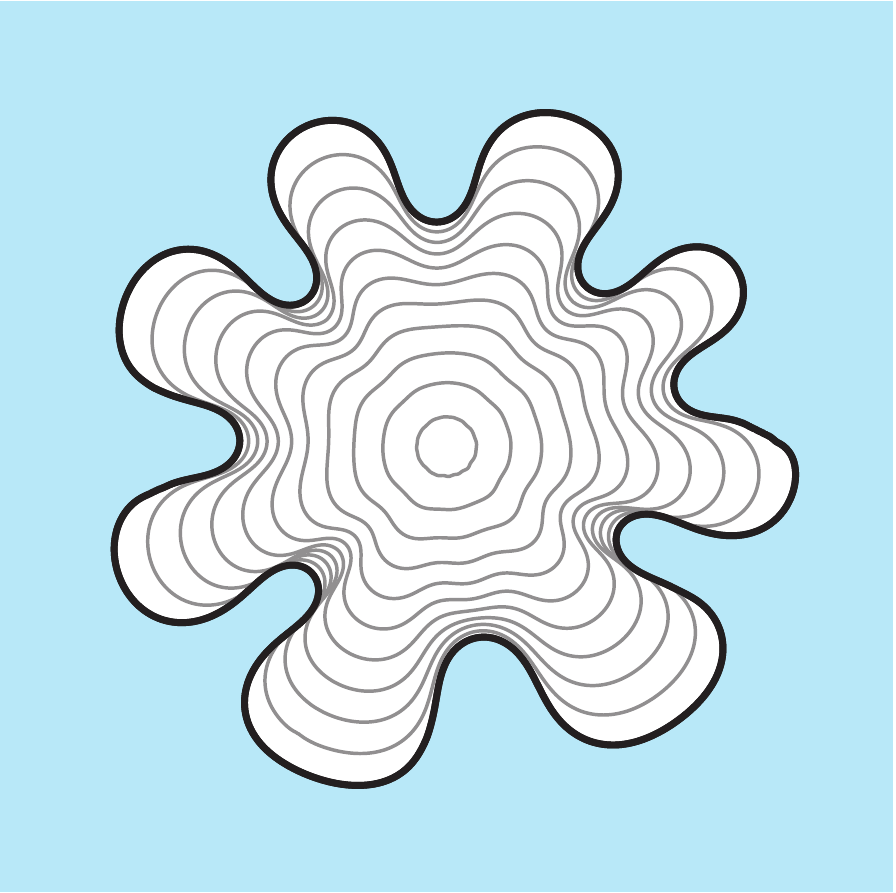} &
		\includegraphics[width=0.22\linewidth]{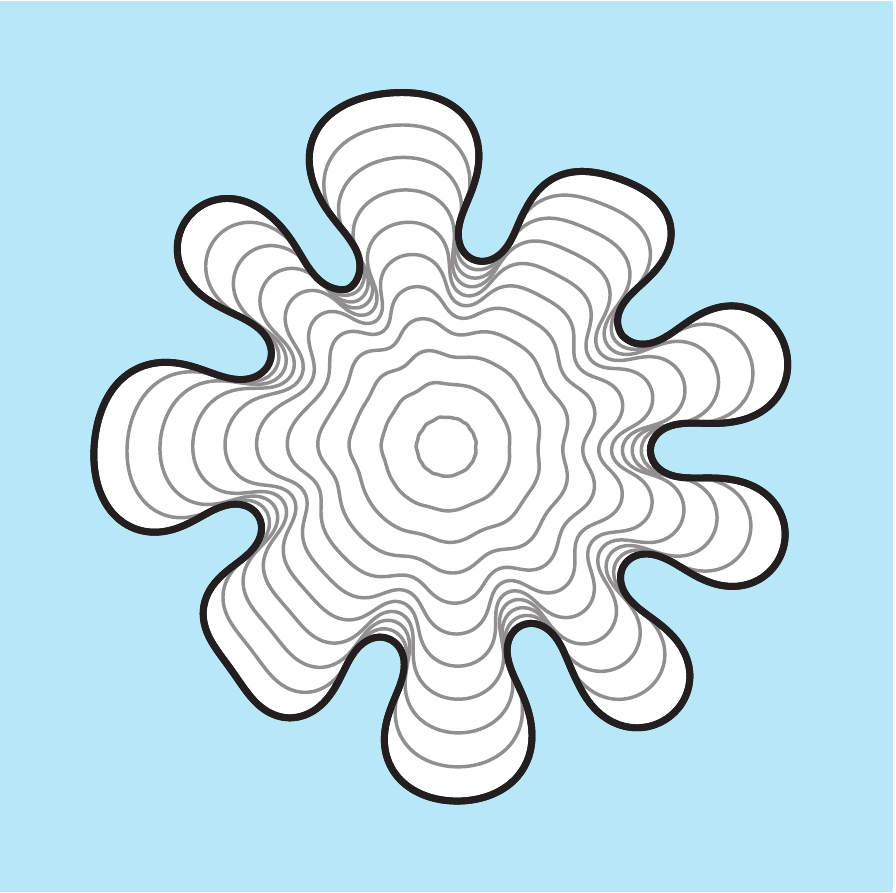}
		\\
		\hline
	\end{tabular}
		\caption{
			Numerical simulations with distance between plates and injection rate of the form (\ref{eq:zhengrates}) for different values of $\alpha_Q$ and $\alpha_b$.
			For row 1, $Q = 1.2$ mL/s,  $b = 0.205 t^{1/7}$ cm$\cdot$s$^{-1/7}$, and (left to right) $\sigma = 0.242$, 0.178, 0.136, and 0.107 g/s$^2$.
			For row 2, $Q = 1.2t^{-1/3}$ mL/s$^{2/3}$, $b = 0.12$ cm and (left to right) $\sigma = 2.257$, 1.654, 1.2644, and 1 g/s$^2$.
			For row 3, $Q = 1.2t^{-1/6}$ mL/s$^{5/6}$, $b = 0.15 t^{1/14}$ cm$\cdot$s$^{-1/14}$, and (left to right) $\sigma = 0.845$, 0.62, 0.474, and 0.374 g/s$^2$.
			For all simulations, the initial condition is \eqref{eq:nFingersIC}. The scale bar represents a length of 5 cm.}
		\label{fig:Figure8}
	\end{figure}
	
In \S~\ref{sec:MinisingInstabilities}, we investigated our first objective for controlling the development of viscous fingers, which involved reducing the fingering pattern when injecting a prescribed amount of viscous fluid over a finite period of time. We now turn our attention to the second objective, which is to control the number of non-splitting fingers that develop in a Hele-Shaw cell.  Numerous theoretical and experimental investigations have been performed to determine strategies for controlling the number of viscous fingers. Using linear stability analysis, \cite{Zheng2015} proposed that if the gap thickness and injection rates are of the form
\begin{equation}
	b = b_1 t^{\alpha_b},
	\quad
	Q = Q_1 t^{\alpha_Q},
	\quad
	\mbox{where}
	\quad
	7 \alpha_b  - 3 \alpha_Q=1,
	\label{eq:zhengrates}
\end{equation}
then the expected number of fingers is
\begin{align}\label{eq:nMax_LiftingPlates2}
	N \approx \sqrt{\frac{1 + \hat{J}}{3}},
\end{align}
where
\begin{align}
	\hat{J} = \frac{6 \mu Q_1^{3/2}}{\pi^{3/2} \sigma (\alpha_Q + 1)^{1/2} b_1^{7/2}},
\end{align}
is a dimensionless control parameter. \cite{Li2009} showed using numerical simulations that when $\alpha_Q = -1/3$ and $\alpha_b = 0$, the interface will tend to $N$-fold symmetric shapes as time increases. Furthermore, the case $\alpha_Q = 0$ and $\alpha_b = 1/7$ has been investigated experimentally by \cite{Zheng2015}, who were able to produce interfaces with different numbers of non-splitting fingers. To date, however, configurations where both $\alpha_Q \ne 0$ and $\alpha_b \ne 0$ have not been considered either experimentally or numerically. In this section, we perform nonlinear numerical simulations to gain insight into the feasibility of controlling the number of non-splitting fingers when imposing a time-dependent gap thickness and/or injection rate according to \eqref{eq:zhengrates}. Note for all simulations in this section, $\omega = 0$ g/(s$^2\cdot$mL) and $\mu = 1/12$ g/(cm$\cdot$s).
	
Performing linear stability analysis on the circular solution to \eqref{eq:Model1}-\eqref{eq:Model6}, we find that
\begin{align} \label{eq:LSA_LiftingPlates}
	\frac{\dot{\gamma}_n}{\gamma_n} = \frac{n - 1}{s_0^2} \left(  \frac{Q}{2 \pi b} - \frac{n(n+1)b^2\sigma}{12 \mu s_0} \right) - \frac{(n+1)}{2 b} \frac{\textrm{d} b}{\textrm{d} t},
\end{align}
and the most unstable mode of perturbation is
\begin{align} \label{eq:nMax_LiftingPlates}
	n_{\max} = \sqrt{\frac{1}{3} \left(  1 + \frac{6 \mu Q s_0}{b^3 \pi \sigma} - \frac{6 \mu s_0^3 }{b^3 \sigma}\frac{\textrm{d} b}{\textrm{d} t} \right) }.
\end{align}
By setting $\textrm{d} b / \textrm{d} t = 0$, \eqref{eq:LSA_LiftingPlates} and \eqref{eq:nMax_LiftingPlates} reduce to \eqref{eq:LSAclassic} and \eqref{eq:UnstableMode}. We can infer from \eqref{eq:LSA_LiftingPlates} that increasing the gap between the plates, $\textrm{d} b / \textrm{d} t > 0$, contributes a negative offset to  $\dot{\gamma}_n / \gamma_n$, resulting in a stabilising effect.
By considering $s_0 = \sqrt{ \int Q / \pi b \textrm{d} t}$ such that $s_0(t) \gg s_0(0)$ and choosing $Q$ and $b$ of the form \eqref{eq:zhengrates}, it follows that $n_{\max}$ will be independent of time and equal to
\begin{align} \label{eq:nFingersLSA}
	n_{\max} = \sqrt{\frac{1 + J}{3}} \quad \textrm{where} \quad J = \frac{(1 + \alpha_Q - \alpha_b)6 \mu Q_1^{3/2}}{\pi^{3/2}\sigma(\alpha_Q + 1)^{3/2} b_1^{7/2}} = \frac{1 + \alpha_Q - \alpha_b}{1 + \alpha_Q}\hat{J}
\end{align}
We note the discrepancy between $J$ and $\hat{J}$ is due to \cite{Zheng2015} possibly ignoring the non-homogeneous term in \eqref{eq:Model3} such that \eqref{eq:LSA_LiftingPlates} and \eqref{eq:nMax_LiftingPlates} reduce to \eqref{eq:LSAclassic} and \eqref{eq:UnstableMode}. However, we can see from \eqref{eq:nFingersLSA} that its inclusion is significant when $\alpha_b \ne 0$.

In addition to linear stability analysis, \cite{Zheng2015} also performed a series of experiments where the gap between the plates satisfies $b \propto t^{1/7}$ for different constant injection rates. In figure \ref{fig:ZhengComparison}, we compare these experiments with the corresponding numerical solution to \eqref{eq:Model1}-\eqref{eq:Model6}. Simulations are performed with initial condition
\begin{align} \label{eq:nFingersIC}
	s(\theta, 0) = 1 + 5 \times 10^{-3} \sum_{n=2}^{20} \cos(n (\theta - 2 \pi \theta_n)),
\end{align}
where $\theta_n$ is a uniformly generated random number between 0 and 1. Simulations are performed on the domain $0 \le r \le 15$ with $0 \le \theta < 2 \pi$ using $800 \times 355$ equally spaced nodes. The interface is evolved until the mean radius of the interface is 10 cm,
which is approximately the point at which the experiments by \cite{Zheng2015} are concluded. In these experiments, it was observed that the interface develops non-splitting fingers, and our numerical simulations reproduce this morphology. In addition to this, we also perform simulations, shown in figure \ref{fig:Figure8}, for different choices of $\alpha_Q$ and $\alpha_b$ that satisfy \eqref{eq:zhengrates} with parameters chosen such that $n_{\max}$ varies from 6 to 9. The injection rate and plate gap width are chosen of the form $Q = Q_1 (t + t_0)^{\alpha_Q}$ and $b = b_1 (t + t_0)^{\alpha_b}$ where $t_0 = 0.4$ s to avoid $Q = \infty$ and $b = 0$ at $t = 0$. We see that for each configuration chosen, we are able to generate interfaces whose number of fingers compare well with the number predicted by linear theory.  Finally, we emphasise again that we are deliberately running our simulations for roughly the same time-scales as \cite{Zheng2015} does in their experiments; for much longer scales, obviously the Hele-Shaw model would break down for $\alpha_b>0$ as the gap between the plates would no longer be small.
	
\begin{figure}
	\centering
	\includegraphics[width=0.5\linewidth]{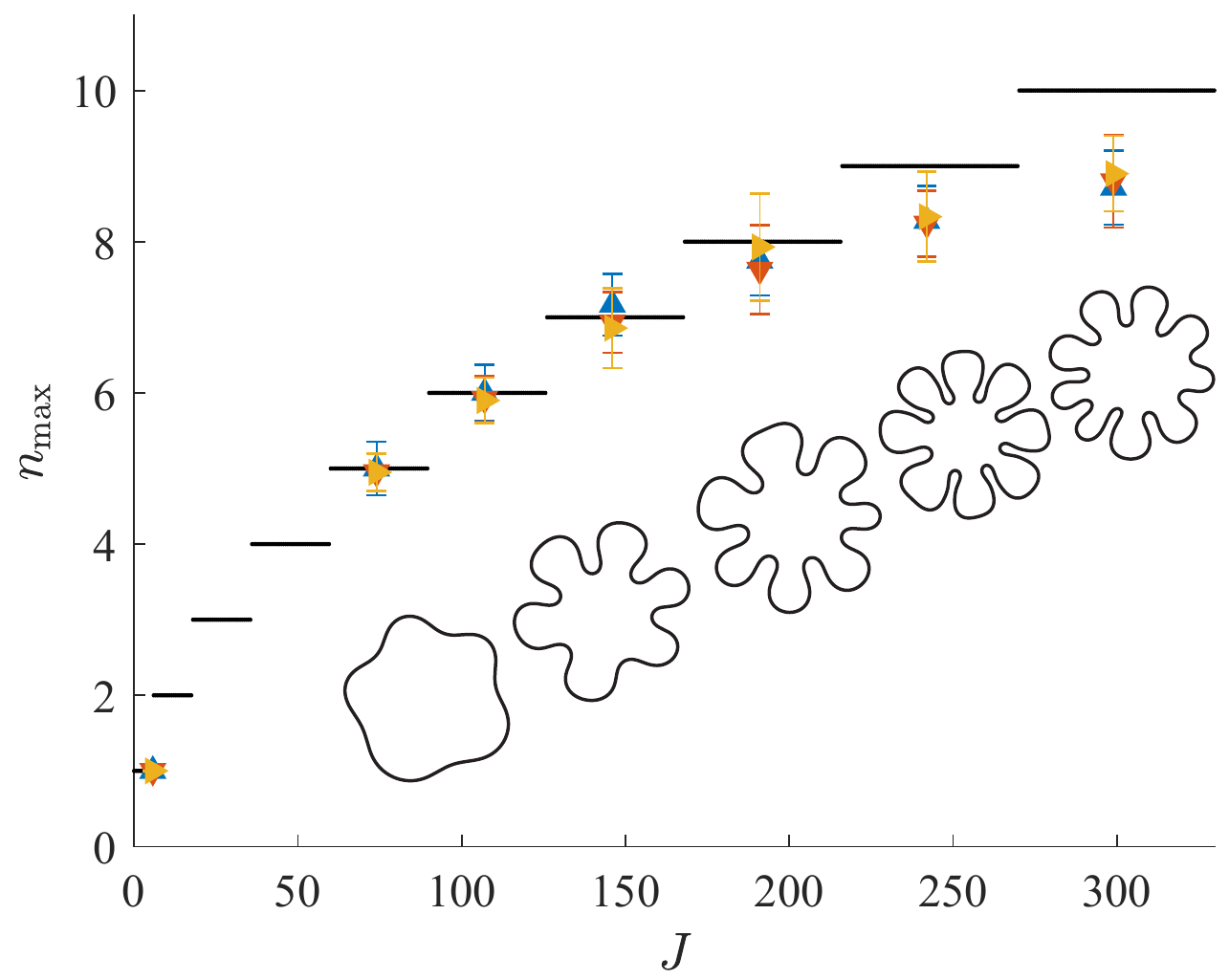}
		\caption{Average number of fingers that develop as a function of the control parameter $J$. Injection rate and plate gap thickness are of the form (\ref{eq:zhengrates}) with: $\alpha_Q =-1/3$ and $\alpha_b=0$ ($\blacktriangleup$, blue); $\alpha_Q = 0$ and $\alpha_b=1/7$ ($\blacktriangledown$, red); and $\alpha_Q = -1/6$ and $\alpha_b=1/14$ ($\blacktriangleright$, yellow). The parameters used are same as figure \ref{fig:Figure8}. The initial condition for all simulations is \eqref{eq:nFingersIC}.  The horizontal (black) lines denote the most unstable mode of perturbation approximated from linear stability analysis by \eqref{eq:nMax_LiftingPlates2} rounded to the nearest integer. Inserts are examples of numerical simulations for different values of $J$.}
		\label{fig:Figure10}
\end{figure}

To understand the relationship between the control parameter, $J$, and predicted number of fingers from (\ref{eq:nFingersLSA}), we perform a series of numerical simulations with different choices of $\alpha_Q$ and $\alpha_b$.  For each combination of parameters, 10 simulations are performed, and the number of fingers at $t = t_f$ are averaged and illustrated in figure~\ref{fig:Figure10}.  Also shown is the closest integer to the most unstable mode (black line). We see that across the values of $J$ considered, the average number of fingers that develop is consistent for each combination of $\alpha_Q$ and $\alpha_b$. In comparison to linear stability analysis, this figure indicates agreement between \eqref{eq:nFingersLSA} and numerical simulations for parameters that give rise to $n_{\max} = 1$, 5, 6, 7, and 8. However, for larger values of $J$, the number of fingers observed from the numerical simulations is slightly less than the number predicted by linear stability analysis.

\begin{figure}
	\centering
	$n_{\max} = 10$\\
	\includegraphics[width=0.19\linewidth]{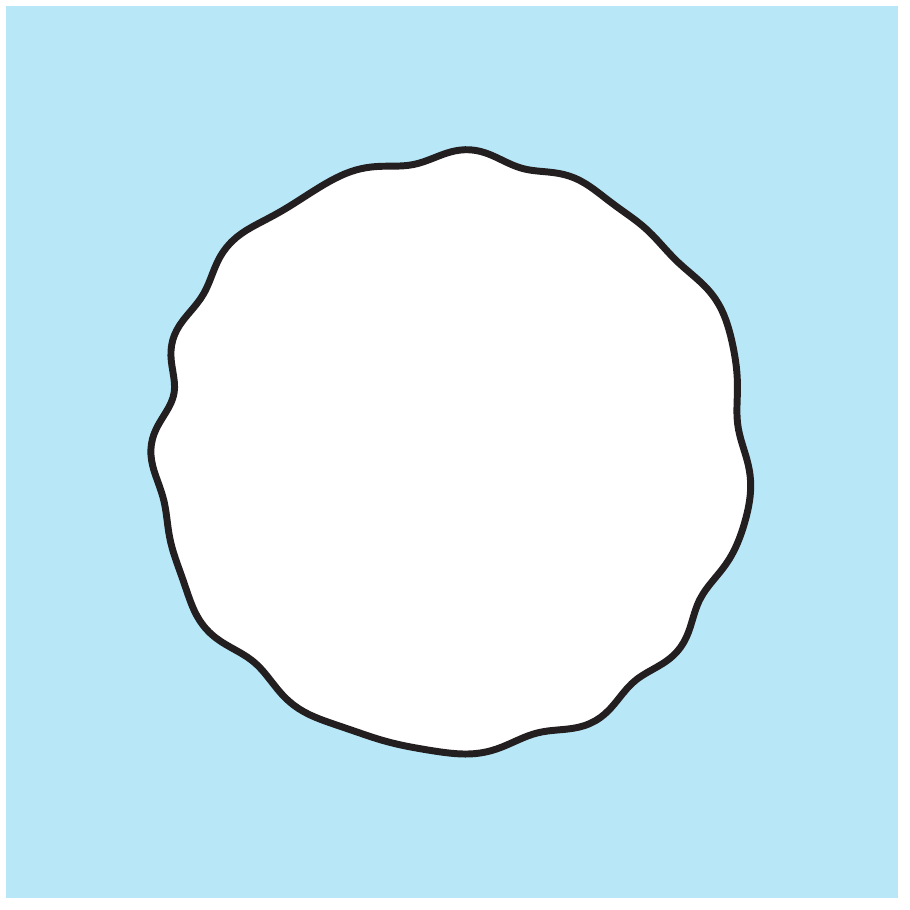}
	\includegraphics[width=0.19\linewidth]{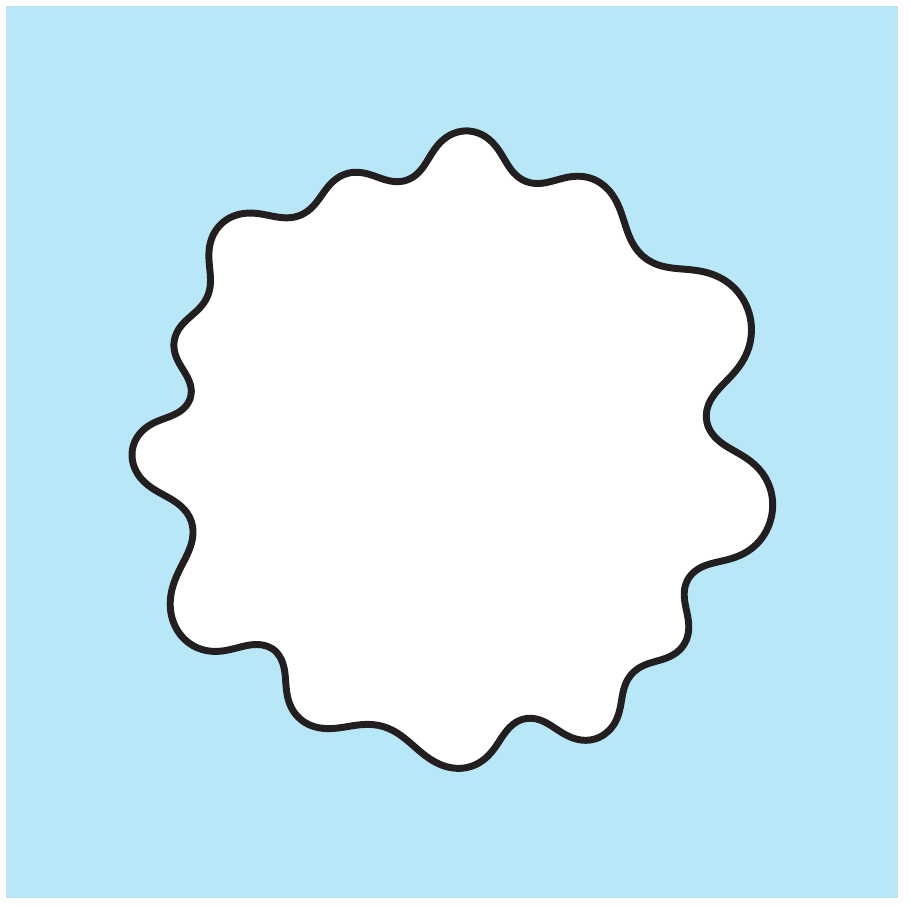}
	\includegraphics[width=0.19\linewidth]{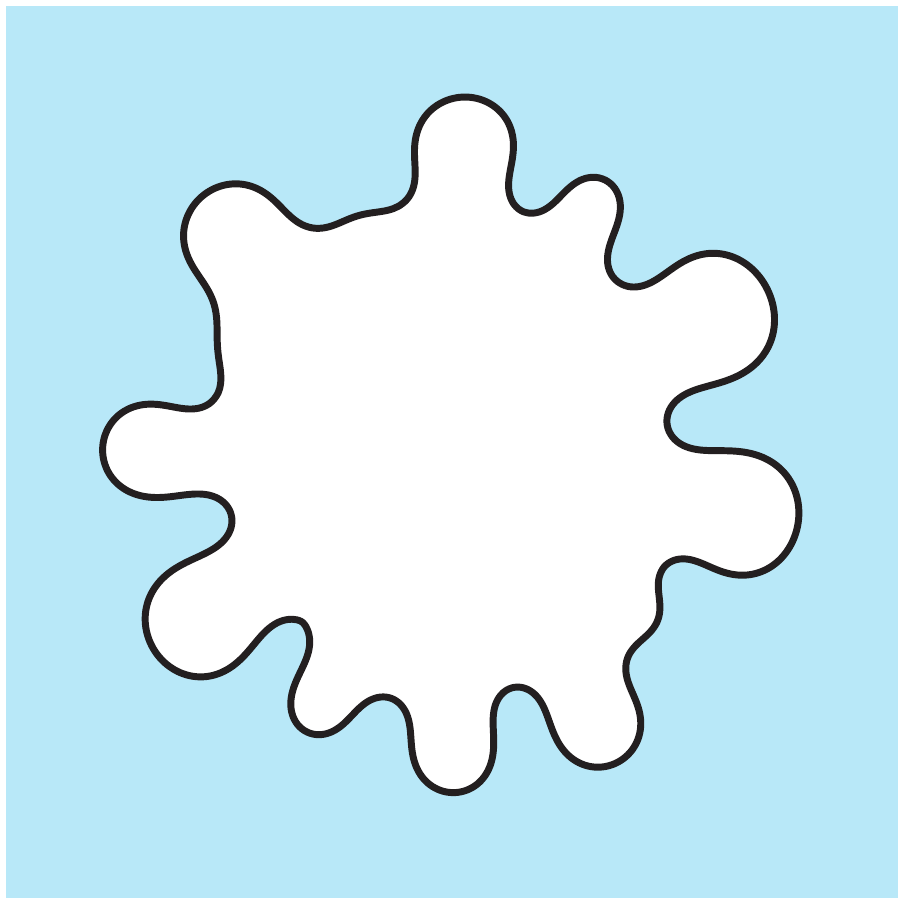}
	\includegraphics[width=0.19\linewidth]{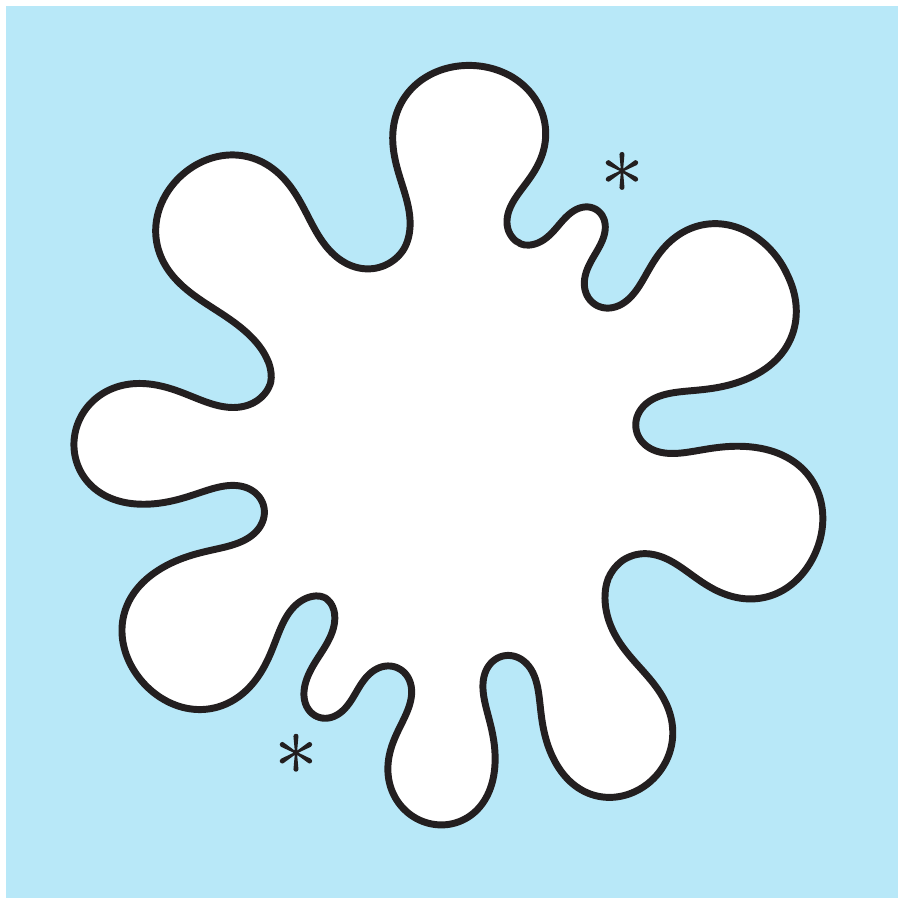}
	\includegraphics[width=0.19\linewidth]{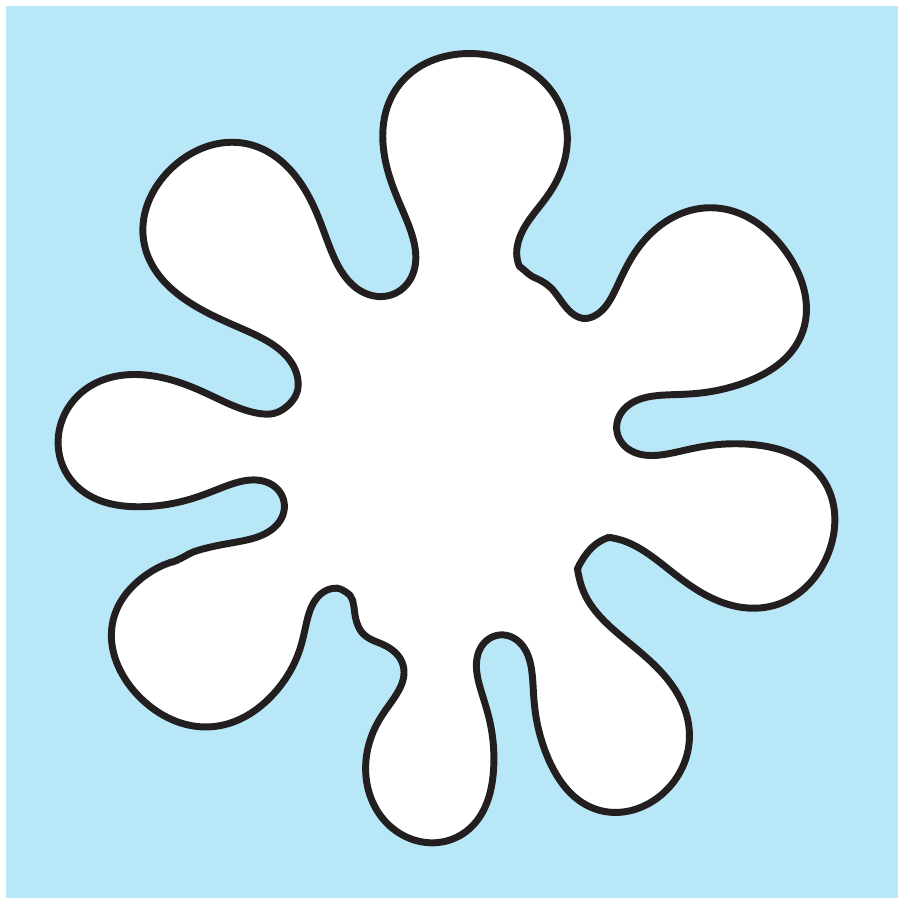}	
	\\ \vspace{0.2cm}
	$n_{\max} = 7$\\
	\includegraphics[width=0.19\linewidth]{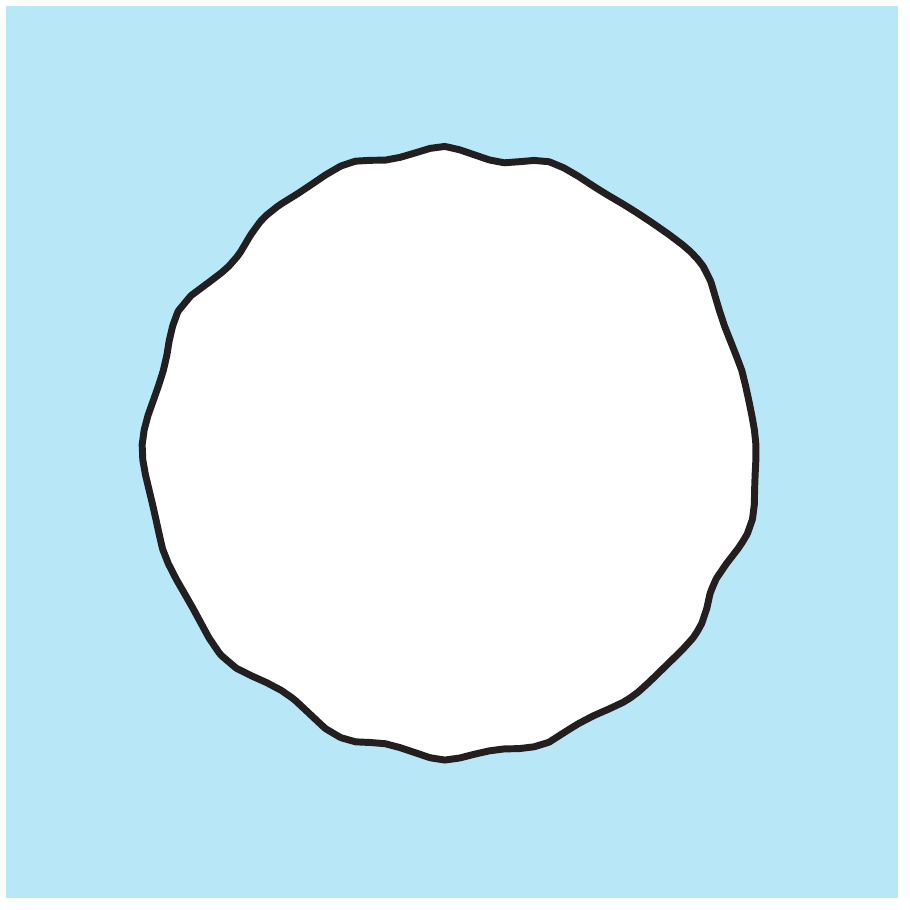}
	\includegraphics[width=0.19\linewidth]{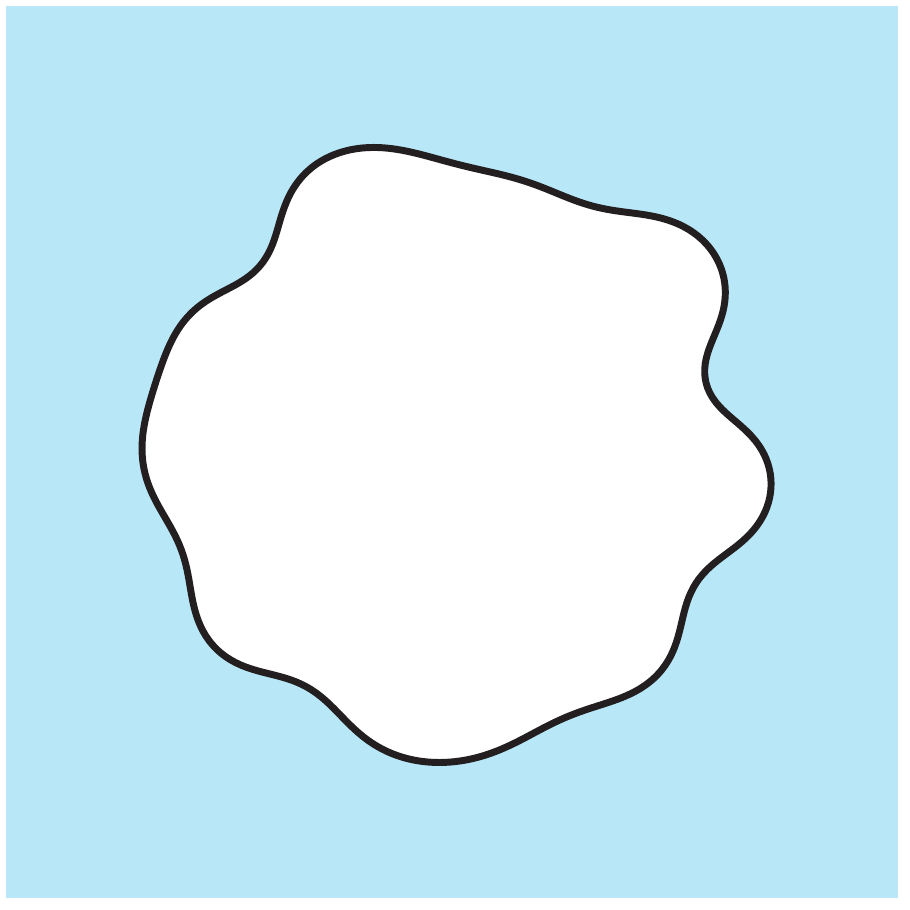}
	\includegraphics[width=0.19\linewidth]{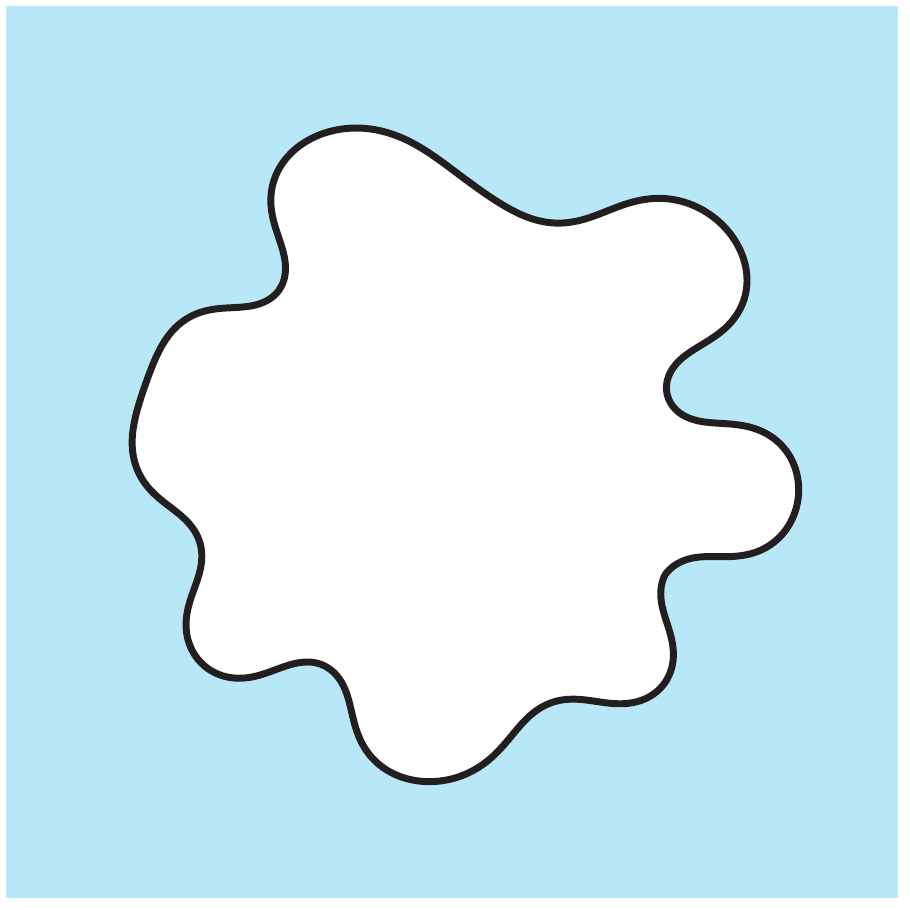}
	\includegraphics[width=0.19\linewidth]{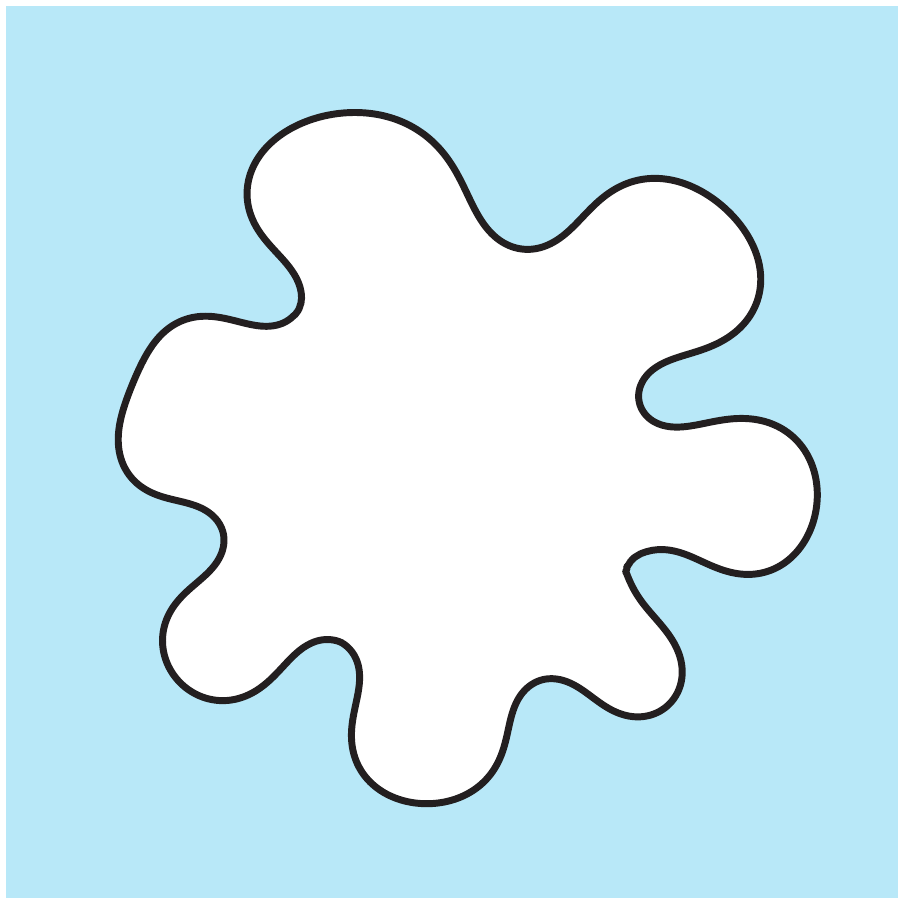}
	\includegraphics[width=0.19\linewidth]{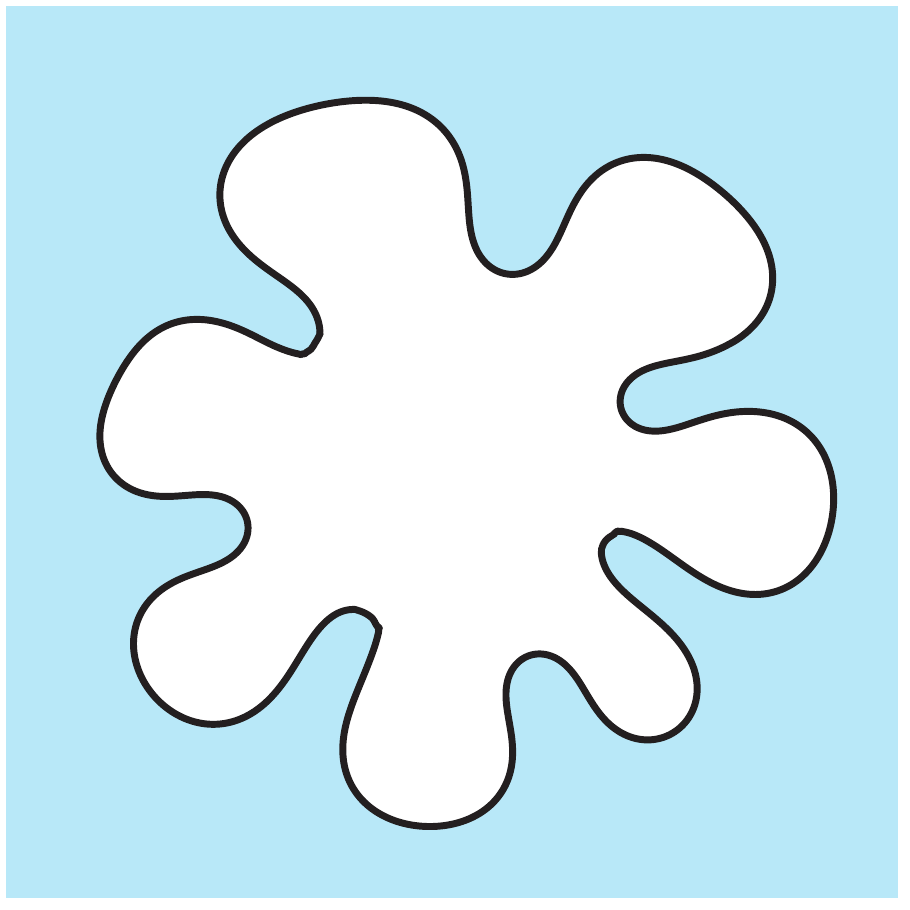}				
		\caption{Evolution of numerical solution to \eqref{eq:Model1}-\eqref{eq:Model6} for different values of $n_{\max}$. For row 1, $b = 0.12$ cm, $Q = 1.2 t^{-1/3}$ mL/s$^{2/3}$, and $\sigma = 0.808$ g/s$^2$. 			The $*$ refer to fingers that retract due to competition with neighbouring fingers. For row 2, $b = 0.12$ cm, $Q = 1.2 t^{-1/3}$ mL/s$^{2/3}$, and $\sigma = 1.654$ g/s$^2$. Solutions for both rows are shown at times (left to right) $t = 0$, 2.3, 6.5 18.7 and 48.4 s, and initial condition is \eqref{eq:nFingersIC}. We note that solutions here have been scaled such that the average radius of the interface is 1 cm.}
		\label{fig:10fingres}
\end{figure}

To further investigate this apparent discrepancy, we examine the behaviour of the solution to \eqref{eq:Model1}-\eqref{eq:Model6} with parameters chosen such that $n_{\max} = 10$, shown in the first row of figure \ref{fig:10fingres}.  For small time, we see 13 fingers developing (second column) and, as time increases, several of these fingers retract resulting in 10 fingers (third column), which is the same as that predicted by linear stability analysis. However, for later times, two of the fingers (denoted by a $*$ in the fourth column) do not appear to grow as fast as their neighbours. As a result, these fingers are `blocked off' and retract, resulting in the interface developing eight fingers by the end of the simulation (fifth column).  By comparison, the second row of figure \ref{fig:10fingres} shows the numerical solution with parameters chosen such that $n_{\max} = 7$.  Again, fingers begin to grow at various rates; however, the interaction between the fingers appears to be not as severe as in the first row and thus the interface maintains seven fingers. From these observations, we infer that for larger values of $J$, there are more fingers that compete with each other as they grow and, in turn, this competition can result in fewer fingers than that predicted by linear stability analysis.
	
\section{Discussion} \label{sec:Conclusion}

We have conducted a numerical investigation into determining how manipulating the geometry of the classic Hele-Shaw cell experiment can be used to control viscous fingering patterns. By utilising a numerical scheme based on the level set method, we have been able to compute nonlinear numerical solutions both when the gap between the plates is spatially- and time-dependent as well in the case in which the plates are rotating.  As a preliminary test of our scheme, we have shown that our numerical solutions of \eqref{eq:Model1}-\eqref{eq:Model6} compare well with a variety of experimental results for different injection rates and plate configurations. Subsequently, we have been able to determine new relationships between these various manipulations and their influence on interfacial instabilities which extends well beyond the limitations of linear stability analysis and previously performed experiments.  We summarise our findings below.

In \S~\ref{sec:TimeDependentInjection}-\ref{sec:TaperedOptimal} we considered several strategies for reducing viscous fingering when a fixed amount inviscid fluid is injected over a finite time interval. In particular, we investigated how imposing a time-dependent injection rate and/or tapering the plate gap (in either the converging or diverging configurations) influences the morphology of the interface. By performing a series of numerical simulations and applying standard metrics for measuring how round the interface is, we have shown that each of these configurations is able to produce a less unstable interface at $t=t_f$ (for certain taper angles) than the standard configuration with parallel plates and constant injection rate. In other words, we can reduce viscous fingering by either tapering plates in the converging ($\alpha>0$) or diverging ($\alpha<0$) directions and suppress the fingering further by imposing a time-dependent injection rate that is chosen to minimise the growth rate of the most unstable mode of perturbation (from linear stability analysis).  Of all of these strategies, our results indicate that injecting at a wisely chosen time-dependent rate when $\alpha>0$ is the most effective configuration for minimising instabilities.  Note that ours are the first fully nonlinear results for these non-standard configurations.

It is interesting to note the different morphological features that are produced by the tapered plate configurations. For instance, when $t\sim t_f$ for the converging configuration ($\alpha>0$), the interface quickly develops numerous short fingers which appear different to the standard Hele-Shaw patterns.  In \S~\ref{sec:TaperedPlates} one example of interest relates to the tapered case with constant injection (third row of figure~\ref{fig:Figure4} and the downward-facing triangles in figure \ref{fig:Figure6}).  Here, as $\bar{Q}$ increases, the isoperimetric ratio $\mathcal{I}$ increases at a rate that is higher than the other configurations, while the increase in circularity ratio $\mathcal{C}$ is not so dramatic.  Thus for higher flow rates we are observing interfaces that are highly complex but whose fjords are not as deep as in the standard pattern.

Another example relates to the converging case with optimal injection (fourth row of figure~\ref{fig:Figure4}).  Here, the morphology is significantly different from the standard case, and in fact this interfacial pattern resembles the short flat-tipped ``stubby" fingers observed both experimentally and numerically by \cite{Pihler2012,Pihler2013}, who considered a Hele-Shaw cell where the top plate is replaced by an elastic membrane.  Other closely related studies, including links with the problems of opening an initially collapsed channel and peeling of a viscous strip, are outlined in \citet{Al2013,Ducloue2017,Juel2018,Lister2013,McCue18,Pihler2014,Pihler2018}.  These observations suggest there is a one-parameter family of solutions ($\alpha$) joining the standard Hele-Shaw problem to one where the pattern formation is similar to that produced by a deformable boundary.  This is perhaps not surprising as the elastic membrane acts like a tapered upper boundary near the interface.

The morphological features for diverging plates are also interesting.  Here, we notice the interface appears significantly different from both the parallel and converging cases, with fewer and longer fingers forming over the duration of each simulation.  In addition, implementing our `optimal' (time-dependent) injection rate has little observable effect on this morphology.

In \S~\ref{sec:RotatingPlates}, we considered the effects of rotating the Hele-Shaw plates on the development of viscous fingers. While the complementary problem for which the viscous fluid is surrounded by an inviscid bubble has previously been studied, the results we present here are the first to consider the case where the bubble is being injected into the viscous fluid while under the influence of a centrifugal force. We have presented evidence, both in the form of linear stability analysis and nonlinear numerical simulations, suggesting the interface will eventually become circular if $\omega > 0$. This is somewhat of an unusual result, since for all the configurations considered in \S~\ref{sec:TimeDependentInjection}-\ref{sec:TaperedOptimal}  we found that both the isoperimetric and circularity ratios are always monotonically increasing in time. Our simulations indicate that as the bubble expands, the centrifugal force causes the dense fluid to propel outward, stabilising the interface. It would be beneficial to undertake an experimental study for the problem we have presented here to confirm that the behaviour of the interface is consistent with our numerical simulations. 

In \S~\ref{sec:NumberOFingers}, we performed nonlinear simulations to determine whether imposing a time-dependent injection rate and/or plate gap can be used to control the number of fingers which develop.  In particular, following the suggestion by \cite{Zheng2015}, we allowed the injection rate and gap thickness to vary according to power laws in time, with exponents $\alpha_Q$ and $\alpha_b$, respectively.  For a range of combinations of $\alpha_Q$, $\alpha_b$ and the control parameter $J$, we tested the hypothesis that (after an initial period in which various modes of perturbation grow or decay) a fixed number of non-splitting fingers emerge, $n_{\max}$, which is equal to the most unstable mode.  Our results support this hypothesis for the cases which predict between $5$ and 8 fingers, as well as the less interesting case in which there are no fingers (a stable interface).  For larger values of $n_{\max}$, the average number of fingers observed in our simulations is slightly less than that predicted by linear stability theory.  In this parameter regime, our explanation for observing fewer than $n_{\max}$ fingers is that, on the time-scale of our simulations, it appears that when $n_{\max}$ is sufficiently large, nonlinear interactions between closely packed fingers can cause a small number of them to retract (see Figure~\ref{fig:Figure10}).  These numerical simulations are consistent with the experimental results of \cite{Leshchiner2010}, who test a time-dependent injection rate ($\alpha_Q =-1/3$ and $\alpha_b=0$) for a single control parameter, and \cite{Zheng2015} who treat the lifting-plate case ($\alpha_Q = 0$ and $\alpha_b = 1/7$) in some detail.  Further, our findings for the lifting-plate case ($\alpha_Q = 0$ and $\alpha_b = 1/7$) are in agreement with the very recent numerical study of \cite{Vaquero2019}, who run their simulations for a larger range of the control parameter $J$, namely up to $J=1000$.

From a theoretical perspective, beyond the lifetime of a normal experiment, there is a question about the ultimate long-time behaviour of our mathematical solutions.  It was shown numerically by \cite{Li2009} that, for the special case $\alpha_Q =-1/3$ and $\alpha_b=0$ (stationary plates with $Q\sim t^{-1/3}$), the interface develops $N$-fold symmetry independent of the initial condition over an extremely long time period, at least for values of $J$ which predict up to nine non-splitting fingers.  The likely reason for this $N$-fold symmetric long-time attractor is that the problem with $\alpha_Q =-1/3$ and $\alpha_b=0$ has self-similar solutions of the form $p=t^{-1/3}P(X,Y)$, where $(X,Y)=(x/t^{1/3},y/t^{1/3})$ \citep{Amar1991,Combescot1991}.  Further, \cite{Vaquero2019} also observes time-dependent solutions approaching a self-similar form for the case $\alpha_Q = 0$ and $\alpha_b = 1/7$ (lifting plates with constant injection); however, here the initial conditions were also $N$-fold symmetric.  In general, it seems there are self-similar solutions of the form $p=-2\sigma/b+t^{-(2\alpha_b+1)/3}P(X,Y)$, where $(X,Y)=(x/t^{(2\alpha_b+1)/3}, y/t^{(2\alpha_b+1)/3})$ and $7\alpha_b-3\alpha_Q=1$, although these have not been computed before.  For this combination of parameters, at this stage it is not clear whether the bubble evolves to an perfect $N$-fold symmetric shape with $N$ fingers predicted by (\ref{eq:nMax_LiftingPlates}) for randomly chosen modes of perturbation in the initial condition (except for the special case $\alpha_Q =-1/3$ and $\alpha_b=0$).  Either way, these questions are worthy of further enquiry.

\section*{Acknowledgements}

The authors acknowledge the support of the Australian Research Council via the Discovery Project DP140100933, as well as the computational resources provided by the High Performance Computing and Research Support Group at QUT.   We thank Anne Juel and Draga Pihler-Puzovi{\'c} for helpful discussions.  Finally, we are grateful to the anonymous referees for their detailed reviews and insight.

\appendix

\section{Numerical scheme} \label{sec:NumericalScheme}

\subsection{The level set method}

To implement the level set method, a level set function, $\phi(\boldsymbol{x}, t)$, is constructed as a signed distance function whose zero level set describes the location of the interface between the viscous and inviscid fluid regions, and
\begin{align}
	\phi > 0 & \quad \text{if } \boldsymbol{x} \in \mathbb{R}^2 \backslash \Omega(t), \\
	\phi < 0 & \quad \text{if } \boldsymbol{x} \in \Omega(t).
\end{align}
If the interface has a normal speed $V_n$, then we wish to construct a function, $F$, such that $F = V_n$ on the interface and is continuous over the entire computational domain. Thus $\phi$ satisfies the level set equation
\begin{align}
	\frac{\partial \phi}{\partial t} + F |\nabla \phi| = 0.		               \label{eq:LevelSetEquation}
\end{align}
We approximate the spatial derivatives in \eqref{eq:LevelSetEquation} using a second order essentially non-oscillatory scheme, and integrate in time using second order Runge-Kutta. We choose a time step size of $\Delta t = 0.25 \times \Delta x / \max|F|$ to ensure stability. Furthermore, to maintain $\phi$ as a signed distance function, re-initialisation is periodically performed by solving
\begin{align}
	\frac{\partial \phi}{\partial \tau} + S(\phi) (|\nabla \phi| - 1) = 0,
\end{align}
to steady state where
\begin{align}
	S(\phi) = \frac{\phi}{\sqrt{\phi^2 + \Delta x^2}},
\end{align}
and $\tau$ is a pseudo time variable.

A major limitation of the level set method is that solutions can suffer from volume loss (or gain). To alleviate this problem, we implement the particle level set method, which combines the Eulerian level set method with a marker particle based Lagrangian approach. The particle level set method, first proposed by \cite{Enright2002}, extends the traditional level set method by placing massless marker particles around the interface. These particles are advected using the same velocity field as the level set function. As the particles do not suffer from mass loss, the level set function can be corrected if the particles are found to incorrectly cross the interface. We refer the reader to \cite{Enright2002} for details on how to implement the particle level set method, as well as examples illustrating its effectiveness.

\subsection{Solving for $F$}

Defining $\boldsymbol{n} = \nabla \phi / |\nabla \phi|$ as the outward facing normal of the interface, we have the expression
\begin{align}
	F =  -\frac{b^2}{12 \mu} \frac{\nabla p \cdot \nabla \phi}{|\nabla \phi|}  \qquad \boldsymbol{x} \notin \Omega(t).	\label{eq:NormalSpeed}
\end{align}
This satisfies \eqref{eq:Model5} on the interface, and provides a continuous expression for $F$ in the viscous fluid region. However, to solve \eqref{eq:LevelSetEquation} we require an expression for $F$ over the entire computational domain. \cite{Moroney2017} proposed that the speed function can be extended into the inviscid fluid region by solving the biharmonic equation
\begin{align} \label{eq:Biharmonic}
	\nabla^4 F = 0  \qquad \boldsymbol{x} \in \Omega(t).
\end{align}
By solving \eqref{eq:Biharmonic}, this ensures that $F = V_n$ on the interface and gives a continuous expression for $F$ away from the interface. The sign of $\phi$ is used to determine nodes inside the interface that need to be included in the biharmonic stencil. As such, the location of the interface does not need to be known explicitly, similar to the level set method itself. This velocity extension process is a variant of a thin plate spline in two dimensions. We refer the reader to \cite{Moroney2017} for further details.

\subsection{Solving for pressure}

To evaluate the speed function $F$, we must first compute the pressure field. We consider \eqref{eq:Model1}-\eqref{eq:Model6} in polar coordinates with $p = p(r, \theta, t)$ and the location of the interface is given by $r = s(\theta, t)$. Thus \eqref{eq:Model3} becomes
\begin{align} \label{eq:LaplacePolar}
	\frac{1}{r} \frac{\partial}{\partial r}  \left( \frac{r b^3}{12 \mu} \frac{\partial p}{\partial r} \right) + \frac{1}{r^2} \diffp{}{\theta} \left( \frac{b^3}{12 \mu} \diffp{p}{\theta} \right) = \frac{\partial b}{\partial t} \qquad r > s(\theta, t).
\end{align}
In order to solve for the pressure at nodes that are not adjacent to the interface, we discretise \eqref{eq:LaplacePolar} using a standard central finite difference scheme. Denoting $\beta = r b^3/12\mu$, the $r$-derivatives in \eqref{eq:LaplacePolar} are approximated via
\begin{align}
	\frac{1}{r} \frac{\partial}{\partial r} \left(  \beta \frac{\partial p}{\partial r} \right)  \to  \frac{1}{r_{i,j} \Delta r} \left(  \beta_{i+1/2,j} \frac{p_{i+1,j} - p_{i,j}}{\Delta r} - \beta_{i-1/2,j}  \frac{p_{i,j} - p_{i-1,j}}{\Delta r} \right), \label{eq:FiniteDifference}
\end{align}
where $\beta_{i+1/2,j} = (\beta_{i+1,j} + \beta_{i,j})/2$ and $\beta_{i-1/2,j} = (\beta_{i-1,j} + \beta_{i,j})/2$. The derivatives in the $\theta$-direction are discretised in a similar fashion.

Special care must be taken when solving for nodes adjacent to the interface. Suppose that the interface is located at $r = r_I$ where $r_{i-1,j} < r_I < r_{i,j}$ where the nodes $r_{i-1,j}$ and $r_{i,j}$ are in the inviscid and viscous fluid regions respectively. When discretising \eqref{eq:LaplacePolar}, we can no longer incorporate $p_{i-1,j}$ into our finite difference stencil as it is not in the domain $\boldsymbol{x} \in \mathbb{R}^2 \backslash \Omega(t)$. Instead, we define a ghost node at $r_I$ whose value is $p_I$. By noting that $\phi$ is a signed distance function, the distance between $r_{i,j}$ and $r_I$ is computed via
\begin{align}
	h = \Delta r \left| \frac{\phi_{i,j}}{\phi_{i-1,j} - \phi_{i,j}} \right|.
\end{align}
As per \cite{Chen1997}, our finite difference stencil becomes
\begin{equation}
	\begin{split}
		\frac{1}{r} \frac{\partial}{\partial r} \left( \beta \frac{\partial p}{\partial r} \right)  &\to  \frac{2}{r_{i,j}(\Delta r + h)} \left( \beta_{i+1/2,j}  \frac{p_{i+1,j} - p_{i,j}}{\Delta r} -\hat{\beta}_{i-1/2,j} \frac{p_{i,j}}{h} \right) \\ &+ \overbrace{\frac{2}{r_{i,j} h (\Delta r + h)} \hat{\beta}_{i-1/2,j} p_I}^{\text{non-homog.}}.
	\end{split}
\end{equation}
Here $\hat{\beta}_{i-1/2,j} = (\beta_{i,j} + \beta_I)/2$ where $\beta_I$ is the value of $\beta$ on the interface, and is computed via linear interpolation using $\beta_{i,j}$ and $\beta_{i-1,j}$. When the node and interface are sufficiently close together such that $h < \Delta r^2$, we set $p_{i,j} = p_I$. A similar procedure is applied if the interface lies between $r_{i,j} < r_I < r_{i+1,j}$ and in the azimuthal direction. The value of $p_I$ is computed from the dynamic boundary condition \eqref{eq:Model4}, where the curvature of the interface is $\kappa = \nabla \cdot \boldsymbol{n}$.

\subsubsection{Far-field boundary condition}

To incorporate the far-field boundary condition \eqref{eq:Model6} into our finite difference stencil, we utilise a Dirichlet to Neumann map. This is implemented by imposing an artificial circular boundary at $r = R$ such that $R > s(\theta, t)$. By only considering the region in domain $r \ge R$, we seek a solution to \eqref{eq:Model3} of the form
\begin{align} \label{eq:BC1}
	\hat{p}(r,\theta,t) = A_0 - \frac{Q}{2 \pi} \log r + \frac{r^2}{4} \frac{\partial b}{\partial t} + \sum_{n = 1}^{\infty} r^{-n} \left( A_n \cos n \theta + B_n \sin n \theta \right),
\end{align}
where $A_0$, $A_n$, and $B_n$ are unknown, and $\hat{p} = (b^3/12\mu) p$.
The expansion \eqref{eq:BC1} assumes that $b$ is spatially uniform in $r \ge R$, and the choice of linearly tapered plate gap \eqref{eq:TaperedPlate} is consistent with this.
Considering the value of pressure on the artificial boundary, suppose that $\hat{p}(R, \theta, t)$ can be represented as a Fourier series
\begin{align} \label{eq:BC2}
	\hat{p}(R, \theta, t) = a_0 + \sum_{n = 1}^{\infty} a_n \cos n \theta + b_n \sin n \theta,
\end{align}
where
\begin{align}
	a_0 &= \frac{1}{2\pi} \int_{0}^{2\pi} \hat{p}(R, \theta,t) \hspace{0.15em} \text{d} \theta, \\
	a_n &= \frac{1}{\pi} \int_{0}^{2\pi} \hat{p}(R, \theta,t) \cos n \theta \hspace{0.15em} \text{d}\theta, \\
	b_n &= \frac{1}{\pi} \int_{0}^{2\pi} \hat{p}(R, \theta,t) \sin n \theta \hspace{0.15em}  \text{d}\theta .
\end{align}
By equating \eqref{eq:BC2} with \eqref{eq:BC1} evaluated at $r = R$, we find that $A_0 = a_0 +(Q/2\pi)\log R - \dot{b}R^2 / 4$, $A_n = R^n a_n$ and $B_n = R^n b_n$.

We differentiate our expression for $\hat{p}$ with respect to $r$ and evaluate it at $r = R$ to give
\begin{align}
	\frac{\partial}{\partial r} \hat{p}(R, \theta_j) &= -\frac{Q}{2 \pi R} + \frac{R}{2} \frac{\partial b}{\partial t} - \sum_{n = 1}^{\infty} \frac{n}{R} \left(  a_n \cos n \theta_j + b_n \sin n \theta_j \right), \\
	&\approx -\frac{Q}{2 \pi R} + \frac{R}{2}\frac{\partial b}{\partial t} - \frac{\Delta \theta}{R \pi} \sum_{k = 1}^{m}   w_{jk}  \hat{p}(R, \theta_{k}),
\end{align}
where
	\begin{align}
		w_{jk} = \sum_{n = 1}^{\infty} n \cos(n ( \theta_j - \theta_k)).
\end{align}
Defining $I$ as the outermost index at which $r = R$, then our expression for $\partial p / \partial r$ is incorporated into our finite difference stencil,
	\begin{dmath}
		\frac{1}{r} \frac{\partial}{\partial r} \left(  \beta \frac{\partial p}{\partial r} \right) \to  \frac{2}{R \Delta r}   \left\lbrace -\beta_{I-1/2,j} \frac{  p_{I,j} - p_{I-1,j}}{\Delta r} +  R  \left[ -\frac{Q}{2 \pi R} + \frac{R}{2} \frac{\partial b}{\partial t} - \frac{b^3}{12 \mu} \frac{\Delta \theta}{R \pi} \sum_{k = 1}^{m} w_{jk} p_{I,k}   \right]   \right\rbrace,
\end{dmath}
\noindent recalling $\beta = r b^3 / 12 \mu$. The finite difference stencil for the derivatives in the $\theta$-direction is not changed. Furthermore, a similar procedure to the one presented here could be used to model the Dirichlet boundary condition $p \sim p_\infty$ as $r \to r$ where $p_\infty$ is prescribed.

\subsection{Numerical validation}

\begin{figure}
	\centering
	\includegraphics[width=0.23\linewidth]{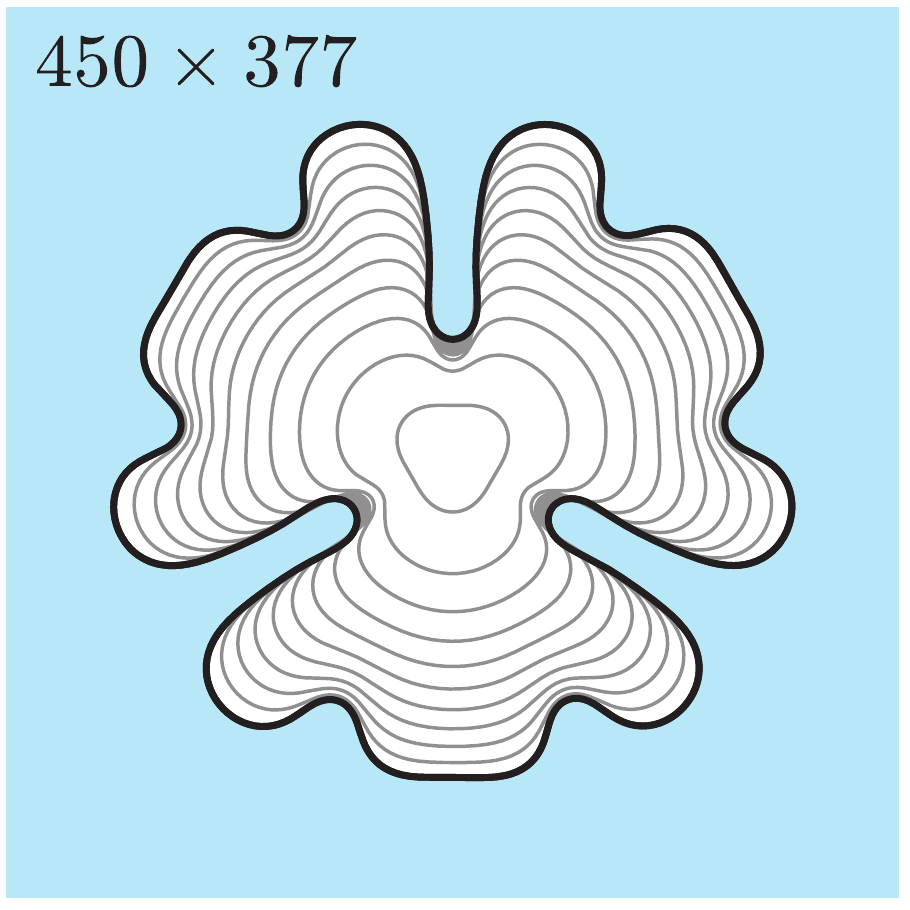}
	\includegraphics[width=0.23\linewidth]{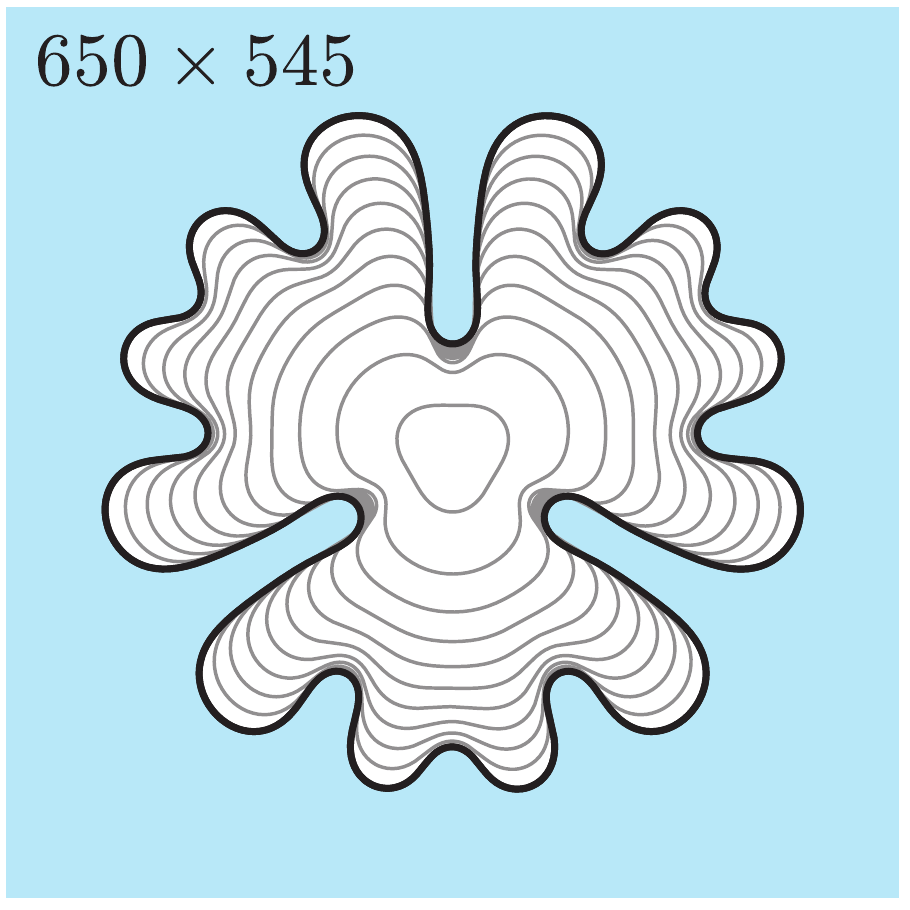}
	\includegraphics[width=0.23\linewidth]{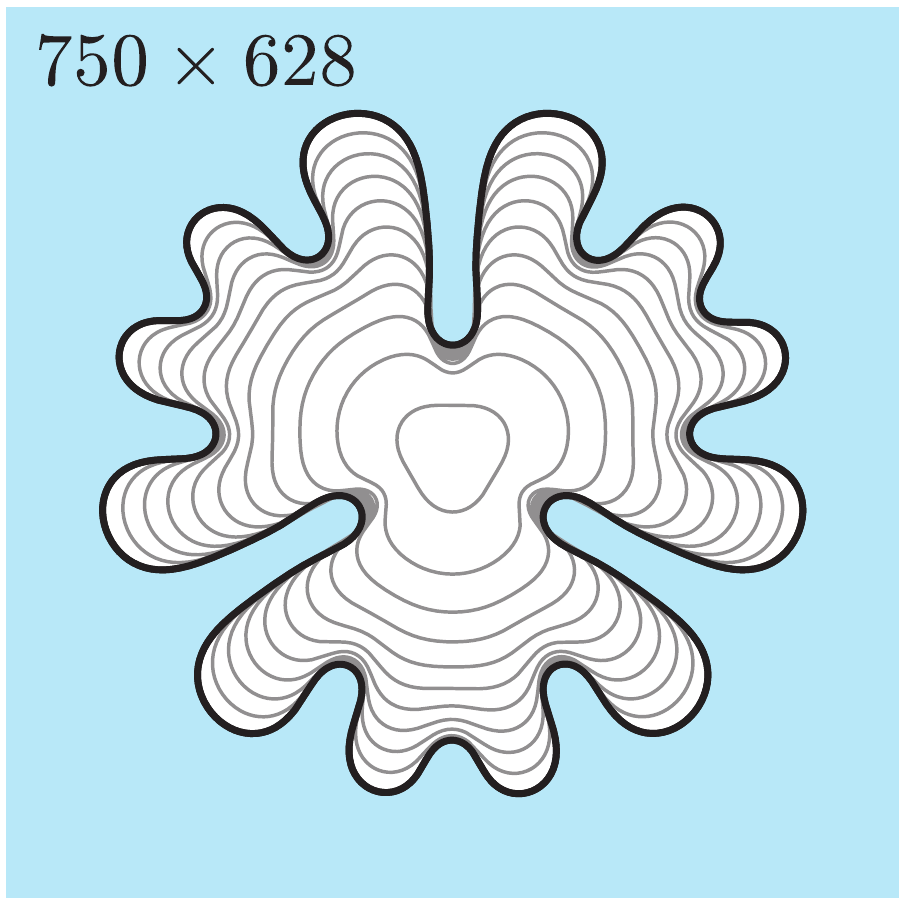}		
	\includegraphics[width=0.23\linewidth]{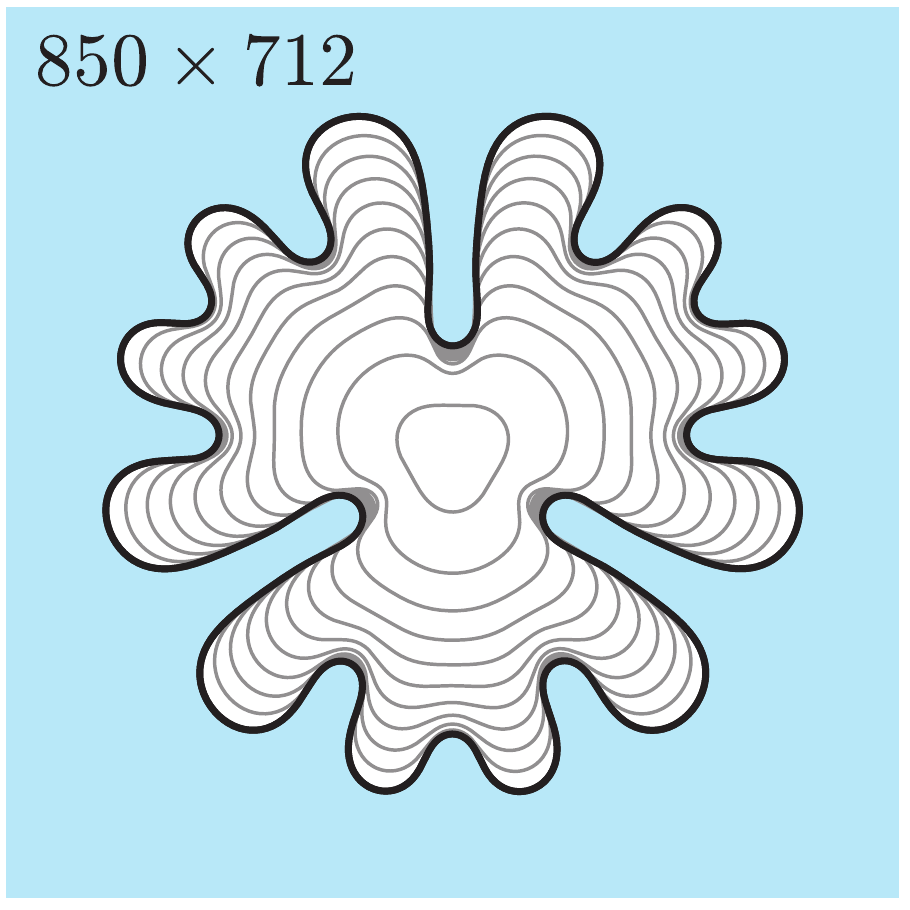}					
		\caption{Convergence test of numerical scheme for the evolution of a bubble with initial condition \eqref{eq:TestIC1} and $Q = 1.25$ mL/s, $b = 0.2$ cm, $\mu = 1$ g/(cm$\cdot$s), $\sigma = 1.5$ g/s$^2$, and $t_f = 12.4$ s. Simulations are performed on the domain $0 \le r \le 7.5$ cm and $0 \le \theta < 2 \pi$.}
		\label{fig:Figure11}
\end{figure}

To establish that our numerical scheme converges as the grid is refined, we consider the initial condition (in cm)
\begin{align} \label{eq:TestIC1}
	s(\theta, 0) = 0.9 + 0.1 \sin 3 \theta,
\end{align}
where $0 \le \theta \le 2 \pi$. Simulations are performed by employing an increasingly refined mesh with a set of parameter values that are in the range of those used elsewhere in this study.  These simulations, shown in figure \ref{fig:Figure11}, indicate that when the grid is sufficiency refined, both the size and shape of the fingers that develop are unchanged, and convergence appears to be achieved using $750 \times 628$ equally spaced nodes. Furthermore, the bubble appears to maintain three-fold symmetry over the duration of the simulation.
	
\begin{figure}
	\centering
	\includegraphics[width=0.5\linewidth]{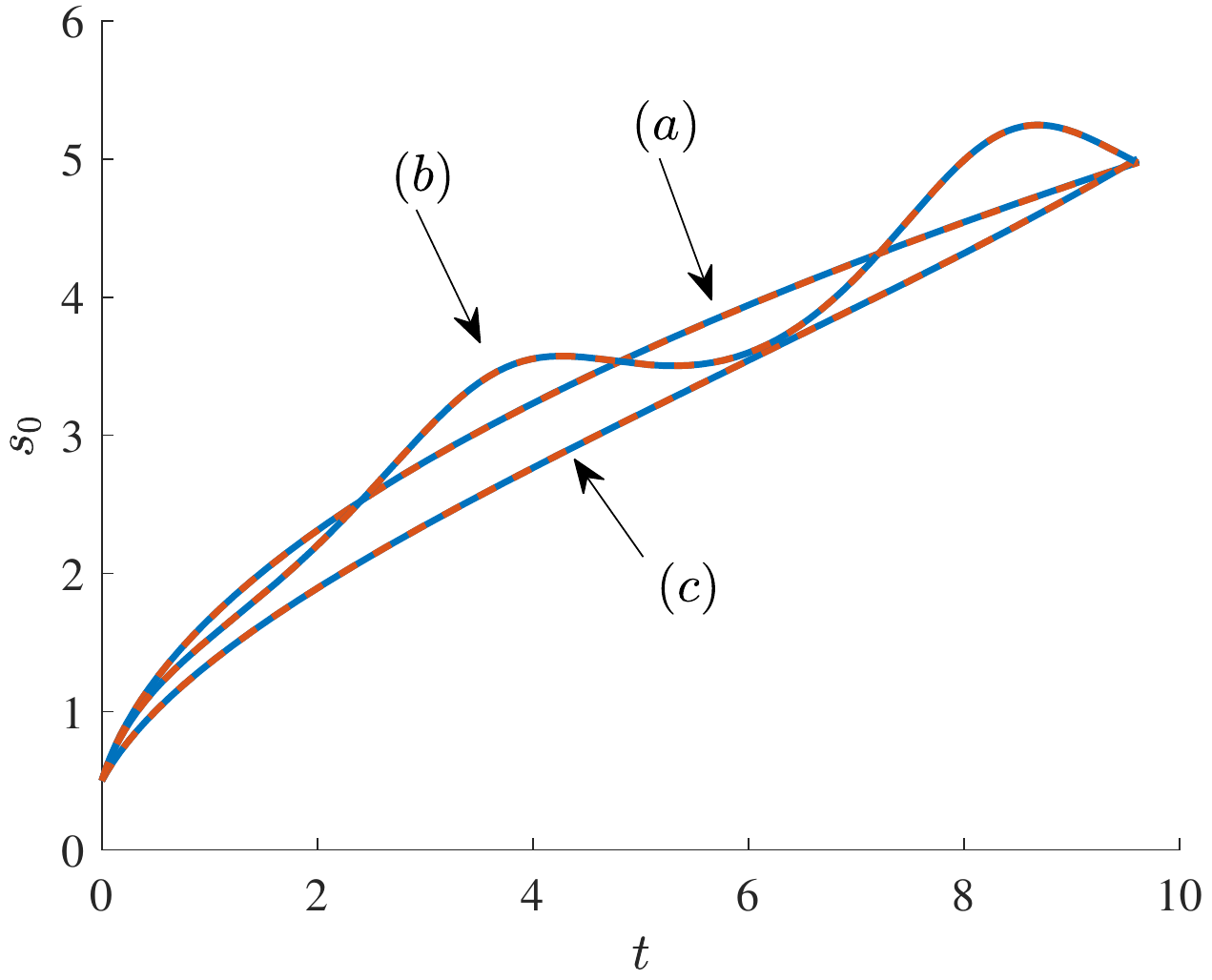}
	\caption{Comparison of the numerical solution to \eqref{eq:Model1}-\eqref{eq:Model6} (solid blue) with solution to \eqref{eq:CircleRadius} (dashed red), where the gap between the plates is $(a)$ $b = 0.2$ cm, $(b)$ $b = 0.2 + 0.04 \sin (4 \pi t / t_f)$ cm, and  $(c)$ of the form \eqref{eq:TaperedPlate} with $b_0 = 0.005$ cm, $r_0 = 7$ cm, and $\alpha = 0.053$. The final time of simulations is $t_f = 9.81$ s and the injection rate is $Q = 1.6$ mL/s.  Simulations are performed on the domain $0 \le r \le 7.5$ cm and $0 \le \theta < 2 \pi$ with $\mu = 1$ g/(cm$\cdot$s), $\sigma = 2$ g/s$^2$ and $750 \times 628$ equally spaced nodes.}
	\label{fig:radiuscomparison}
\end{figure}

We also wish to determine that our numerical scheme is able to accurately describe the behaviour of the interface for the different plate configurations considered in this article. To do so, we perform simulations where the interface is initially a circle of radius 0.5 cm, and compare the evolution of the radius with the solution to
\begin{align} \label{eq:CircleRadius}
	\dot{s}_0 = \frac{Q}{2 \pi b(s_0)s_0} - \frac{s_0}{2b(s_0)}\frac{\partial b}{\partial t}.
\end{align}		
We consider three configurations. The first is the classic configuration in which the plates are parallel and stationary. The second involves parallel plates with the distance between the two plates evolving according to (in cm)
\begin{align}
	b(t) = 0.2 + 0.04 \sin \left( \frac{4 \pi t}{t_f} \right).
\end{align}
The third configuration is for stationary plates that are linearly tapered according to \eqref{eq:TaperedPlate}. Comparing the numerical solution to \eqref{eq:Model1}-\eqref{eq:Model6} with the solution to \eqref{eq:CircleRadius} in figure \ref{fig:radiuscomparison}, we observe good agreement at this scale, suggesting that our numerical scheme accurately describes the evolution of a bubble when the gap between the plates is either spatially or temporally dependent.

\end{document}